\documentclass[aps,prl,twocolumn,groupedaddress, floatfix,longbibliography,nofootinbib,byrevtex]{revtex4}
\usepackage{graphicx}
\usepackage{dcolumn}
\usepackage{bm}
\usepackage[utf8]{inputenc}
\usepackage{amsmath,amsfonts,calc}
\usepackage[pagebackref = true, colorlinks, linkcolor = Green, citecolor = Blue, bookmarksdepth=2, linktocpage=true]{hyperref}
\usepackage[capitalize]{cleveref}
\usepackage{comment}
\usepackage[normalem]{ulem}
\usepackage[usenames,dvipsnames]{color}
\usepackage{graphicx}
\usepackage{scalerel}
\usepackage{mathtools}
\usepackage{stackengine,wasysym}
\makeatletter
\newsavebox\myboxA
\newsavebox\myboxB
\newlength\mylenA
\makeatother
\begin{document}
\title{
On the static effective Hamiltonian of a rapidly driven nonlinear system}
\author{Jayameenakshi Venkatraman}
\thanks{These two authors contributed equally}
\email{jaya.venkat@yale.edu, xu.xiao@yale.edu}
\affiliation{Department of Applied Physics and Physics, Yale University, New Haven, CT 06520, USA}
\author{Xu Xiao}
\thanks{These two authors contributed equally}
\email{jaya.venkat@yale.edu, xu.xiao@yale.edu}
\affiliation{Department of Applied Physics and Physics, Yale University, New Haven, CT 06520, USA}
\author{Rodrigo G. Corti\~nas}
\affiliation{Department of Applied Physics and Physics, Yale University, New Haven, CT 06520, USA}
\author{Alec Eickbusch}
\affiliation{Department of Applied Physics and Physics, Yale University, New Haven, CT 06520, USA}
\author{Michel H. Devoret}
\email{michel.devoret@yale.edu}
\affiliation{Department of Applied Physics and Physics, Yale University, New Haven, CT 06520, USA}
\date{\today}
\begin{abstract}
We present a recursive formula for the computation of the static effective Hamiltonian of a system under a fast-oscillating drive. Our analytical result is well-suited to symbolic calculations performed by a computer and can be implemented to arbitrary order, thus overcoming limitations of existing time-dependent perturbation methods and allowing computations that were impossible before. We also provide a simple diagrammatic tool for calculation and treat illustrative examples. By construction, our method applies directly to both quantum and classical systems; the difference is left to a low-level subroutine. This aspect sheds light on the relationship between seemingly disconnected independently developed methods in the literature and has direct applications in quantum engineering.
\end{abstract}
\maketitle 
Driven nonlinear systems display a rich spectrum of phenomena which includes bifurcation, chaos, and topological order \cite{dykman2012,guckenheimer2013}. Their behaviour is often counterintuitive and, beyond fundamental interest \cite{zurek1999}, yields important  applications. A classic example is the Kapitza pendulum \cite{landau1976}. This system inverts its equilibrium position against gravity when driven by an appropriate fast-oscillating force and serves as a model for dynamical stabilization of mechanical systems \cite{bechhoefer2021}. The potency for novel applications transcends classical physics; recently the dynamical stabilization of a Schr\"odinger cat manifold \cite{leghtas2015,lescanne2020,grimm2020}, despite its famous fragility \cite{haroche2006}, has opened new perspectives for large-scale quantum computation \cite{chamberland2020}. In the promising field of quantum simulation, Floquet engineering of potentials \cite{goldman2014} in ultracold atom experiments has permitted the realization of novel quantum systems with exotic properties unachievable otherwise \cite{wintersperger2020}. Other important related phenomena are discrete time crystals \cite{yao2017} and many-body dynamical localization \cite{nandkishore2015}, just to name a few.

In general, driven nonlinear systems do not admit closed form solutions for their time-evolution. But remarkably, under a rapid drive, their dynamics can be mapped to that generated by a time-independent effective Hamiltonian. This ``Kamiltonian" describes a \textit{slow} dynamics of the system, corrected only perturbatively by a \textit{fast} micromotion. Over the last century, different perturbation methods have been developed to construct such effective Hamiltonians and have succeeded in explaining several important nonlinear dynamical phenomena \cite{nayfeh2008,dykman2012,eckardt2017,blais2021,zeuch2020}. However, these perturbation methods can hardly be carried out beyond the lowest orders in practice and a clear understanding of the connection between many of these methods is missing \cite{rahav2003,bukov2017}. The differences are exacerbated by the wide disparity in starting points of the classical \cite{krylov1937,bogoliubov1961,nayfeh2008} and quantum methods \cite{casas2001,mirrahimi2015,eckardt2015,mikami2016,eckardt2017,xiao2021,zeuch2020,petrescu2020}.

In this article, we construct a time-independent Kamiltonian perturbatively by seeking a pertinent \emph{canonical} transformation. The small parameter of the expansion is the ratio of the typical energy of the driven system to the frequency of the driving force. We present a recursive formula for the Kamiltonian that allows its calculation to arbitrary order and is well-suited for symbolic manipulation. It can be applied indifferently to the classical and quantum cases, the change involving only a low-level subroutine of the symbolic algorithm. Our result unifies existing methods that have been developed solely in either the classical or quantum regimes.

We start with the equations governing time-evolution of the classical or quantum state vector $\rho$ under the action of a time-dependent Hamiltonian $H(t)$ that we write jointly as
\begin{align}
\label{eq:Moyal2}
    \partial_t{\rho} &= \{\!\!\{H,\rho\}\!\!\},
\end{align}
where the double bracket can be understood as
\begin{align}\label{eq:Lie-product}
\{\!\!\{H,\rho\}\!\!\} \rightarrow 
 \begin{cases}
  \{\tilde{H}, \tilde \rho\} & \mathrm{classical\; (Liouville)},\\
 \frac{1}{i\hbar}[\hat{H}, \hat{\rho}]& \mathrm{quantum\; (von\; Neumann)}.
 \end{cases}
\end{align}

Here, we have adopted the standard notation $\{\square, \square\}$ for the Poisson bracket over phase-space coordinates $q$ and $p$ and $[\square, \square]$ for the Hilbert space commutator. The state vector $\rho$ can be taken to be either a phase-space distribution $ \tilde \rho(q,p)$ or the density operator $\hat{\rho} = \sum_{x^{\prime}, x^{\prime\prime} }\rho_{x^\prime x^{\prime\prime}} |x^{\prime} \rangle \langle x^{\prime \prime}|$. Its time-evolution is governed by the Hamiltonian $H$ which is either the phase-space Hamiltonian $\tilde{H}(q, p, t)$ or the operator $\hat{H} ({q}, {p}, t)$.  We note that one can also interpret $\{\!\!\{\square, \square\}\!\!\}$ as the Moyal bracket \cite{groenewold1946,moyal1949,zachos2005}, in which case \cref{eq:Moyal2} describes the dynamics of the phase-space Wigner distribution.

In this formalism agnostic to the nature of the system, we seek a canonical transformation $\rho\!\rightarrow\!\varrho$ such that the time-evolution of $\varrho$ is governed, in the transformed frame, by the sought-after time-independent Kamiltonian. We thus consider the Lie transformation generated by a time-dependent generator $S$ and parametrized by $\epsilon$,
\begin{align}\label{eq:LT}
    \begin{split}
  \varrho &= e^{\epsilon L_S}\rho=\sum_{k=0} \frac{\epsilon^kL^k_S}{k!} \rho;\\
     &= \rho + \epsilon\{\!\!\{S, \rho\}\!\!\}  + \frac{\epsilon^2}{2!} \{\!\!\{ S, \{\!\!\{S, \rho \}\!\!\}\! \}\!\!\} +\cdots,
 \end{split}
\end{align}
where $L_S \square \! = \!\{\!\!\{S, \square\}\!\!\}$ is the Lie derivative \cite{dragt1976} generated by $S$. Here $S$ is either a real phase-space function $\tilde{S}(q, p, t)$ or an Hermitian operator $\hat{S}(q, p, t)$. Equivalently, the transformed state $\varrho$ is the solution to the differential equation $\partial_\epsilon \varrho = \!\{\!\!\{S, \varrho\}\!\!\}$, with initial condition $\varrho(\epsilon\!=\!0)\!=\!\rho$.

\begin{figure}[t!]
\includegraphics{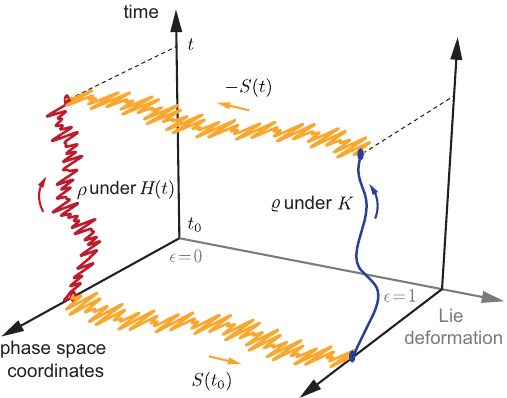}
\caption{Time-evolution of the state vector in the transformed and un-transformed frames. The red curve represents the complicated time-evolution of $\rho$ under the time-dependent $H$. The blue curve represents the simpler time-evolution of $\varrho$ under the time-independent $K$. The transformation is exact. Under a sufficiently fast oscillating drive, the fast micromotion captured by $S$ can be neglected and $K$ can be taken to generate the time-evolution of $\rho$ in the un-transformed frame.}
\label{fig:flow}
\end{figure}
In the transformed representation, the dynamics obeys formally (\ref{eq:Moyal2}) as $\partial_t{\varrho}\! =\! \{\!\!\{K,\varrho\}\!\!\}$, with the Kamiltonian $K$ given by
\begin{align}\label{eq:K}
    K = e^{L_S} H + \int_{0}^{1}d\epsilon \,e^{\epsilon L_S} \dot{S};
\end{align}
see Supplementary material Section A for the derivation. Note that in the quantum case, \cref{eq:K} reduces to the familiar expression $\hat{K} = \hat{U}^{\dagger} (\hat{H} - i \hbar \partial_t) \hat{U}$ with $\hat{U} = e^{-\hat{S}/i \hbar}.$

We now carry out a perturbative expansion generated by $S$, while imposing that $K$ is rendered time-independent. The transformation of the time-evolution from $\rho \rightarrow \varrho$ is represented schematically in \cref{fig:flow} and yields
 \begin{align}
 \begin{split}
 \label{eq:W-Lie2}
    \rho &= \mathcal{T}e^{\int_{t_0}^t dt^\prime\, L_{H(t^\prime)}}\rho_0\\
     &= e^{L_{-S(t)}}   e^{L_K (t-t_0)} e^{L_{S(t_0)}} \rho_0,
\end{split}
 \end{align}
 where $\mathcal T$ is the time-ordering operator and $\rho_0$ is the initial state. The time-evolution of $\rho$ under $H$ (a Lie transformation generated by $H$ and parametrized by $t$) can be understood as being decomposed into three successive Lie transformations generated by $S(t_0)$,  $K$, and $-S(t)$. Under this decomposition, the time-ordering operator drops out in the time-evolution under $K$, providing an important simplification. 
 
 To carry out the perturbative expansion, we consider the Hamiltonian
 \begin{align}
     H(t) = \sum_{m \in \mathbb{Z}} H_{m} e^{i m \omega t}
 \end{align}
 with period $T = 2 \pi/ \omega$. For the perturbative treatment to be valid, the rate of evolution under any one $H_m$ needs to be much smaller than $\omega$. In the case of an unbounded Hamiltonian, either quantum or classical, the corresponding space will require truncation. We focus on the case of a periodic drive for simplicity but we note that our treatment can be generalized to include quasiperiodic or non-monochromatic drives; see Supplement section B III. We take as an ansatz for $S$ and $K$
\begin{align}
\label{eq:S}
S&=\sum_{n\in \mathbb{N}} S^{(n)},& {K} &= \sum_{n \in \mathbb{N}}{K}^{(n)},
\end{align}
where we take $S^{(0)} = 0$ and the $n^{\mathrm{th}}$ terms to be of order $n$ in the expansion parameter, here taken to be $1/\omega$. Substituting Eqs. (\ref{eq:S}) into \cref{eq:K} separates the problem into orders of $1/\omega$. At each order, $K^{(n)}$ can further be expressed as a sum of terms generated by a Lie series as in \cref{eq:LT}, which we write as $K^{(n)} = \sum_{k} K^{(n)}_{[k]}$. Demanding $K$  to be time-independent to all orders, we find, after a few lines of algebra, the following coupled recursive formulas:
\begin{subequations}\label{eq:recursive}
\begin{align}
\label{eq:recursive-K}
{K}_{[k]}^{(n)}\;\;\; &= \begin{cases}
H & n = k =0\\
\dot{S}^{(n+1)} + L_{S^{(n)}} H & k=1\\
\sum\limits_{m=0}^{n-1} \frac{1}{k}L_{S^{(n-m)}} {K}_{[k-1]}^{(m)} & 1<k\le n+1\\
0 & \text{otherwise},
\end{cases}\\[4pt]
{S}^{(n+1)} &= \begin{cases}\label{eq:recursive-S}
-\int dt\,\mathrm{\textbf{osc}} (H) & n=0\\[2pt]
-\int dt\,\mathrm{\textbf{osc}} \Big(L_{S^{(n)}} H  & \; \\
\;\;+ \sum\limits_{k>1}^{n+1} \sum \limits_{m=0}^{n-1} \frac{1}{k}L_{S^{(n-m)}} {K}_{[k-1]}^{(m)}\Big) & n > 0,
\end{cases}
\end{align}
where $\mathrm{\textbf{osc}} (f) \coloneqq f - \overline{f}$, and $\overline{f} = \frac{1}{T}\int_{0}^{T} dt \, f$. \color{black} Note that $H$ is taken to be of order zero in the perturbation parameter, but this hypothesis can be relaxed in a more elaborate treatment; see Supplement section B III.
\end{subequations}

By construction, taking the time-derivative of \cref{eq:recursive-S}, substituting the result into \cref{eq:recursive-K}, and summing over $k$ yields a time-independent $K^{(n)}$. All in all, the computations of $K$ and $S$ are interleaved so that the computation of $K^{(n)}$ requires as an input the value of $S^{(m \le n)}$. Demanding the time-independence of $K^{(n)}$ fixes $\dot{S}^{(n+1)}$, allowing the recursion to be carried out to the next order. The coupled recursive formula in \cref{eq:recursive} constructs, as announced, $S$ and $K$ order-by-order.
 
\begin{figure}[t!]
\includegraphics[width = \columnwidth]{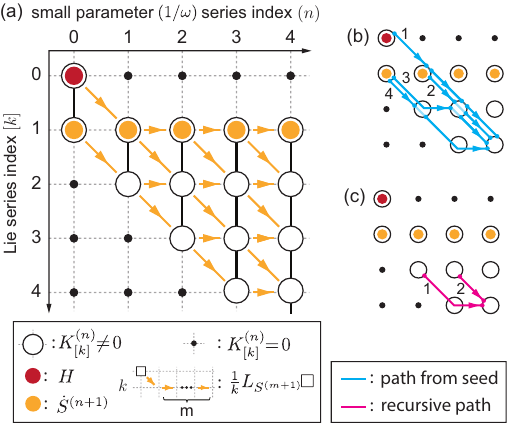}
\caption{(a) Grid for the diagrammatic construction of $K$ and $S$. Colored circles represent the seeds generating the series to all orders. (b) As an example, all the paths contributing to the calculation of $K^{(3)}_{[3]}$ are highlighted. (c) Here, only the subpaths contributing to the recursive expression of the same term are highlighted.}
\label{fig:grid}
\end{figure}
The mathematical structure of the recursive formula \cref{eq:recursive} is shown diagrammatically in \cref{fig:grid}, as we now explain. The figure consists of a grid indexed by the integers $n$ and $k$. The grid supports a graph. Each node $(n, k)$ of the graph corresponds to a summand $K^{(n)}_{[k]}$, and the colored ones represent the ``seeds'' of the calculation. The summand $K^{(n)}_{[k]}$ is itself a sum of terms, each corresponding to a path connecting the node $(n, k)$ to a seed. Evaluating a path corresponds to taking Lie derivatives over $H$ or $\dot{S}^{(n+1)}$ as dictated by the seed color. The rule is that each Lie derivative is specified by a valid subpath, which must start ``downwards'' and, when followed by $m$ horizontal edges at row $k$, contributes with $L_{S^{(m+1)}}/k$. Finally, if the considered node is itself colored, either $H$ or $\dot{S}^{(n+1)}$ must be added to the sum. We note that our grid construction is inspired by \cite{deprit1969},
where the construction is limited to completely classical and time-independent systems.

Let us discuss, as an example, how $K^{(3)}_{[3]}$ is evaluated from the figure. As indicated by panel \cref{fig:grid}(b), $K^{(3)}_{[3]}$ contains only four terms corresponding to the concatenations of the valid subpaths (in blue). The sum reads
\begin{align*}
\begin{split}
    K^{(3)}_{[3]} ={}&\frac{L_{S^{(1)}}}{1}\frac{L_{S^{(1)}}}{2}\frac{L_{S^{(1)}}}{3} H + \frac{L_{S^{(1)}}}{2}\frac{L_{S^{(1)}}}{3} \dot{S}^{(2)} \\
&{}+\frac{L_{S^{(1)}}}{2} \frac{L_{S^{(2)}}}{3} \dot{S}^{(1)} +\frac{L_{S^{(2)}}}{2} \frac{L_{S^{(1)}}}{3} \dot{S}^{(1)},
\end{split}
\end{align*}
where the terms are ordered as enumerated in the figure.

Alternatively, one could have expressed $K^{(3)}_{[3]}$ recursively by directly applying \cref{eq:recursive-K}. The computation of $K^{(3)}_{[3]}$ then involves only the two pink subpaths shown in \cref{fig:grid}(c) and yields
\begin{align*}
    K^{(3)}_{[3]} = \frac{L_{S^{(2)}}}{3} K^{(1)}_{[2]} + \frac{L_{S^{(1)}}}{3} K^{(2)}_{[2]}.
\end{align*}

At this stage, once all entries of the $n^{\mathrm{th}}$ column are computed, the calculation proceeds by demanding the time-independence of $K^{(n)}$ computed as their row-sum over column $n$ and represented by the vertical bold lines in \cref{fig:grid}(a). This is required by \cref{eq:recursive-S}. For the column $n=3$ the algorithm yields
\begin{align*}
    S^{(4)} = -\int dt \, \mathrm{\textbf{osc}} \left(L_{S^{(3)}} H + K^{(3)}_{[2]} + K^{(3)}_{[3]} + K^{(3)}_{[4]} \right),
\end{align*}
which is a necessary ingredient to compute $K^{(5)}$ and so, the calculation proceeds. 

The formulation is further illustrated by treating as concrete examples the classical and quantum Kapitza pendulum and the driven Duffing oscillator in Section B of the Supplementary material. Moreover, we also present in section B III the remarkable agreement with recent experiments \cite{eickbusch2021} that independently measures the effective Hamiltonian of a driven transmon-cavity superconducting circuit and has been only analyzed numerically until now.

We stress again that \cref{eq:recursive} is agnostic to the choice of Lie bracket in \cref{eq:Lie-product}. A Lie-based formulation is thus well-suited to unify seemingly disconnected perturbation methods, in particular those that, to be linked, require the quantum-classical correspondence to be made explicit.

Exploiting this property, we now turn to discuss the connection between several time-dependent perturbation methods developed independently. We consider their common starting point to  be the additive ansatz
\begin{align}
\label{eq:ansatz}
Z=\mathcal Z+\zeta(\mathcal Z,^c\!\!\mathcal Z),
\end{align}
where $Z$ is the state variable, $^c \!\mathcal Z$ is the conjugate variable to $\mathcal Z$ and $\zeta$ is a correction. For time-varying problems, it is customary to take $\mathcal Z$ to describe the slow dynamics and the correction to describe the fast dynamics with vanishing time-average ($\bar{\zeta}\! =\! 0$). To compute $\zeta$, different methods proceed in vastly different ways: the classical ones rely on partial derivatives of phase-space functions \cite{krylov1937,bogoliubov1961,landau1976,rahav2003}; this draws a line separating them from the quantum methods which rely on matrix products \cite{rahav2003,eckardt2015,casas2001,mirrahimi2015,mikami2016,xiao2021}. These procedural differences hide their shared structure.

We uncover the connection between these methods by realizing that the procedural differences stem from \emph{premature} specifications of a particular Lie bracket.
Identifying this feature allows us to relate the correction $\zeta$, specified by each method, to the generator $S$ of the Lie transformation as $\zeta\!\! =\!\! (e^{L_{-S}}\!\! -\!\! I)\mathcal Z$. 
In other words, the ansatz in \cref{eq:ansatz} corresponds to the additive representation of the exponential map in \cref{eq:LT}. It follows that, if carried out to all orders, these methods correspond to invertible canonical transformations. See Section C of the Supplementary material for a detailed discussion on the relationship between the different ansätzs.

\begin{figure}[t!]
\includegraphics{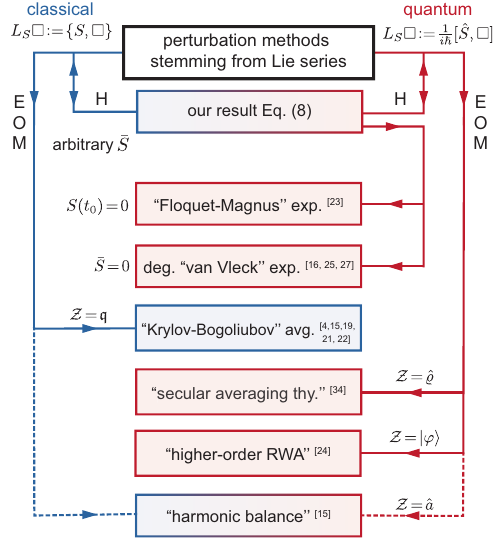}
\caption{Relationship between various perturbation methods for systems submitted to a time-dependent oscillatory drive.  All the methods in the tree can be seen as falling under the umbrella of Lie series. The methods can be divided between classical (blue lines-left) and quantum (red lines-right) ones. Other forks separate the methods based on the equation of motion (EOM) from the methods based on the Hamiltonian (H). The dashed lines refer to a class of methods whose effective Hamiltonians have not yet been identified and they are left for future work. The methods labeled by a $\mathcal{Z}$ symbol are additive ansatz-based unlike the methods containing $S$ which can be thought as multiplicative. For the Hamiltonian methods, the associated integration constant for $S$ in \cref{eq:recursive-S} is specified as a condition. The acronym RWA stands for rotating wave approximation. The double arrow refers to a bidirectional relationship.}
\label{fig:tree}
\end{figure}

In \cref{fig:tree}, we show how we can collect seemingly disconnected perturbation methods and unify them under the umbrella of Lie series. The two main branches, colored in red (right) and blue (left), group the quantum and classical methods. The ones based on equations of motion (EOM) correspond to different choices of $Z$ in \cref{eq:ansatz}. Among the quantum methods, secular averaging theory (SAT) \cite{buishvili1979} corresponds to $Z\! \!=\!\! \hat{\rho}$. The perturbative expansion can also be developed at the level of the wave function $Z\!\!=\! |\phi \rangle$, allowing for the derivation of higher-order rotating wave approximations (RWA) \cite{mirrahimi2015}. In this case, the ansatz is $\mathcal{Z}\!\! =\! |\varphi\rangle$ and $\zeta\!\!=\!\hat{\delta} |\varphi \rangle$ and it can be mapped to our approach by taking $\zeta\!\! =\!\! (e^{-\hat{S}/ i \hbar}\! - 1) |\varphi \rangle$. This last relation can be understood by noting that in the quantum case \cref{eq:LT} reduces to $e^{L_S} \hat \rho\! =\! e^{-\hat{S}/i\hbar} \hat \rho e^{\hat{S}/i\hbar}$. Classically, the Krylov-Bogoliubov (KB) method \cite{krylov1937,bogoliubov1961,rahav2003,landau1976,nayfeh2008} averages the equation of motion of the position coordinate $\mathcal Z \!\!=\! \mathfrak q$ and $\zeta \!\! =\! \zeta (\mathfrak q, \mathfrak p, t)$, where $\mathfrak p$ is the conjugate momentum, and it maps to our approach with $\zeta \!\! =\!\! (e^{L_S} - I) \mathfrak q$ (the change of sign in $S$ is simply a change from the active representation used so far to the passive representation used in KB).

Besides the EOM methods, we also include the most common quantum Hamiltonian methods in the genealogy of \cref{fig:tree}. They are characterized by the utilization of Floquet theorem \cite{shirley1965}, which guarantees the existence of a unitary transformation rendering the Kamiltonian time-independent. They map naturally to the exponential representation discussed in this work. We find that using the freedom in the integration constant of \cref{eq:recursive-S}, as specified in \cref{fig:tree}, we recover the so-called Floquet-Magnus expansion $(S(t_0) = 0)$ \cite{casas2001} or the van Vleck expansion ($\bar{S} = 0$) \cite{rahav2003,eckardt2015,xiao2021}. This gives formal ground to the observations made in \cite{bukov2017,mikami2016} on the connection between these two methods.

Note that, with the exception of the Floquet-Magnus expansion, all the aforementioned methods were limited to the lowest orders. Instead, our symbolic formula can be readily used to carry out the calculation to arbitrary order. We illustrate this by taking as an example the widely employed van Vleck expansion.  We have explicitly written a symbolic algebra algorithm, made available in \cite{vv2021}, and used it to explicitly display the expansion up to order five, automatedly (see Section D of the Supplementary material). To the best of our knowledge, the expansion could only be found to order three in the literature \cite{mikami2016} until now. 

In this article, we developed a perturbation method to treat rapidly driven nonlinear systems yielding a double coupled recursive formula well-suited for symbolic computation. The construction can thus be carried out in an automated way to arbitrary order. We achieve this by constructing a canonical transformation that explicitly decouples the relevant dynamics, governed by a time-independent effective Hamiltonian, from the complicated micromotion. Our treatment is completely agnostic to the classical or quantum nature of the problem and sheds light on the longstanding discussion of the relationship between well-known perturbation methods developed independently of each other. We remark that our formula can be generalized to treat Hamiltonians of arbitrary order in the perturbation parameter. This is particularly powerful when treating parametric processes in driven nonlinear bosonic oscillators \cite{xiao20212}, paving the way for Hamiltonian engineering in superconducting quantum circuits. Finally, we note that there is a strong relationship between time-periodic (Floquet) and space-periodic (Bloch) systems, and thus, our recursive formula can be adapted to compute the Schrieffer-Wolff transformation \cite{bravyi2011} to arbitrary order and potentially get new results in quantum many-body problems. 

We acknowledge Shoumik Chowdhury, Mark Dykman, Steve Girvin, Leonid Glazman, Pratyush Sarkar, and Yaxing Zhang for insightful discussions. We also thank Hideo Aoki and Sota Kitamura for a discussion on their classification of terms in the van Vleck expansion. This research was supported by the National Science Foundation (NSF) under grant number 1941583, the Air Force Office of Scientific Research (AFOSR) under grant number FA9550-19-1-0399, and the Army Research Office (ARO) under grant number W911NF-18-1-0212. The views and conclusions contained in this document are those of the authors and should not be interpreted as representing the official policies, either expressed or implied, of the NSF, the AFOSR, the ARO, or the U.S. Government. The U.S. Government is authorized to reproduce and distribute reprints
for Government purposes notwithstanding any copyright notation herein.
\bibliography{bibmain}

\begin{thebibliography}{39}
\expandafter\ifx\csname natexlab\endcsname\relax\def\natexlab#1{#1}\fi
\expandafter\ifx\csname bibnamefont\endcsname\relax
  \def\bibnamefont#1{#1}\fi
\expandafter\ifx\csname bibfnamefont\endcsname\relax
  \def\bibfnamefont#1{#1}\fi
\expandafter\ifx\csname citenamefont\endcsname\relax
  \def\citenamefont#1{#1}\fi
\expandafter\ifx\csname url\endcsname\relax
  \def\url#1{\texttt{#1}}\fi
\expandafter\ifx\csname urlprefix\endcsname\relax\def\urlprefix{URL }\fi
\providecommand{\bibinfo}[2]{#2}
\providecommand{\eprint}[2][]{\url{#2}}

\bibitem[{\citenamefont{Dykman}(2012)}]{dykman2012}
\bibinfo{author}{\bibfnamefont{M.}~\bibnamefont{Dykman}},
  \emph{\bibinfo{title}{Fluctuating nonlinear oscillators: from nanomechanics
  to quantum superconducting circuits}} (\bibinfo{publisher}{Oxford University
  Press}, \bibinfo{year}{2012}).

\bibitem[{\citenamefont{Guckenheimer and Holmes}(2013)}]{guckenheimer2013}
\bibinfo{author}{\bibfnamefont{J.}~\bibnamefont{Guckenheimer}}
  \bibnamefont{and} \bibinfo{author}{\bibfnamefont{P.}~\bibnamefont{Holmes}},
  \emph{\bibinfo{title}{Nonlinear oscillations, dynamical systems, and
  bifurcations of vector fields}}, vol.~\bibinfo{volume}{42}
  (\bibinfo{publisher}{Springer Science \& Business Media},
  \bibinfo{year}{2013}).

\bibitem[{\citenamefont{Zurek and Paz}(1999)}]{zurek1999}
\bibinfo{author}{\bibfnamefont{W.~H.} \bibnamefont{Zurek}} \bibnamefont{and}
  \bibinfo{author}{\bibfnamefont{J.~P.} \bibnamefont{Paz}}, in
  \emph{\bibinfo{booktitle}{Epistemological and Experimental Perspectives on
  Quantum Physics}} (\bibinfo{publisher}{Springer}, \bibinfo{year}{1999}), pp.
  \bibinfo{pages}{167--177}.

\bibitem[{\citenamefont{Landau and Lifshitz}(1976)}]{landau1976}
\bibinfo{author}{\bibfnamefont{L.~D.} \bibnamefont{Landau}} \bibnamefont{and}
  \bibinfo{author}{\bibfnamefont{E.~M.} \bibnamefont{Lifshitz}},
  \emph{\bibinfo{title}{Mechanics: Volume 1}}, vol.~\bibinfo{volume}{1}
  (\bibinfo{publisher}{Butterworth-Heinemann}, \bibinfo{year}{1976}).

\bibitem[{\citenamefont{Bechhoefer}(2021)}]{bechhoefer2021}
\bibinfo{author}{\bibfnamefont{J.}~\bibnamefont{Bechhoefer}},
  \emph{\bibinfo{title}{Control Theory for Physicists}}
  (\bibinfo{publisher}{Cambridge University Press}, \bibinfo{year}{2021}).

\bibitem[{\citenamefont{Leghtas et~al.}(2015)\citenamefont{Leghtas, Touzard,
  Pop, Kou, Vlastakis, Petrenko, Sliwa, Narla, Shankar, Hatridge
  et~al.}}]{leghtas2015}
\bibinfo{author}{\bibfnamefont{Z.}~\bibnamefont{Leghtas}},
  \bibinfo{author}{\bibfnamefont{S.}~\bibnamefont{Touzard}},
  \bibinfo{author}{\bibfnamefont{I.~M.} \bibnamefont{Pop}},
  \bibinfo{author}{\bibfnamefont{A.}~\bibnamefont{Kou}},
  \bibinfo{author}{\bibfnamefont{B.}~\bibnamefont{Vlastakis}},
  \bibinfo{author}{\bibfnamefont{A.}~\bibnamefont{Petrenko}},
  \bibinfo{author}{\bibfnamefont{K.~M.} \bibnamefont{Sliwa}},
  \bibinfo{author}{\bibfnamefont{A.}~\bibnamefont{Narla}},
  \bibinfo{author}{\bibfnamefont{S.}~\bibnamefont{Shankar}},
  \bibinfo{author}{\bibfnamefont{M.~J.} \bibnamefont{Hatridge}},
  \bibnamefont{et~al.}, \bibinfo{journal}{Science}
  \textbf{\bibinfo{volume}{347}}, \bibinfo{pages}{853} (\bibinfo{year}{2015}).

\bibitem[{\citenamefont{Lescanne et~al.}(2020)\citenamefont{Lescanne, Villiers,
  Peronnin, Sarlette, Delbecq, Huard, Kontos, Mirrahimi, and
  Leghtas}}]{lescanne2020}
\bibinfo{author}{\bibfnamefont{R.}~\bibnamefont{Lescanne}},
  \bibinfo{author}{\bibfnamefont{M.}~\bibnamefont{Villiers}},
  \bibinfo{author}{\bibfnamefont{T.}~\bibnamefont{Peronnin}},
  \bibinfo{author}{\bibfnamefont{A.}~\bibnamefont{Sarlette}},
  \bibinfo{author}{\bibfnamefont{M.}~\bibnamefont{Delbecq}},
  \bibinfo{author}{\bibfnamefont{B.}~\bibnamefont{Huard}},
  \bibinfo{author}{\bibfnamefont{T.}~\bibnamefont{Kontos}},
  \bibinfo{author}{\bibfnamefont{M.}~\bibnamefont{Mirrahimi}},
  \bibnamefont{and} \bibinfo{author}{\bibfnamefont{Z.}~\bibnamefont{Leghtas}},
  \bibinfo{journal}{Nature Physics} \textbf{\bibinfo{volume}{16}},
  \bibinfo{pages}{509} (\bibinfo{year}{2020}).

\bibitem[{\citenamefont{Grimm et~al.}(2020)\citenamefont{Grimm, Frattini, Puri,
  Mundhada, Touzard, Mirrahimi, Girvin, Shankar, and Devoret}}]{grimm2020}
\bibinfo{author}{\bibfnamefont{A.}~\bibnamefont{Grimm}},
  \bibinfo{author}{\bibfnamefont{N.~E.} \bibnamefont{Frattini}},
  \bibinfo{author}{\bibfnamefont{S.}~\bibnamefont{Puri}},
  \bibinfo{author}{\bibfnamefont{S.~O.} \bibnamefont{Mundhada}},
  \bibinfo{author}{\bibfnamefont{S.}~\bibnamefont{Touzard}},
  \bibinfo{author}{\bibfnamefont{M.}~\bibnamefont{Mirrahimi}},
  \bibinfo{author}{\bibfnamefont{S.~M.} \bibnamefont{Girvin}},
  \bibinfo{author}{\bibfnamefont{S.}~\bibnamefont{Shankar}}, \bibnamefont{and}
  \bibinfo{author}{\bibfnamefont{M.~H.} \bibnamefont{Devoret}},
  \bibinfo{journal}{Nature} \textbf{\bibinfo{volume}{584}},
  \bibinfo{pages}{205} (\bibinfo{year}{2020}).

\bibitem[{\citenamefont{Haroche and Raimond}(2006)}]{haroche2006}
\bibinfo{author}{\bibfnamefont{S.}~\bibnamefont{Haroche}} \bibnamefont{and}
  \bibinfo{author}{\bibfnamefont{J.-M.} \bibnamefont{Raimond}},
  \emph{\bibinfo{title}{Exploring the quantum: atoms, cavities, and photons}}
  (\bibinfo{publisher}{Oxford university press}, \bibinfo{year}{2006}).

\bibitem[{\citenamefont{Chamberland et~al.}(2020)\citenamefont{Chamberland,
  Noh, Arrangoiz-Arriola, Campbell, Hann, Iverson, Putterman, Bohdanowicz,
  Flammia, Keller et~al.}}]{chamberland2020}
\bibinfo{author}{\bibfnamefont{C.}~\bibnamefont{Chamberland}},
  \bibinfo{author}{\bibfnamefont{K.}~\bibnamefont{Noh}},
  \bibinfo{author}{\bibfnamefont{P.}~\bibnamefont{Arrangoiz-Arriola}},
  \bibinfo{author}{\bibfnamefont{E.~T.} \bibnamefont{Campbell}},
  \bibinfo{author}{\bibfnamefont{C.~T.} \bibnamefont{Hann}},
  \bibinfo{author}{\bibfnamefont{J.}~\bibnamefont{Iverson}},
  \bibinfo{author}{\bibfnamefont{H.}~\bibnamefont{Putterman}},
  \bibinfo{author}{\bibfnamefont{T.~C.} \bibnamefont{Bohdanowicz}},
  \bibinfo{author}{\bibfnamefont{S.~T.} \bibnamefont{Flammia}},
  \bibinfo{author}{\bibfnamefont{A.}~\bibnamefont{Keller}},
  \bibnamefont{et~al.}, \bibinfo{journal}{arXiv preprint arXiv:2012.04108}
  (\bibinfo{year}{2020}).

\bibitem[{\citenamefont{Goldman and Dalibard}(2014)}]{goldman2014}
\bibinfo{author}{\bibfnamefont{N.}~\bibnamefont{Goldman}} \bibnamefont{and}
  \bibinfo{author}{\bibfnamefont{J.}~\bibnamefont{Dalibard}},
  \bibinfo{journal}{Phys. Rev. X} \textbf{\bibinfo{volume}{4}},
  \bibinfo{pages}{031027} (\bibinfo{year}{2014}).

\bibitem[{\citenamefont{Wintersperger et~al.}(2020)\citenamefont{Wintersperger,
  Braun, {\"U}nal, Eckardt, Di~Liberto, Goldman, Bloch, and
  Aidelsburger}}]{wintersperger2020}
\bibinfo{author}{\bibfnamefont{K.}~\bibnamefont{Wintersperger}},
  \bibinfo{author}{\bibfnamefont{C.}~\bibnamefont{Braun}},
  \bibinfo{author}{\bibfnamefont{F.~N.} \bibnamefont{{\"U}nal}},
  \bibinfo{author}{\bibfnamefont{A.}~\bibnamefont{Eckardt}},
  \bibinfo{author}{\bibfnamefont{M.}~\bibnamefont{Di~Liberto}},
  \bibinfo{author}{\bibfnamefont{N.}~\bibnamefont{Goldman}},
  \bibinfo{author}{\bibfnamefont{I.}~\bibnamefont{Bloch}}, \bibnamefont{and}
  \bibinfo{author}{\bibfnamefont{M.}~\bibnamefont{Aidelsburger}},
  \bibinfo{journal}{Nature Physics} \textbf{\bibinfo{volume}{16}},
  \bibinfo{pages}{1058} (\bibinfo{year}{2020}).

\bibitem[{\citenamefont{Yao et~al.}(2017)\citenamefont{Yao, Potter, Potirniche,
  and Vishwanath}}]{yao2017}
\bibinfo{author}{\bibfnamefont{N.~Y.} \bibnamefont{Yao}},
  \bibinfo{author}{\bibfnamefont{A.~C.} \bibnamefont{Potter}},
  \bibinfo{author}{\bibfnamefont{I.-D.} \bibnamefont{Potirniche}},
  \bibnamefont{and}
  \bibinfo{author}{\bibfnamefont{A.}~\bibnamefont{Vishwanath}},
  \bibinfo{journal}{Phys. Rev. Lett.} \textbf{\bibinfo{volume}{118}},
  \bibinfo{pages}{030401} (\bibinfo{year}{2017}).

\bibitem[{\citenamefont{Nandkishore and Huse}(2015)}]{nandkishore2015}
\bibinfo{author}{\bibfnamefont{R.}~\bibnamefont{Nandkishore}} \bibnamefont{and}
  \bibinfo{author}{\bibfnamefont{D.~A.} \bibnamefont{Huse}},
  \bibinfo{journal}{Annu. Rev. Condens. Matter Phys.}
  \textbf{\bibinfo{volume}{6}}, \bibinfo{pages}{15} (\bibinfo{year}{2015}).

\bibitem[{\citenamefont{Nayfeh}(2008)}]{nayfeh2008}
\bibinfo{author}{\bibfnamefont{A.~H.} \bibnamefont{Nayfeh}},
  \emph{\bibinfo{title}{Perturbation methods}} (\bibinfo{publisher}{John Wiley
  \& Sons}, \bibinfo{year}{2008}).

\bibitem[{\citenamefont{Eckardt}(2017)}]{eckardt2017}
\bibinfo{author}{\bibfnamefont{A.}~\bibnamefont{Eckardt}},
  \bibinfo{journal}{Reviews of Modern Physics} \textbf{\bibinfo{volume}{89}},
  \bibinfo{pages}{011004} (\bibinfo{year}{2017}).

\bibitem[{\citenamefont{Blais et~al.}(2021)\citenamefont{Blais, Grimsmo,
  Girvin, and Wallraff}}]{blais2021}
\bibinfo{author}{\bibfnamefont{A.}~\bibnamefont{Blais}},
  \bibinfo{author}{\bibfnamefont{A.~L.} \bibnamefont{Grimsmo}},
  \bibinfo{author}{\bibfnamefont{S.~M.} \bibnamefont{Girvin}},
  \bibnamefont{and} \bibinfo{author}{\bibfnamefont{A.}~\bibnamefont{Wallraff}},
  \bibinfo{journal}{Rev. Mod. Phys.} \textbf{\bibinfo{volume}{93}},
  \bibinfo{pages}{025005} (\bibinfo{year}{2021}).

\bibitem[{\citenamefont{Zeuch et~al.}(2020)\citenamefont{Zeuch, Hassler, Slim,
  and DiVincenzo}}]{zeuch2020}
\bibinfo{author}{\bibfnamefont{D.}~\bibnamefont{Zeuch}},
  \bibinfo{author}{\bibfnamefont{F.}~\bibnamefont{Hassler}},
  \bibinfo{author}{\bibfnamefont{J.~J.} \bibnamefont{Slim}}, \bibnamefont{and}
  \bibinfo{author}{\bibfnamefont{D.~P.} \bibnamefont{DiVincenzo}},
  \bibinfo{journal}{Annals of Physics} \textbf{\bibinfo{volume}{423}},
  \bibinfo{pages}{168327} (\bibinfo{year}{2020}), ISSN
  \bibinfo{issn}{0003-4916}.

\bibitem[{\citenamefont{Rahav et~al.}(2003)\citenamefont{Rahav, Gilary, and
  Fishman}}]{rahav2003}
\bibinfo{author}{\bibfnamefont{S.}~\bibnamefont{Rahav}},
  \bibinfo{author}{\bibfnamefont{I.}~\bibnamefont{Gilary}}, \bibnamefont{and}
  \bibinfo{author}{\bibfnamefont{S.}~\bibnamefont{Fishman}},
  \bibinfo{journal}{Phys. Rev. A} \textbf{\bibinfo{volume}{68}},
  \bibinfo{pages}{013820} (\bibinfo{year}{2003}).

\bibitem[{\citenamefont{Bukov}(2017)}]{bukov2017}
\bibinfo{author}{\bibfnamefont{M.}~\bibnamefont{Bukov}}, Ph.D. thesis,
  \bibinfo{school}{Boston University} (\bibinfo{year}{2017}).

\bibitem[{\citenamefont{Krylov and Bogoliubov}(1937)}]{krylov1937}
\bibinfo{author}{\bibfnamefont{N.~M.} \bibnamefont{Krylov}} \bibnamefont{and}
  \bibinfo{author}{\bibfnamefont{N.~N.} \bibnamefont{Bogoliubov}},
  \emph{\bibinfo{title}{Introduction to non-linear mechanics (In Russian)}}
  (\bibinfo{publisher}{Ac. of Sci. Ukr. SSR}, \bibinfo{year}{1937}),
  \bibinfo{note}{[English translation by Princeton University Press, Princeton,
  1947]}.

\bibitem[{\citenamefont{Bogoliubov and Mitropolsky}(1961)}]{bogoliubov1961}
\bibinfo{author}{\bibfnamefont{N.~N.} \bibnamefont{Bogoliubov}}
  \bibnamefont{and} \bibinfo{author}{\bibfnamefont{Y.~A.}
  \bibnamefont{Mitropolsky}}, \emph{\bibinfo{title}{{Asymptotic methods in the
  theory of non-linear oscillations (In Russian)}}} (\bibinfo{year}{1961}),
  \bibinfo{note}{[English translation by Hindustan Pub., Delhi, 1961; Gordon
  and Breach, New York, 1961]}.

\bibitem[{\citenamefont{Casas et~al.}(2001)\citenamefont{Casas, Oteo, and
  Ros}}]{casas2001}
\bibinfo{author}{\bibfnamefont{F.}~\bibnamefont{Casas}},
  \bibinfo{author}{\bibfnamefont{J.~A.} \bibnamefont{Oteo}}, \bibnamefont{and}
  \bibinfo{author}{\bibfnamefont{J.}~\bibnamefont{Ros}},
  \bibinfo{journal}{Journal of Physics A, Mathematical and General}
  \textbf{\bibinfo{volume}{34}}, \bibinfo{pages}{3379} (\bibinfo{year}{2001}),
  ISSN \bibinfo{issn}{0305-4470}.

\bibitem[{\citenamefont{Mirrahimi and Rouchon}(2015)}]{mirrahimi2015}
\bibinfo{author}{\bibfnamefont{M.}~\bibnamefont{Mirrahimi}} \bibnamefont{and}
  \bibinfo{author}{\bibfnamefont{P.}~\bibnamefont{Rouchon}},
  \emph{\bibinfo{title}{Dynamics and control of open quantum systems}}
  (\bibinfo{year}{2015}).

\bibitem[{\citenamefont{Eckardt and Anisimovas}(2015)}]{eckardt2015}
\bibinfo{author}{\bibfnamefont{A.}~\bibnamefont{Eckardt}} \bibnamefont{and}
  \bibinfo{author}{\bibfnamefont{E.}~\bibnamefont{Anisimovas}},
  \bibinfo{journal}{New journal of physics} \textbf{\bibinfo{volume}{17}},
  \bibinfo{pages}{093039} (\bibinfo{year}{2015}).

\bibitem[{\citenamefont{Mikami et~al.}(2016)\citenamefont{Mikami, Kitamura,
  Yasuda, Tsuji, Oka, and Aoki}}]{mikami2016}
\bibinfo{author}{\bibfnamefont{T.}~\bibnamefont{Mikami}},
  \bibinfo{author}{\bibfnamefont{S.}~\bibnamefont{Kitamura}},
  \bibinfo{author}{\bibfnamefont{K.}~\bibnamefont{Yasuda}},
  \bibinfo{author}{\bibfnamefont{N.}~\bibnamefont{Tsuji}},
  \bibinfo{author}{\bibfnamefont{T.}~\bibnamefont{Oka}}, \bibnamefont{and}
  \bibinfo{author}{\bibfnamefont{H.}~\bibnamefont{Aoki}},
  \bibinfo{journal}{Physical Review B} \textbf{\bibinfo{volume}{93}},
  \bibinfo{pages}{144307} (\bibinfo{year}{2016}).

\bibitem[{\citenamefont{Xiao et~al.}(2021)\citenamefont{Xiao, Doucet, Noh,
  Ranzani, Simmonds, Govia, and Kamal}}]{xiao2021}
\bibinfo{author}{\bibfnamefont{Z.}~\bibnamefont{Xiao}},
  \bibinfo{author}{\bibfnamefont{E.}~\bibnamefont{Doucet}},
  \bibinfo{author}{\bibfnamefont{T.}~\bibnamefont{Noh}},
  \bibinfo{author}{\bibfnamefont{L.}~\bibnamefont{Ranzani}},
  \bibinfo{author}{\bibfnamefont{R.}~\bibnamefont{Simmonds}},
  \bibinfo{author}{\bibfnamefont{L.}~\bibnamefont{Govia}}, \bibnamefont{and}
  \bibinfo{author}{\bibfnamefont{A.}~\bibnamefont{Kamal}},
  \bibinfo{journal}{arXiv preprint arXiv:2103.09260}  (\bibinfo{year}{2021}).

\bibitem[{\citenamefont{Petrescu et~al.}(2020)\citenamefont{Petrescu,
  Malekakhlagh, and T\"ureci}}]{petrescu2020}
\bibinfo{author}{\bibfnamefont{A.}~\bibnamefont{Petrescu}},
  \bibinfo{author}{\bibfnamefont{M.}~\bibnamefont{Malekakhlagh}},
  \bibnamefont{and} \bibinfo{author}{\bibfnamefont{H.~E.}
  \bibnamefont{T\"ureci}}, \bibinfo{journal}{Phys. Rev. B}
  \textbf{\bibinfo{volume}{101}}, \bibinfo{pages}{134510}
  (\bibinfo{year}{2020}).

\bibitem[{\citenamefont{Groenewold}(1946)}]{groenewold1946}
\bibinfo{author}{\bibfnamefont{H.~J.} \bibnamefont{Groenewold}}, in
  \emph{\bibinfo{booktitle}{On the Principles of Elementary Quantum Mechanics}}
  (\bibinfo{publisher}{Springer}, \bibinfo{year}{1946}), pp.
  \bibinfo{pages}{1--56}.

\bibitem[{\citenamefont{Moyal}(1949)}]{moyal1949}
\bibinfo{author}{\bibfnamefont{J.~E.} \bibnamefont{Moyal}},
  \bibinfo{journal}{Mathematical Proceedings of the Cambridge Philosophical
  Society} \textbf{\bibinfo{volume}{45}}, \bibinfo{pages}{99–124}
  (\bibinfo{year}{1949}).

\bibitem[{\citenamefont{Zachos et~al.}(2005)\citenamefont{Zachos, Fairlie, and
  Curtright}}]{zachos2005}
\bibinfo{author}{\bibfnamefont{C.}~\bibnamefont{Zachos}},
  \bibinfo{author}{\bibfnamefont{D.}~\bibnamefont{Fairlie}}, \bibnamefont{and}
  \bibinfo{author}{\bibfnamefont{T.}~\bibnamefont{Curtright}}
  (\bibinfo{year}{2005}).

\bibitem[{\citenamefont{Dragt and Finn}(1976)}]{dragt1976}
\bibinfo{author}{\bibfnamefont{A.~J.} \bibnamefont{Dragt}} \bibnamefont{and}
  \bibinfo{author}{\bibfnamefont{J.~M.} \bibnamefont{Finn}},
  \bibinfo{journal}{Journal of Mathematical Physics}
  \textbf{\bibinfo{volume}{17}}, \bibinfo{pages}{2215} (\bibinfo{year}{1976}).

\bibitem[{\citenamefont{Deprit}(1969)}]{deprit1969}
\bibinfo{author}{\bibfnamefont{A.}~\bibnamefont{Deprit}},
  \bibinfo{journal}{Celestial mechanics} \textbf{\bibinfo{volume}{1}},
  \bibinfo{pages}{12} (\bibinfo{year}{1969}).

\bibitem[{\citenamefont{Eickbusch et~al.}(2021)\citenamefont{Eickbusch, Sivak,
  Ding, Elder, Jha, Venkatraman, Royer, Girvin, Schoelkopf, and
  Devoret}}]{eickbusch2021}
\bibinfo{author}{\bibfnamefont{A.}~\bibnamefont{Eickbusch}},
  \bibinfo{author}{\bibfnamefont{V.}~\bibnamefont{Sivak}},
  \bibinfo{author}{\bibfnamefont{A.~Z.} \bibnamefont{Ding}},
  \bibinfo{author}{\bibfnamefont{S.~S.} \bibnamefont{Elder}},
  \bibinfo{author}{\bibfnamefont{S.~R.} \bibnamefont{Jha}},
  \bibinfo{author}{\bibfnamefont{J.}~\bibnamefont{Venkatraman}},
  \bibinfo{author}{\bibfnamefont{B.}~\bibnamefont{Royer}},
  \bibinfo{author}{\bibfnamefont{S.}~\bibnamefont{Girvin}},
  \bibinfo{author}{\bibfnamefont{R.~J.} \bibnamefont{Schoelkopf}},
  \bibnamefont{and} \bibinfo{author}{\bibfnamefont{M.~H.}
  \bibnamefont{Devoret}}, \bibinfo{journal}{arXiv preprint arXiv:2111.06414}
  (\bibinfo{year}{2021}).

\bibitem[{\citenamefont{Buishvili and Menabde}(1979)}]{buishvili1979}
\bibinfo{author}{\bibfnamefont{L.~L.} \bibnamefont{Buishvili}}
  \bibnamefont{and} \bibinfo{author}{\bibfnamefont{M.~G.}
  \bibnamefont{Menabde}}, \bibinfo{journal}{Theoretical and Mathematical
  Physics} \textbf{\bibinfo{volume}{50}}, \bibinfo{pages}{2435}
  (\bibinfo{year}{1979}).

\bibitem[{\citenamefont{Shirley}(1965)}]{shirley1965}
\bibinfo{author}{\bibfnamefont{J.~H.} \bibnamefont{Shirley}},
  \bibinfo{journal}{Physical Review} \textbf{\bibinfo{volume}{138}},
  \bibinfo{pages}{B979} (\bibinfo{year}{1965}).

\bibitem[{vv2(2021)}]{vv2021}
\bibinfo{howpublished}{\url{https://github.com/xiaoxuisaac/vanVleck-recursion}}
  (\bibinfo{year}{2021}).

\bibitem[{\citenamefont{Xiao et~al.}(2022)\citenamefont{Xiao, Venkatraman,
  Cortiñas, Eickbusch, Chowdhury, and Devoret}}]{xiao20212}
\bibinfo{author}{\bibfnamefont{X.}~\bibnamefont{Xiao}},
  \bibinfo{author}{\bibfnamefont{J.}~\bibnamefont{Venkatraman}},
  \bibinfo{author}{\bibfnamefont{R.~G.} \bibnamefont{Cortiñas}},
  \bibinfo{author}{\bibfnamefont{A.}~\bibnamefont{Eickbusch}},
  \bibinfo{author}{\bibfnamefont{S.}~\bibnamefont{Chowdhury}},
  \bibnamefont{and} \bibinfo{author}{\bibfnamefont{M.~H.}
  \bibnamefont{Devoret}}, \bibinfo{journal}{In Preparation}
  (\bibinfo{year}{2022}).

\bibitem[{\citenamefont{Bravyi et~al.}(2011)\citenamefont{Bravyi, DiVincenzo,
  and Loss}}]{bravyi2011}
\bibinfo{author}{\bibfnamefont{S.}~\bibnamefont{Bravyi}},
  \bibinfo{author}{\bibfnamefont{D.~P.} \bibnamefont{DiVincenzo}},
  \bibnamefont{and} \bibinfo{author}{\bibfnamefont{D.}~\bibnamefont{Loss}},
  \bibinfo{journal}{Annals of Physics} \textbf{\bibinfo{volume}{326}},
  \bibinfo{pages}{2793} (\bibinfo{year}{2011}), ISSN \bibinfo{issn}{0003-4916}.

\end{thebibliography}


\begin{thebibliography}{27}
\expandafter\ifx\csname natexlab\endcsname\relax\def\natexlab#1{#1}\fi
\expandafter\ifx\csname bibnamefont\endcsname\relax
  \def\bibnamefont#1{#1}\fi
\expandafter\ifx\csname bibfnamefont\endcsname\relax
  \def\bibfnamefont#1{#1}\fi
\expandafter\ifx\csname citenamefont\endcsname\relax
  \def\citenamefont#1{#1}\fi
\expandafter\ifx\csname url\endcsname\relax
  \def\url#1{\texttt{#1}}\fi
\expandafter\ifx\csname urlprefix\endcsname\relax\def\urlprefix{URL }\fi
\providecommand{\bibinfo}[2]{#2}
\providecommand{\eprint}[2][]{\url{#2}}

\bibitem[{\citenamefont{Dragt and Finn}(1976)}]{dragt1976}
\bibinfo{author}{\bibfnamefont{A.~J.} \bibnamefont{Dragt}} \bibnamefont{and}
  \bibinfo{author}{\bibfnamefont{J.~M.} \bibnamefont{Finn}},
  \bibinfo{journal}{Journal of Mathematical Physics}
  \textbf{\bibinfo{volume}{17}}, \bibinfo{pages}{2215} (\bibinfo{year}{1976}).

\bibitem[{\citenamefont{Schur}(1891)}]{schur1891}
\bibinfo{author}{\bibfnamefont{F.}~\bibnamefont{Schur}},
  \bibinfo{journal}{Mathematische Annalen} \textbf{\bibinfo{volume}{38}},
  \bibinfo{pages}{263} (\bibinfo{year}{1891}).

\bibitem[{\citenamefont{Wilcox}(1967)}]{wilcox1967}
\bibinfo{author}{\bibfnamefont{R.~M.} \bibnamefont{Wilcox}},
  \bibinfo{journal}{Journal of Mathematical Physics}
  \textbf{\bibinfo{volume}{8}}, \bibinfo{pages}{962} (\bibinfo{year}{1967}).

\bibitem[{\citenamefont{Rossmann}(2006)}]{rossmann2006}
\bibinfo{author}{\bibfnamefont{W.}~\bibnamefont{Rossmann}},
  \emph{\bibinfo{title}{Lie groups: an introduction through linear groups}},
  vol.~\bibinfo{volume}{5} (\bibinfo{publisher}{Oxford University Press on
  Demand}, \bibinfo{year}{2006}).

\bibitem[{\citenamefont{Grabowski}(1993)}]{grabowski1993}
\bibinfo{author}{\bibfnamefont{J.}~\bibnamefont{Grabowski}},
  \bibinfo{journal}{Annals of Global Analysis and Geometry}
  \textbf{\bibinfo{volume}{11}}, \bibinfo{pages}{213} (\bibinfo{year}{1993}),
  ISSN \bibinfo{issn}{1572-9060}.

\bibitem[{\citenamefont{Cannas~da Silva}(2001)}]{dasilva2001}
\bibinfo{author}{\bibfnamefont{A.}~\bibnamefont{Cannas~da Silva}},
  \emph{\bibinfo{title}{Lectures on symplectic geometry}}, vol.
  \bibinfo{volume}{1764} of \emph{\bibinfo{series}{Lecture Notes in
  Mathematics}} (\bibinfo{publisher}{Springer-Verlag, Berlin},
  \bibinfo{year}{2001}), ISBN \bibinfo{isbn}{3-540-42195-5}.

\bibitem[{\citenamefont{Landau and Lifshitz}(1976)}]{landau1976}
\bibinfo{author}{\bibfnamefont{L.~D.} \bibnamefont{Landau}} \bibnamefont{and}
  \bibinfo{author}{\bibfnamefont{E.~M.} \bibnamefont{Lifshitz}},
  \emph{\bibinfo{title}{Mechanics: Volume 1}}, vol.~\bibinfo{volume}{1}
  (\bibinfo{publisher}{Butterworth-Heinemann}, \bibinfo{year}{1976}).

\bibitem[{\citenamefont{Rahav et~al.}(2003)\citenamefont{Rahav, Gilary, and
  Fishman}}]{rahav2003}
\bibinfo{author}{\bibfnamefont{S.}~\bibnamefont{Rahav}},
  \bibinfo{author}{\bibfnamefont{I.}~\bibnamefont{Gilary}}, \bibnamefont{and}
  \bibinfo{author}{\bibfnamefont{S.}~\bibnamefont{Fishman}},
  \bibinfo{journal}{Phys. Rev. A} \textbf{\bibinfo{volume}{68}},
  \bibinfo{pages}{013820} (\bibinfo{year}{2003}).

\bibitem[{\citenamefont{Dirac}(1930)}]{dirac1930}
\bibinfo{author}{\bibfnamefont{P.~A.~M.} \bibnamefont{Dirac}},
  \emph{\bibinfo{title}{The principles of quantum mechanics}},
  \bibinfo{number}{27} (\bibinfo{publisher}{Oxford university press},
  \bibinfo{year}{1930}).

\bibitem[{\citenamefont{McCoy}(1932)}]{mccoy1932}
\bibinfo{author}{\bibfnamefont{N.~H.} \bibnamefont{McCoy}},
  \bibinfo{journal}{Proceedings of the National Academy of Sciences of the
  United States of America} \textbf{\bibinfo{volume}{18}}, \bibinfo{pages}{674}
  (\bibinfo{year}{1932}).

\bibitem[{\citenamefont{Groenewold}(1946)}]{groenewold1946}
\bibinfo{author}{\bibfnamefont{H.~J.} \bibnamefont{Groenewold}}, in
  \emph{\bibinfo{booktitle}{On the principles of elementary quantum mechanics}}
  (\bibinfo{publisher}{Springer}, \bibinfo{year}{1946}), pp.
  \bibinfo{pages}{1--56}.

\bibitem[{\citenamefont{Smith et~al.}(2020)\citenamefont{Smith, Kou, Xiao,
  Vool, and Devoret}}]{smith2020}
\bibinfo{author}{\bibfnamefont{W.~C.} \bibnamefont{Smith}},
  \bibinfo{author}{\bibfnamefont{A.}~\bibnamefont{Kou}},
  \bibinfo{author}{\bibfnamefont{X.}~\bibnamefont{Xiao}},
  \bibinfo{author}{\bibfnamefont{U.}~\bibnamefont{Vool}}, \bibnamefont{and}
  \bibinfo{author}{\bibfnamefont{M.~H.} \bibnamefont{Devoret}},
  \bibinfo{journal}{npj Quantum Information} \textbf{\bibinfo{volume}{6}},
  \bibinfo{pages}{8} (\bibinfo{year}{2020}), ISSN \bibinfo{issn}{2056-6387}.

\bibitem[{\citenamefont{Eickbusch et~al.}(2021)\citenamefont{Eickbusch, Sivak,
  Ding, Elder, Jha, Venkatraman, Royer, Girvin, Schoelkopf, and
  Devoret}}]{eickbusch2021}
\bibinfo{author}{\bibfnamefont{A.}~\bibnamefont{Eickbusch}},
  \bibinfo{author}{\bibfnamefont{V.}~\bibnamefont{Sivak}},
  \bibinfo{author}{\bibfnamefont{A.~Z.} \bibnamefont{Ding}},
  \bibinfo{author}{\bibfnamefont{S.~S.} \bibnamefont{Elder}},
  \bibinfo{author}{\bibfnamefont{S.~R.} \bibnamefont{Jha}},
  \bibinfo{author}{\bibfnamefont{J.}~\bibnamefont{Venkatraman}},
  \bibinfo{author}{\bibfnamefont{B.}~\bibnamefont{Royer}},
  \bibinfo{author}{\bibfnamefont{S.}~\bibnamefont{Girvin}},
  \bibinfo{author}{\bibfnamefont{R.~J.} \bibnamefont{Schoelkopf}},
  \bibnamefont{and} \bibinfo{author}{\bibfnamefont{M.~H.}
  \bibnamefont{Devoret}}, \bibinfo{journal}{arXiv preprint arXiv:2111.06414}
  (\bibinfo{year}{2021}).

\bibitem[{\citenamefont{Heeres et~al.}(2017)\citenamefont{Heeres, Reinhold,
  Ofek, Frunzio, Jiang, Devoret, and Schoelkopf}}]{heeres2017}
\bibinfo{author}{\bibfnamefont{R.~W.} \bibnamefont{Heeres}},
  \bibinfo{author}{\bibfnamefont{P.}~\bibnamefont{Reinhold}},
  \bibinfo{author}{\bibfnamefont{N.}~\bibnamefont{Ofek}},
  \bibinfo{author}{\bibfnamefont{L.}~\bibnamefont{Frunzio}},
  \bibinfo{author}{\bibfnamefont{L.}~\bibnamefont{Jiang}},
  \bibinfo{author}{\bibfnamefont{M.~H.} \bibnamefont{Devoret}},
  \bibnamefont{and} \bibinfo{author}{\bibfnamefont{R.~J.}
  \bibnamefont{Schoelkopf}}, \bibinfo{journal}{Nature Communications}
  \textbf{\bibinfo{volume}{8}}, \bibinfo{pages}{94} (\bibinfo{year}{2017}),
  ISSN \bibinfo{issn}{2041-1723}.

\bibitem[{\citenamefont{Verney et~al.}(2019)\citenamefont{Verney, Lescanne,
  Devoret, Leghtas, and Mirrahimi}}]{verney2019}
\bibinfo{author}{\bibfnamefont{L.}~\bibnamefont{Verney}},
  \bibinfo{author}{\bibfnamefont{R.}~\bibnamefont{Lescanne}},
  \bibinfo{author}{\bibfnamefont{M.~H.} \bibnamefont{Devoret}},
  \bibinfo{author}{\bibfnamefont{Z.}~\bibnamefont{Leghtas}}, \bibnamefont{and}
  \bibinfo{author}{\bibfnamefont{M.}~\bibnamefont{Mirrahimi}},
  \bibinfo{journal}{Phys. Rev. Applied} \textbf{\bibinfo{volume}{11}},
  \bibinfo{pages}{024003} (\bibinfo{year}{2019}).

\bibitem[{\citenamefont{Zhang et~al.}(2019)\citenamefont{Zhang, Lester, Gao,
  Jiang, Schoelkopf, and Girvin}}]{zhang2019}
\bibinfo{author}{\bibfnamefont{Y.}~\bibnamefont{Zhang}},
  \bibinfo{author}{\bibfnamefont{B.~J.} \bibnamefont{Lester}},
  \bibinfo{author}{\bibfnamefont{Y.~Y.} \bibnamefont{Gao}},
  \bibinfo{author}{\bibfnamefont{L.}~\bibnamefont{Jiang}},
  \bibinfo{author}{\bibfnamefont{R.~J.} \bibnamefont{Schoelkopf}},
  \bibnamefont{and} \bibinfo{author}{\bibfnamefont{S.~M.}
  \bibnamefont{Girvin}}, \bibinfo{journal}{Phys. Rev. A}
  \textbf{\bibinfo{volume}{99}}, \bibinfo{pages}{012314}
  (\bibinfo{year}{2019}).

\bibitem[{\citenamefont{Meurer et~al.}(2017)\citenamefont{Meurer, Smith,
  Paprocki, {\v{C}}ert{\'\i}k, Kirpichev, Rocklin, Kumar, Ivanov, Moore, Singh
  et~al.}}]{meurer2017}
\bibinfo{author}{\bibfnamefont{A.}~\bibnamefont{Meurer}},
  \bibinfo{author}{\bibfnamefont{C.~P.} \bibnamefont{Smith}},
  \bibinfo{author}{\bibfnamefont{M.}~\bibnamefont{Paprocki}},
  \bibinfo{author}{\bibfnamefont{O.}~\bibnamefont{{\v{C}}ert{\'\i}k}},
  \bibinfo{author}{\bibfnamefont{S.~B.} \bibnamefont{Kirpichev}},
  \bibinfo{author}{\bibfnamefont{M.}~\bibnamefont{Rocklin}},
  \bibinfo{author}{\bibfnamefont{A.}~\bibnamefont{Kumar}},
  \bibinfo{author}{\bibfnamefont{S.}~\bibnamefont{Ivanov}},
  \bibinfo{author}{\bibfnamefont{J.~K.} \bibnamefont{Moore}},
  \bibinfo{author}{\bibfnamefont{S.}~\bibnamefont{Singh}},
  \bibnamefont{et~al.}, \bibinfo{journal}{PeerJ Computer Science}
  \textbf{\bibinfo{volume}{3}}, \bibinfo{pages}{e103} (\bibinfo{year}{2017}).

\bibitem[{\citenamefont{Stein et~al.}(2021)}]{sage2021}
\bibinfo{author}{\bibfnamefont{W.}~\bibnamefont{Stein}} \bibnamefont{et~al.},
  \emph{\bibinfo{title}{{S}age {M}athematics {S}oftware ({V}ersion 8.6.0)}},
  \bibinfo{organization}{The Sage Development Team} (\bibinfo{year}{2021}),
  \bibinfo{note}{{\tt http://www.sagemath.org}}.

\bibitem[{\citenamefont{Blais et~al.}(2021)\citenamefont{Blais, Grimsmo,
  Girvin, and Wallraff}}]{Blais_2021}
\bibinfo{author}{\bibfnamefont{A.}~\bibnamefont{Blais}},
  \bibinfo{author}{\bibfnamefont{A.~L.} \bibnamefont{Grimsmo}},
  \bibinfo{author}{\bibfnamefont{S.}~\bibnamefont{Girvin}}, \bibnamefont{and}
  \bibinfo{author}{\bibfnamefont{A.}~\bibnamefont{Wallraff}},
  \bibinfo{journal}{Reviews of Modern Physics} \textbf{\bibinfo{volume}{93}}
  (\bibinfo{year}{2021}), ISSN \bibinfo{issn}{1539-0756}.

\bibitem[{\citenamefont{Koch et~al.}(2007)\citenamefont{Koch, Yu, Gambetta,
  Houck, Schuster, Majer, Blais, Devoret, Girvin, and Schoelkopf}}]{koch2007}
\bibinfo{author}{\bibfnamefont{J.}~\bibnamefont{Koch}},
  \bibinfo{author}{\bibfnamefont{T.~M.} \bibnamefont{Yu}},
  \bibinfo{author}{\bibfnamefont{J.}~\bibnamefont{Gambetta}},
  \bibinfo{author}{\bibfnamefont{A.~A.} \bibnamefont{Houck}},
  \bibinfo{author}{\bibfnamefont{D.~I.} \bibnamefont{Schuster}},
  \bibinfo{author}{\bibfnamefont{J.}~\bibnamefont{Majer}},
  \bibinfo{author}{\bibfnamefont{A.}~\bibnamefont{Blais}},
  \bibinfo{author}{\bibfnamefont{M.~H.} \bibnamefont{Devoret}},
  \bibinfo{author}{\bibfnamefont{S.~M.} \bibnamefont{Girvin}},
  \bibnamefont{and} \bibinfo{author}{\bibfnamefont{R.~J.}
  \bibnamefont{Schoelkopf}}, \bibinfo{journal}{Phys. Rev. A}
  \textbf{\bibinfo{volume}{76}}, \bibinfo{pages}{042319}
  (\bibinfo{year}{2007}).

\bibitem[{\citenamefont{Minev et~al.}(2021)\citenamefont{Minev, Leghtas,
  Mundhada, Christakis, Pop, and Devoret}}]{minev2021}
\bibinfo{author}{\bibfnamefont{Z.~K.} \bibnamefont{Minev}},
  \bibinfo{author}{\bibfnamefont{Z.}~\bibnamefont{Leghtas}},
  \bibinfo{author}{\bibfnamefont{S.~O.} \bibnamefont{Mundhada}},
  \bibinfo{author}{\bibfnamefont{L.}~\bibnamefont{Christakis}},
  \bibinfo{author}{\bibfnamefont{I.~M.} \bibnamefont{Pop}}, \bibnamefont{and}
  \bibinfo{author}{\bibfnamefont{M.~H.} \bibnamefont{Devoret}},
  \bibinfo{journal}{npj Quantum Information} \textbf{\bibinfo{volume}{7}},
  \bibinfo{pages}{1} (\bibinfo{year}{2021}).

\bibitem[{\citenamefont{Xiao et~al.}(2022)\citenamefont{Xiao, Venkatraman,
  Cortiñas, Eickbusch, Chowdhury, and Devoret}}]{xiao20212}
\bibinfo{author}{\bibfnamefont{X.}~\bibnamefont{Xiao}},
  \bibinfo{author}{\bibfnamefont{J.}~\bibnamefont{Venkatraman}},
  \bibinfo{author}{\bibfnamefont{R.~G.} \bibnamefont{Cortiñas}},
  \bibinfo{author}{\bibfnamefont{A.}~\bibnamefont{Eickbusch}},
  \bibinfo{author}{\bibfnamefont{S.}~\bibnamefont{Chowdhury}},
  \bibnamefont{and} \bibinfo{author}{\bibfnamefont{M.~H.}
  \bibnamefont{Devoret}}, \bibinfo{journal}{In Preparation}
  (\bibinfo{year}{2022}).

\bibitem[{\citenamefont{Mirrahimi and Rouchon}(2015)}]{mirrahimi2015}
\bibinfo{author}{\bibfnamefont{M.}~\bibnamefont{Mirrahimi}} \bibnamefont{and}
  \bibinfo{author}{\bibfnamefont{P.}~\bibnamefont{Rouchon}},
  \bibinfo{journal}{Lecture notes}  (\bibinfo{year}{2015}).

\bibitem[{\citenamefont{Grozdanov and Rakovi{\'c}}(1988)}]{grozdanov1988}
\bibinfo{author}{\bibfnamefont{T.}~\bibnamefont{Grozdanov}} \bibnamefont{and}
  \bibinfo{author}{\bibfnamefont{M.}~\bibnamefont{Rakovi{\'c}}},
  \bibinfo{journal}{Physical Review A} \textbf{\bibinfo{volume}{38}},
  \bibinfo{pages}{1739} (\bibinfo{year}{1988}).

\bibitem[{xia(2021)}]{xiao2021}
\bibinfo{howpublished}{\url{https://github.com/xiaoxuisaac/vanVleck-recursion}}
  (\bibinfo{year}{2021}).

\bibitem[{\citenamefont{Eckardt and Anisimovas}(2015)}]{eckardt2015}
\bibinfo{author}{\bibfnamefont{A.}~\bibnamefont{Eckardt}} \bibnamefont{and}
  \bibinfo{author}{\bibfnamefont{E.}~\bibnamefont{Anisimovas}},
  \bibinfo{journal}{New journal of physics} \textbf{\bibinfo{volume}{17}},
  \bibinfo{pages}{093039} (\bibinfo{year}{2015}).

\bibitem[{\citenamefont{Mikami et~al.}(2016)\citenamefont{Mikami, Kitamura,
  Yasuda, Tsuji, Oka, and Aoki}}]{mikami2016}
\bibinfo{author}{\bibfnamefont{T.}~\bibnamefont{Mikami}},
  \bibinfo{author}{\bibfnamefont{S.}~\bibnamefont{Kitamura}},
  \bibinfo{author}{\bibfnamefont{K.}~\bibnamefont{Yasuda}},
  \bibinfo{author}{\bibfnamefont{N.}~\bibnamefont{Tsuji}},
  \bibinfo{author}{\bibfnamefont{T.}~\bibnamefont{Oka}}, \bibnamefont{and}
  \bibinfo{author}{\bibfnamefont{H.}~\bibnamefont{Aoki}},
  \bibinfo{journal}{Physical Review B} \textbf{\bibinfo{volume}{93}},
  \bibinfo{pages}{144307} (\bibinfo{year}{2016}).

\end{thebibliography}
\end{document}


\title{ Supplementary material: \\ On the static effective Hamiltonian of a rapidly driven nonlinear system}
\author{Jayameenakshi Venkatraman}
\thanks{These two authors contributed equally}
\email{jaya.venkat@yale.edu, xu.xiao@yale.edu}
\affiliation{Department of Applied Physics and Physics, Yale University, New Haven, CT 06520, USA}
\author{Xu Xiao}
\thanks{These two authors contributed equally}
\email{jaya.venkat@yale.edu, xu.xiao@yale.edu}
\affiliation{Department of Applied Physics and Physics, Yale University, New Haven, CT 06520, USA}
\author{Rodrigo G. Corti\~nas}
\affiliation{Department of Applied Physics and Physics, Yale University, New Haven, CT 06520, USA}
\author{Alec Eickbusch}
\affiliation{Department of Applied Physics and Physics, Yale University, New Haven, CT 06520, USA}
\author{Michel H. Devoret}
\email{michel.devoret@yale.edu}
\affiliation{Department of Applied Physics and Physics, Yale University, New Haven, CT 06520, USA}
\date{\today}
\maketitle
\color{black}
\section{Transformation of the Hamiltonian under a canonical transformation}
\label{sec:appA}
\emph{Claim}: Under a Lie transformation of the state $\rho \rightarrow \varrho = e^{L_S} \rho$, where $\partial_t \rho = \{\!\!\{ H, \rho \}\!\!\}$, the transformed state $\varrho$ is governed by the equation $\partial_t \varrho = \{\!\!\{K, \varrho\}\!\!\}$, where
\begin{align}
\label{eq:K}
    K = e^{L_S} H + \int_{0}^{1} d\epsilon\, e^{\epsilon L_S} \dot{S}.
\end{align}
\emph{Proof:}  Let $\varrho = e^{L_S} \rho$, so that $\rho = e^{L_{-S}} \varrho$. We take the equation governing the time-evolution of $\rho$ to be
\begin{subequations}
    \begin{align}
        \partial_t \rho = \{\!\!\{H, \rho\}\!\!\},
    \end{align}
implying that
    \begin{align}
    \label{eq:varrho-eom}
        \partial_t (e^{L_{-S}}\varrho) = \{\!\!\{H, e^{L_{-S}}\varrho\}\!\!\}
    \end{align}
and using the product rule, we get
    \begin{align}
    \label{eq:varrho-eom-2}
     \partial_t \varrho =  \{\!\!\{e^{L_{S}} H, \varrho\}\!\!\} - e^{L_{S}}\partial_t (e^{L_{-S}}) \varrho,
    \end{align}
where we have used $e^{L_S} \{\!\!\{F, G\}\!\!\} = \{\!\!\{e^{L_{S}} F, e^{L_{S}} G\}\!\!\}$ \cite{dragt1976}.
\end{subequations}

Now, \cref{eq:K} is indeed true if
\begin{align}
\label{eq:to-prove-2}
    e^{L_{S}}\partial_t (e^{L_{-S}}) \varrho = -L_{\int_0^{1} d\epsilon\, e^{\epsilon L_{S}} \dot{S}} \varrho,
\end{align}
which we now prove.
To evaluate the left hand side of \cref{eq:to-prove-2}, we invoke the following identity \cite{schur1891,wilcox1967,rossmann2006,grabowski1993}:
\begin{subequations}
    \begin{align}
    \label{eq:derivative-exp}
        e^{-\square} \frac{d}{d t} e^{\square} = \left(\frac{1 - e^{-\mathrm{ad}_{\square}}}{\mathrm{ad}_{\square}} \frac{ d \square}{dt}\right),
    \end{align}
where the right hand side of \cref{eq:derivative-exp} can be further expanded using 
    \begin{align*}
        \frac{1 - e^{-\mathrm{ad}_{\square}}}{\mathrm{ad}_{\square}} = \sum_{k = 0}^{\infty} \frac{(-1)^k}{(k+1)!} {\mathrm{ad}^k_{\square}}.
    \end{align*}
In \cref{eq:derivative-exp}, we have defined $\mathrm{ad}_{\square_1} \square_2 = \lfloor \square_1, \square_2 \rfloor = \square_1 \square_2 - \square_2 \square_1$, where $  \lfloor \square_1, \square_2 \rfloor$ represents the Lie bracket associated with the \textit{operators} of the Lie algebra \cite{schur1891,wilcox1967,rossmann2006,grabowski1993}. Comparing \cref{eq:derivative-exp} with \cref{eq:to-prove-2} and taking $\square \leftarrow (-L_S)$, \cref{eq:to-prove-2} becomes
    \begin{align}
    \begin{split}
    \label{eq:partial-t}
        e^{L_S} \frac{\partial}{\partial t} e^{L_{-S}} &= -\left( \sum_{k = 0}^{\infty} \frac{1}{(k+1)!} {\mathrm{ad}^k_{L_{S}}} L_{\dot{S}}\right) \\
        &= -\Bigg(L_{\dot{S}} + \frac{1}{2!} \lfloor L_{S}, L_{\dot{S}} \rfloor + \frac{1}{3!} \lfloor L_{S}, \lfloor L_S, L_{\dot{S}} \rfloor \rfloor  \Bigg. \\ & \qquad  + \cdots  \Bigg).
    \end{split}
    \end{align}
Note that we converted the total time-derivative in \cref{eq:derivative-exp} to a partial time-derivative in \cref{eq:partial-t} since we work in the active representation where the coordinates are static.
\end{subequations}

Now, we use an important property of the Lie bracket \cite{dragt1976,dasilva2001}:
\begin{align}
\label{eq:identity}
\begin{split}
     \lfloor L_F, L_G \rfloor &= L_{\{\!\!\{F, G\}\!\!\}} \\
     &= L_{L_F G}
\end{split}
\end{align}
to get 
\begin{align}
\label{eq:identity-2}
    \lfloor L_F, \lfloor L_F, \lfloor \cdots \lfloor L_F, L_G \rfloor \rfloor \rfloor \rfloor = L_{L_F^n G}.
\end{align}

\Cref{eq:identity} as written is simply the Jacobi identity and is generalized to \cref{eq:identity-2} by induction. Employing \cref{eq:identity-2} in \cref{eq:partial-t} at each order gives
\begin{align}
\label{eq:result}
\begin{split}
    e^{L_S} \frac{\partial}{\partial t} e^{L_{-S}} &= - \left(L_{\dot{S}} + L_{\frac{1}{2!}L_{S} \dot{S}} + L_{\frac{1}{3!} L_{S}^2 \dot{S}} + \cdots \right) \\
    &= -\left(L_{\sum_{n = 0}^{\infty} \frac{1}{(n+1)!} L_{S}^n \dot{S}} \right) \\
    &= -\left(L_{\int_0^1 d\epsilon\, e^{\epsilon L_S}  \dot{S}} \right).
\end{split}
\end{align}
This completes the proof.

\section{Applications of the recursive formula}
\subsection{Classical and Quantum Kapitza Pendulum}
\label{subsec:Kapitza}
We take as an illustration of our agnostic formulation the case of a rigid pendulum of length $l$ and mass $m$ whose pivot undergoes a sinusoidal motion of amplitude $r$ along the vertical. This driven pendulum, known as the Kapitza pendulum \cite{landau1976}, serves as a model for the dynamical stabilization of mechanical systems, and is described by the classical Hamiltonian
\begin{align}
\begin{split}
\label{eq:Kapitza}
    H(t) = \frac{p_{\varphi}^2}{2 J} - J \omega_o^2 \cos{\varphi} - J\frac{r}{l}\omega^2  \cos{\varphi} \cos{\omega t},
\end{split}
\end{align}
where $J= m l^2$ is the moment of inertia and $\omega_o = \sqrt{\frac{g}{l}}$ is its small oscillation frequency. In \cref{eq:Kapitza}, $\varphi$ is the angle between the pendulum and the vertical and  $p_{\varphi}$ is the angular momentum so that $\{\varphi, p_{\varphi} \} = 1$. Decomposing the Hamiltonian following the notation in Eq. 6 we have
\begin{align}
    H_0 &= \frac{p_{\varphi}^2}{2 J} - J \omega_o^2 \cos{\varphi}, &
    H_{\pm 1} &= -J\frac{r}{2l}\omega^2  \cos{\varphi}.
\end{align}

Usual treatments of the Kapitza pendulum are performed on its equations of motion \cite{landau1976,rahav2003}, which are separated into slow and fast coordinates. Averaging over the latter yields an effective potential in the slow coordinates. Here, we recover well-known results following an alternative route that allows to extend the calculation to higher orders. To demonstrate the coupled construction of $S$ and $K$ in Eq. 8, we proceed order-by-order. The leading order reads
\begin{align}
    K^{(0)} = H + \dot{S}^{(1)}.
\end{align}
By demanding $S^{(1)}$ to be the primitive of $-\boldsymbol{\mathrm{osc}}(H)$, i.e. $S^{(1)} = -\int dt\, \boldsymbol{\mathrm{osc}}(H)$, we find $K^{(0)}$ to be simply the time-average of $H$:
\begin{align}
\begin{split}
\label{eq:K-0}
    K^{(0)} = \overline{H}  &= \frac{1}{T} \int_0^{T} H(t) dt\\ &=  \frac{p_{\varphi}^2}{2 J} - J \omega_o^2 \cos{\varphi}.
\end{split}
\end{align}
\color{black}
Note that, for simplicity, we have set the integration constant of $S^{(1)}$ to be zero and so we shall do for higher orders.

To order one, we have
\begin{align}
    K^{(1)} = \dot{S}^{(2)} + L_{S^{(1)}} H + \frac{1}{2} L_{S^{(1)}} \dot{S}^{(1)}.
\end{align}
By demanding $ S^{(2)} = -\int dt\, \boldsymbol{\mathrm{osc}} (L_{S^{(1)}} H + \frac{1}{2} L_{S^{(1)}} \dot{S}^{(1)})$, we find
\begin{align}
\label{eq:PB}
    K^{(1)} = -\overline{ \bigg\{\int dt\, \boldsymbol{\mathrm{osc}}(H), \boldsymbol{\mathrm{osc}}(H)\bigg\} } = 0,
\end{align}
where $\{\square_1, \square_2\}$ in \cref{eq:PB} is the Poisson bracket between phase space functions $\square_1$ and $\square_2$.

Following this procedure, we compute
\begin{align}
\label{eq:K2}
    K^{(2)} = -J\frac{r^2}{l^2}\frac{\omega^2}{8} \cos 2\varphi.
\end{align}

For the perturbative treatment to be valid the typical rate of evolution under any one of the terms in \cref{eq:Kapitza} must be small with respect to the perturbation frequency $\omega$. The typical rate of evolution under $H_0$ is naturally $\omega_o$ while that under $H_{\pm 1}$ is $\sim\sqrt{\frac{r}{2l}}\;\omega $ and, thus, the perturbative conditions read $\omega_o\ll\omega$ and $r \ll 2l$.

We observe that when the second order correction \cref{eq:K2} becomes comparable to the potential energy contribution of the zeroth order, the system develops a secondary stable position at $\varphi = \pi$ \cite{landau1976}. This happens for the well-known condition $\frac{r}{l} \frac{\omega}{\omega_o} > \sqrt{2}$. 

Following this approach it is easy to carry the calculation further. Up to order four we find
\begin{align}
\label{eq:K-cl}
\begin{split}
    K^{(3)} &=0\\
    K^{(4)} &= \frac{r^2}{l^2}\frac{3}{8 J} p_{\varphi}^2 (1 -  \cos{2\varphi}) + 
    \frac{r^2}{l^2} \frac{J \omega_o^2}{4} (\cos{\varphi} - \cos{3 \varphi}).
\end{split}
\end{align}
We see that the fourth order correction introduces a nontrivial modulation to the potential but that no interesting additional stable solution arise due to the perturbative hierarchy of the prefactors.

\emph{The quantum calculation} proceeds analogously by changing the Lie bracket, i.e., the sub-routine in the algorithm. In this case, the Hamiltonian is given by Dirac's canonical quantization recipe \cite{dirac1930} and reads
\begin{align}
\label{eq:Kapitza-qu}
\op{H}(t) = \frac{\op{p}_{\varphi}^2}{2 J} - J \omega_o^2 \cos{\op{\varphi}} - \frac{r}{l}J\omega^2  \cos{\op{\varphi}} \cos{\omega t},
\end{align}
where $\frac{1}{i\hbar}[\op{\varphi}, \op{p}_{\varphi}] = 1$.
To order four the Kamiltonian reads
\begin{align}
\label{eq:K-qu}
\begin{split}
    \op{K}^{(0)} &=  \frac{\op p_{\varphi}^2}{2 J} - J \omega_o^2 \cos{\op \varphi} \\
    \op{K}^{(1)} &= 0 \\
    \op{K}^{(2)} &= -\frac{r^2}{l^2}\frac{J\omega^2}{8} \cos 2\op \varphi\\
    \op{K}^{(3)} &= 0\\
    \op{K}^{(4)} &= \frac{r^2}{l^2} \frac{3}{8 J} \op p_{\varphi}^2  (1 -  \cos{2\op \varphi}) + i \hbar \frac{r^2}{l^2}\frac{3}{4 J} \op p_\varphi \sin {2 \op \varphi}\\ & \qquad - \hbar^2 \frac{r^2}{l^2} \frac{13}{32 J} \cos{2 \op\varphi} +
    \frac{r^2}{l^2} \frac{J \omega_o^2}{4} (\cos{ \op \varphi} - \cos{3 \op \varphi}),
\end{split}
\end{align}
where we have chosen the normal-ordered form with respect to $\op p_{\varphi}$ in order to expose the corrections to the classical expression $(\text{they are proportional to powers of } \hbar)$. A symmetrized form displays the Hermitian character of $\op K^{(4)}$ more directly:
\begin{align}
\begin{split}
  \op{K}^{(4)} =& \frac{r^2}{l^2} \frac{3}{16 J} \left(\op p_{\varphi}^2  (1 -  \cos{2\op \varphi}) + (1 -  \cos{2\op \varphi})\op p_{\varphi}^2 \right) \\
  & - \hbar^2 \frac{r^2}{l^2} \frac{13}{32 J} \cos{2 \op\varphi} +
    \frac{r^2}{l^2} \frac{J \omega_o^2}{4} (\cos{ \op \varphi} - \cos{3 \op \varphi}).
\end{split}
\end{align}
Note that $\op K^{(4)}$, unlike $\op K^{(2)}$, can no longer be constructed from its classical counterpart under Weyl quantization \cite{mccoy1932}. This is a direct consequence of Groenewold's theorem \cite{groenewold1946}.

Finally, we also remark that in the limit $\omega_o\rightarrow 0$ this model provides a $\cos(2\hat \varphi)$ Hamiltonian that has received attention in the context of protected superconducting qubits \cite{smith2020}.
\subsection{Frequency and Kerr renormalization of a driven Duffing oscillator}
\label{subsec:Duff}
In this section, we compute the renormalization of the spectrum of a Duffing oscillator under a drive. This system is an archetype of superconducting circuits, and these effects play an important role for engineering high-fidelity quantum control \cite{eickbusch2021} and gates \cite{heeres2017} in quantum computation.

We consider the quantum Hamiltonian 
\begin{align}
    \frac{\hat H}{\hbar} = \omega_o \hat{q}^\dagger\hat{q} + g_4(\hat{q}+\hat{q}^\dagger)^4 - 2i\Omega\cos(\omega_dt)(\hat{q} - \hat{q}^\dagger),
\end{align}
where $\hat{q}$ and $\hat{q}^\dagger$ are creation and annihilation operators that obey $[\op{q}, \op{q}^\dagger] = 1$, $\omega_o$ and $g_4\ll \omega_o$ are the frequency and nonlinearity of the Duffing oscillator, $\omega_d$ is the frequency of the drive, and $\Omega$ is the drive amplitude. While the drive can be at any frequency, for the sake of concreteness, we take $\omega_d / \omega_o = 1.21 \approx 6/5$.

To ready the Hamiltonian for the perturbative expansion we perform two unitary transformations. The first one is a displacement transformation $\hat U_D = \exp(\alpha(t)\hat{q}^\dagger - \alpha^*(t)\hat{q})$ with $\alpha = i\Omega e^{-i\omega_d t}/(\omega_d - \omega_o)-i\Omega e^{i\omega_d t}/(\omega_d + \omega_o)$ corresponding to the linear response of the oscillator to the drive. The second one is a transformation to a rational harmonic of the drive near the oscillator resonant frequency. We take the frame rotating at $5\omega_d/6 \sim \omega_o$ given by $\hat U_R = \exp(-i\hat{q}^\dagger\hat{q}\frac{5}{6}\omega_dt)$. The transformed Hamiltonian reads
\begin{align}\label{eq:H-tran}
\begin{split}
    \frac{\op H}{\hbar} = \delta \hat{q}^\dagger \hat{q}+ g_4 (&\hat{q} e^{- i \frac{5}{6}\omega_d t} + \hat{q}^{\dagger} e^{+i\frac{5}{6}\omega_d  t} \\&\;\;+ \Pi e^{-i \omega_d t} +  \Pi^* e^{i \omega_d t} )^4,
\end{split}
\end{align}
where $\delta = \omega_o - 5\omega_d/6  \approx -0.008 \omega_o$ and $\Pi = 2i\Omega\omega_d/(\omega_d^2-\omega_o^2)$ is the effective drive amplitude. Note that by going to this rotating frame, all the frequency scales, i.e., $\delta$ and $g_4$, are much smaller than the fundamental frequency $\omega = \omega_d/6$ of the system and thus the problem satisfies the perturbation condition.
\begin{figure}[t!]
\includegraphics[width =\columnwidth]{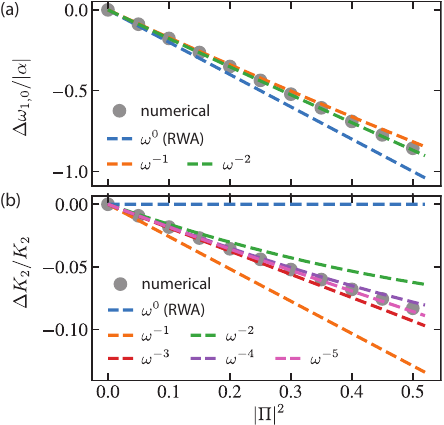}
\caption{Energy renormalizations of a driven Duffing oscillator with $g_4 = -0.001 \omega_{o}$ and $\omega_d = 1.21\omega_o$. Grey dots are obtained from Floquet numerics and the dashed lines are from the perturbative expansion truncated to the indicated orders. (a) The ac Stark shift of the qubit $\Delta \omega_{1,0}$ ($|\Pi|$-dependent frequency renormalization) is plotted in units of qubit anharmonicity $\alpha = 2K_2$ and as a function of drive strength. (b) The $|\Pi|$-dependent Kerr renormalization  $\Delta K_2$ is plotted in units of bare Kerr and as a function of drive strength.} 
\label{fig:ss}
\end{figure}
The leading order the effective Hamiltonian is the time-average of \cref{eq:H-tran} and yields the well-known RWA Hamiltonian
\begin{align*}
    \frac{\hat{K}^{(0)}}{\hbar} = (\delta+12g_4+ 24g_4|\Pi|^2) \hat{q}^\dagger \hat{q}+ 6g_4\hat{q}^{\dagger 2}\hat{q}^2. 
\end{align*}
Note that we have taken $S^{(0)} = 0$ and have set the arbitrary integration constant of $\hat{S}$ at all orders to be zero for simplicity.

For a qubit manifold comprising the ground and first excited states of the oscillator, $\omega_{1,0} = \delta + 12g_4$ is the bare qubit frequency in the rotating frame, $K_2 = 6 g_4$ is usually referred to as the bare Kerr nonlinearity, and $ \tilde{\omega}^{(0)}_{0,1} = \omega_{1,0} +24 g_4|\Pi|^2$ gives the leading order renormalized frequency of the qubit as a function of drive amplitude.

With the recursive formula, we compute $\hat{K}$ up to $\mathcal{O}(1/\omega^5)$ as

\begin{align}
  \frac{\hat{K}}{\hbar} =  \sum_{n = 1}^{5} K_n \hat{q}^{\dagger n} \hat{q}^n,
\end{align}
where $K_1, K_2, \cdots, K_n$ are all functions of $\omega_o, g_4, \Pi$ and $\delta$.
For concreteness, we focus on capturing the renormalization of the frequency and the anharmonicity of the oscillator due to the drive and nonlinearity.
\color{black}
The recursive algorithm yields the next two orders correction as 
\begin{align}\label{eq:ss}
\begin{split}
\tilde{\omega}_{1,0} &= \tilde{\omega}^{(0)}_{0,1} + \frac{g_4^2}{\omega} 531|\Pi|^4 + \frac{g_4^3}{\omega^2}  21832 |\Pi|^6 \\ & \quad + \frac{g_4^2\delta}{\omega^2}665|\Pi|^6+\Delta_{\hbar} +  \mathcal{O}\left(\frac{1}{\omega^3}\right),
\end{split}
\end{align}
where $\Delta_{\hbar}$ denotes the quantum correction arising from the non-commutativity of $\op{a}$ and $\op{a}^{\dagger}$. They read:
\begin{align}
\begin{split}
    \Delta_{\hbar} &=  \frac{g_4^2}{\omega} \left(|\Pi|^2 625 - 58 \right) 
\\ & \quad + \frac{g_4^3}{\omega^2}  \left(|\Pi|^4 43258 +  |\Pi|^2 13815 + 573 \right) \\
&\quad+  \frac{g_4^2\delta}{\omega^2}\left(|\Pi|^2 907 + 12 \right) +\mathcal{O}\left(\frac{1}{\omega^3}\right).
\end{split}
\end{align}

In \cref{fig:ss} (a), the perturbative result is compared with the numerical diagonalization of the Floquet Hamiltonian \cite{verney2019,zhang2019} and shows excellent agreement.

The drive renormalizes not only the frequency of the qubit but also, and more dramatically, its nonlinearity. We compute the correction to the bare Kerr nonlinearity and obtain as a result to third order
\begin{align}
\begin{split}
    \tilde{K}_2 &=  K_2 + \frac{g_4^2}{\omega}|\Pi|^2 312 + \frac{g_4^3}{\omega^2}|\Pi|^4 21629\\
    &\quad + \frac{g_4^4}{\omega^3}|\Pi|^6 1517277 + K_{\hbar} + \mathcal{O}\left(\frac{1}{\omega^4} \right),
    \end{split}
\end{align}
where $K_{\hbar}$ denote the quantum corrections arising from the non-commutativity of $\op{q}$ and $\op{q}^{\dagger}$. They read:
\begin{align}
\begin{split}
    K_{\hbar} &=- \frac{g_4^2}{\omega} 61  + \frac{g_4^3}{\omega^2}  \left(|\Pi|^2 17919 +  1007 \right) + \frac{g_4^2\delta}{\omega^2} \big(|\Pi|^2 453 \\
    &\quad +  12 \big)  + \frac{g_4^4}{\omega^3} \big(|\Pi|^4 2723568+|\Pi|^2 655974 +  20629 \big)\\
    &\quad +\frac{g_4^3\delta}{\omega^3} \big(|\Pi|^4 53383 +|\Pi|^2 46418 +  403 \big)\\
    &\quad+\frac{g_4^2\delta^2}{\omega^3} \big(|\Pi|^2 427 + 2 \big)+ \mathcal{O}\left(\frac{1}{\omega^4} \right).
\end{split}
\end{align}

\color{black}
We have explicitly taken an example that requires an expansion to even higher order to fit the numerical diagonalization in the plotting range chosen for \cref{fig:ss} (b). It is nonetheless easy to compute the extra 34 terms composing the fourth and fifth order corrections to Kerr using our formula (Eq. (8) in the main text) and a computer algebra software \cite{meurer2017,sage2021}. Note that the range of $\Pi$ in \cref{fig:ss} has been chosen to match the parameters of current state-of-the-art experiments \cite{Blais_2021}. We plot in \cref{fig:ss} the analytical corrections up to fifth order to get an acceptable approximation of the numerical data. Our results suggest the convergence of the method applied to the Duffing problem in the studied range.
\subsection{Modeling the effective Hamiltonian of a driven superconducting circuit}
\label{subsec:sc}
In this section, we compute the effective Hamiltonian of a driven superconducting circuit. We are interested in obtaining the renormalization of coefficients in the case of a qubit coupled to resonator in the presence of photons. This framework is powerful for predicting system dynamics in the presence of strong oscillator drives, and has been essential to achieve high-fidelity quantum control \cite{eickbusch2021}.

\begin{figure}[t!]
\includegraphics[width =\columnwidth]{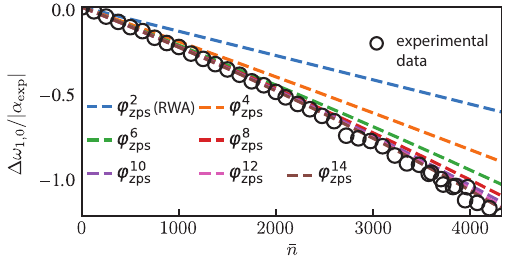}
\caption{Parameter-free model of the ac Stark shift data of the qubit mode in a transmon-cavity superconducting circuit as a function of cavity occupation number $\bar{n}$.  Open black circles represent experimentally measured data. Dashed lines represent theory prediction for different orders of $\varphi_{\mathrm{zps}}^2$. Theory prediction converges to the measured data at sufficiently high order in the perturbation parameter $\varphi_{\mathrm{zps}}^2$. Further system and measurement setup details can be found in \cite{eickbusch2021}.}
\label{fig:ss-exp}
\end{figure}

\begin{figure}[t!]
\includegraphics[width =\columnwidth]{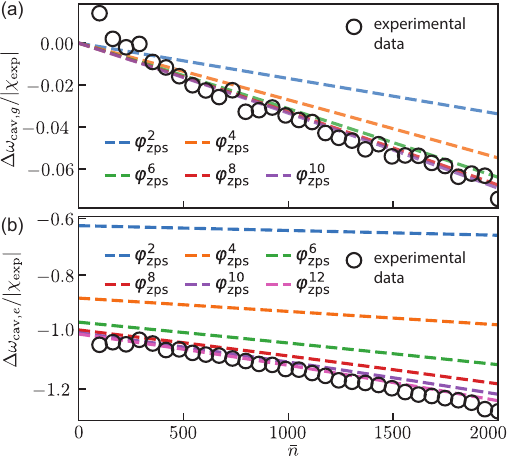}
\caption{Parameter-free model of the cavity frequency data as a function of cavity photon number $\bar{n}$ with the qubit prepared in (a)$|g\rangle$ and (b)$|e\rangle$. Open black circles represent experimentally measured data. Dashed lines represent theory prediction with zero fit parameters. Theory prediction converges to the measured data at sufficiently high order in the perturbation parameter $\varphi_{\mathrm{zps}}^2$. Further system and measurement setup details can be found in \cite{eickbusch2021}.} 
\label{fig:ss-exp-cav}
\end{figure}
We consider the versatile transmon superconducting qubit \cite{koch2007}, that is linearly coupled to a high-Q cavity mode. We write this system Hamiltonian in normal modes as
\begin{align}
\label{eq:H-sc}
    \frac{\op{H}}{\hbar} = \omega_a \op{a}^{\dagger} \op{a} + \omega_q \op{q}^{\dagger} \op{q} - \frac{E_{\mathrm{J}}}{\hbar}\left[ \cos   (\hat{\varphi}) + \frac{\hat{\varphi}^2}{2} \right].
\end{align}
Here $\op a$ and $\op q$ denote annihilation operators associated with the cavity and qubit modes and $\omega_a/ 2\pi$ and $\omega_q/2 \pi$ denote their respective frequencies. The operator
$\hat \varphi = \varphi_a (\op a + \op{a}^{\dagger}) + \varphi_q (\op q + \op{q}^{\dagger}),$ where the parameters $\varphi_a, \varphi_q$ represent the zero point spread of the phase of the transmon in each mode. These spreads are determined as $\varphi_i^2 = p_i \hbar \omega_i / 2 E_{\mathrm{J}}$ for $i = a, q,$ where $p_i$ represents the participation ratio of $i$ \cite{minev2021}. The Josephson energy is denoted $E_{\mathrm{J}}$ and the function $\cos(\hat{\varphi}) + \hat{\varphi}^2/2$ contains the anharmonic part of the Josephson potential.  Going into a rotating frame induced by the Hamiltonian $\hbar \omega_a \op{a}^{\dagger} \op{a} + \hbar \omega_q \op{q}^{\dagger} \op{q}$ and performing a Taylor series expansion yields:
\begin{align}
\label{eq:tra-cav}
\begin{split}
    \frac{\op{H}}{\hbar \omega_q} &= \sum_{n = 4}^{\infty} \frac{g_n}{\omega_q} (\lambda_a (\op a e^{-i \omega_a t} + \op{a}^{\dagger} e^{i \omega_a t}) + \\ & \qquad \qquad \lambda_q (\op q e^{-i \omega_q t}+ \op{q}^{\dagger} e^{i \omega_q t}))^n.
\end{split}
\end{align}
Here $g_n = (-1)^{n/2} E_{\mathrm{J}} \varphi_{\mathrm{zps}}^n /n!$ for even $n$ and $0$ for odd $n$, with $\varphi_{\mathrm{zps}} = \sqrt{\hbar \omega_q/2 E_{\mathrm{J}}}$ and $\lambda_i^2 = p_i \omega_i/ \omega_q$ for $i = a, q$. The physical significance of these parameters is  $g_n$ is the strength of an $n$-way nonlinear mixing interaction involving excitations of the cavity-mode $\hat a$ and qubit-mode $\hat q$, each associated with weight factors $\lambda_a$ and $\lambda_q$ which characterize the modes' participation in the interaction \cite{xiao20212}.

We seek a time-independent effective Hamiltonian that distills from \cref{eq:tra-cav} the effective cavity-qubit interaction. We note that there are three key differences between \cref{eq:tra-cav} and the previous examples we have treated so far.
\begin{inparaenum}[i)]
    \item \label{subitem:1} The first and most nontrivial distinction is that here we choose the small parameter of our expansion to be $\varphi_{\mathrm{zps}}^2$ instead of $1/\omega$. This choice emerges naturally by observing that the coefficient $g_n / \omega_q$ of each summand in \cref{eq:tra-cav} is order $(n-2)/2$ in $\varphi_{\mathrm{zps}}^2$ ($g_n$ is nonzero only for even $n$). With $\varphi_{\mathrm{zps}}^2$ as the small parameter, the grid acquires a simple but nontrivial modification from what is shown in Figure 2 of the main text. Every node at row $0$ and column $n \ge 1$ acquires a red seed to represent the contribution of the Hamiltonian in \cref{eq:tra-cav} at order $n$ in $\varphi_{\mathrm{zps}}^2.$
    \item There are two quantum modes in \cref{eq:tra-cav} in contrast with previously treated systems with only one mode; in fact, the simpler \cref{eq:H-tran} can be recovered from \cref{eq:tra-cav} by setting $g_n = 0$ for $n > 4$, and by further setting the cavity-mode $\hat{a}$ as a parameter, valid for a stiff pump.
    \item \Cref{eq:tra-cav} has two fundamental frequencies: $\omega_a$ and $\omega_d$ in contrast with the monochromatic drive in previously treated systems.
  \end{inparaenum}

Applying the recursive formula to order $7$, we compute a model for the effective Hamiltonian as
\begin{align}
\label{eq:Heff}
\frac{\hat {K}}{\hbar} = \sum_{m = 0}^{7} \sum_{n = 0}^{7} \frac{K_{m, n}}{\hbar} \hat{q}^{\dagger m} \hat{q}^m \hat{a}^{\dagger n} \hat{a}^n  + \mathcal{O}\left(\left(\varphi_{\mathrm{zps}}^2\right)^8\right),
\end{align}
where each $K_{m, n}$ is a function of $\varphi_{\mathrm{zps}}^2, \lambda_a, \lambda_q, \omega_q, \omega_a.$ Their functional form has been suppressed for brevity.

We test our model by comparing the effective Hamiltonian parameters measured in a recent experiment \cite{eickbusch2021} consisting of a transmon qubit coupled to a high-Q superconducting cavity. The coefficients of \cref{eq:tra-cav} are fully determined by the following independently calibrated parameters measured experimentally in the absence of drives:
\begin{align}
\label{eq:exp-paras}
\begin{split}
    \frac{\omega_{q, \mathrm{exp}}}{2 \pi} &= 6.657\, \mathrm{GHz}, \\ \frac{\omega_{a, \mathrm{exp}}}{2 \pi} &=  5.261\, \mathrm{GHz},\\
     \frac{-\chi_{\mathrm{exp}}}{2\pi} &= \frac{K_{1, 1, \mathrm{exp}}}{h} =  -31.2\, \mathrm{kHz}, \\
    \frac{-\alpha_{\mathrm{exp}}}{2\pi} &=  \frac{2 K_{2, 0, \mathrm{exp}}}{h} = -193.29\, \mathrm{MHz}.
\end{split}
\end{align}
Here $\chi_{\mathrm{exp}}$ and $\alpha_{\mathrm{exp}}$ correspond to the measured qubit-cavity cross-kerr coefficient and the qubit anharmonicity respectively. The subscript $\mathrm{exp}$ indicates an experimentally measured quantity; see Supplement of \cite{eickbusch2021} for characterization details.
By a numerical diagonalization of \cref{eq:H-sc}, we map \cref{eq:exp-paras} to $E_{\mathrm{J}}/ 2\pi = 32.33~ \mathrm{GHz}$, $\omega_q/ 2\pi = 6.843~\mathrm{GHz}$,  $\omega_a/ 2\pi = 5.261 ~\mathrm{GHz}$, $\lambda_a =  0.0073$, and $\lambda_q = -0.9998$.
Now, we outline two different experiments detailed in \cite{eickbusch2021} whose results we explain analytically for the first time here.

The first one is a measurement of the transmon's ac Stark shift, its photon-number dependent frequency renormalization, as a function of the mean photon number $\bar{n} = |\alpha|^2$ for a coherent state $|\alpha\rangle$ prepared in the cavity. In the effective Hamiltonian $\hat{K}$ in \cref{eq:Heff}, the measurement represents $\langle e, \alpha| \hat{K} | e, \alpha \rangle - \langle g, \alpha| \hat{K} | g, \alpha \rangle = \sum_{n = 0}^7 K_{1,n} \bar{n}^n/ \hbar$ as a function of $\bar{n} = |\alpha|^2$. Open black circles in \cref{fig:ss-exp} represent experimentally measured data with an independent calibration of $\bar{n}$. Dashed lines represent the theoretical prediction, with different colors representing the expansion to different orders in the perturbation parameter $\varphi_{\mathrm{zps}}^2$. The lines converge to measured data as we perform the  expansion to higher orders.

The second one is a measurement of the cavity frequency as a function of the cavity photon number $\bar{n}$ with the qubit prepared in (a) $|g\rangle$ and (b) $|e\rangle$. Since this frequency renormalization is insensitive to the phase of the cavity field, one can consider the Fock state $|n = \bar{n}\rangle$, instead of a coherent state $|\alpha\rangle$, to model this measurement. Taking this alternative path for ease of calculation and for its illustrative value, we model this measurement as the renormalized cross-Kerr coefficient with the cavity in a Fock state $|n\rangle$, i.e., $\langle i, n+1| \hat{K} | i, n+1 \rangle - \langle i, n| \hat{K} | i, n \rangle $, where in (a) $|i\rangle = |g\rangle$ and (b) $|i\rangle = |e\rangle$. Open black circles in \cref{fig:ss-exp-cav} represent the measured data. Again, dashed lines represent the theoretical prediction, with different colors representing the expansion to different orders in the perturbation parameter $\varphi_{\mathrm{zps}}^2.$

With this example, we demonstrate the generality of the recursive formula: by applying it to a multi-mode, non-monochromatic Hamiltonian, and developing the expansion for a perturbation parameter that is different from $1/\omega$, we are able to predict and explain to a high accuracy the effective Hamiltonian at large intra-cavity photon numbers, matching what is measured in an experiment. In particular, for this experimental system consisting of a strongly anharmonic transmon, an expansion to high order is required to capture the dynamics and it was missing until now.
\color{black}

\section{On the equivalence between various equation of motion-based methods and the Lie method}
Here we extend the discussion on the equivalence between the exponential ansatz Eq. (3) and the additive ansatz Eq. (9). As an illustration of the equivalence, we develop two representative examples. For the first example, we take on ourselves the task of developing the effective static Kamiltonian corresponding to the \textit{classical} Krylov-Bogoliubov (KB) averaging method \cite{landau1976,rahav2003}. We show that this result based on an additive ansatz agrees with our recursive formula. In the second example, we consider the \emph{quantum} higher-order rotating wave approximation method \cite{mirrahimi2015} where once again we recover the expected Kamiltonian from Eq. (9).

We hope that these two examples help to illustrate that the difference between the classical and quantum methods stems from specializations of the Lie bracket, and that they highlight the structural correspondence exploited in this work.

\subsection*{Classical Krylov-Bogoliubov (KB) averaging method}
\label{subsec:KB}
We start with a briefing on the KB averaging procedure developed in \cite{rahav2003}. We consider the Hamilton's equations of motion (EOM)
\begin{align}\label{eq:kb-eom}
    \begin{cases}
    \dot{q} = \partial_p {H}\\
    \dot{p} = - \partial_q {H}
    \end{cases}
\end{align}
for a time-periodic Hamiltonian $H(t+T)\! =\! H(t)$.
The key principle of the KB method is to look for solutions of the form
\begin{align}\label{eq:kb-ansatz}
        q = \mathfrak q + \zeta_q(\mathfrak q, \mathfrak p, t),\quad  p = \mathfrak p + \zeta_p(\mathfrak q, \mathfrak p, t),
\end{align}
where  $\mathfrak q$ and $\mathfrak p$ solve the ``averaged" EOMs
\begin{align}
\begin{cases}
        \dot{\mathfrak q} = \sum_{n\in\mathbb N} A^{(n)}(\mathfrak q, \mathfrak p)\\
        \dot{\mathfrak p} = \sum_{n\in\mathbb N} B^{(n)}(\mathfrak q, \mathfrak p),
\end{cases}
\end{align}
where $A^{(n)}$, $B^{(n)}$  are of order $1/\omega^n$. To separate the fast dynamics of $q$, one demands the time average of $\zeta_q$ to vanish,
\begin{align}
    \bar{\zeta_q} = \frac{1}{2 \pi} \int_0^{2\pi} \zeta_q(\mathfrak q, \mathfrak p, t)= 0.
\end{align} Here the time integration is only over the explicit time dependency on $\zeta_q$. Note that $\bar{\zeta_p}$ is assumed to be non-zero in general otherwise it will over-constrain the computation \cite{grozdanov1988}. 

Plugging \cref{eq:kb-ansatz} into \cref{eq:kb-eom}, one then obtains
\begin{align}\label{eq:kb-eom-expand}
    \begin{cases}
    \dot{\mathfrak q} + \frac{\partial\zeta_q}{\partial t}+ \frac{\partial\zeta_q}{\partial \mathfrak q}\dot{\mathfrak q}+ \frac{\partial\zeta_q}{\partial \mathfrak p}\dot{\mathfrak p} = \partial_p {H}(\mathfrak q + \zeta_q, \mathfrak p + \zeta_p, t)\\[3pt]
    \dot{\mathfrak p} + \frac{\partial\zeta_p}{\partial t}+ \frac{\partial\zeta_p}{\partial \mathfrak q}\dot{\mathfrak q}+ \frac{\partial\zeta_p}{\partial \mathfrak p}\dot{\mathfrak p} = -\partial_q {H}(\mathfrak q + \zeta_q, \mathfrak p + \zeta_p, t)
    \end{cases}
\end{align}

By further Taylor expanding the right hand side of the above equations around $q =\mathfrak q$, $p=\mathfrak p $ and expanding $\zeta_q = \sum_{n>0}\zeta_q^{(n)}$ and $\zeta_p = \sum_{n>0}\zeta_p^{(n)}$ as series in $1/\omega$, one can collect terms in \cref{eq:kb-eom-expand} at each order and compute $\zeta_q^{(n)}, \zeta_p^{(n)}, A^{(n)}$ and $B^{(n)}$ iteratively. For example, to leading order, one has 
\begin{align*}\label{eq:kb-eom-expand-0}
    \begin{cases}
    A^{(0)} + \frac{\partial\zeta_q^{(1)}}{\partial t}= \frac{\partial {H}(\mathfrak q, \mathfrak p, t)}{\partial \mathfrak p}\\[3pt]
    B^{(0)} + \frac{\partial\zeta_p^{(1)}}{\partial t}= -\frac{\partial {H}(\mathfrak q, \mathfrak p, t)}{\partial \mathfrak p}.
    \end{cases}
\end{align*}
Demanding $\zeta_q^{(1)}$ and $\zeta_p^{(1)}$ to balance out the explicit time-dependence of the corresponding equation, one obtains
\begin{align*}
\begin{split}
    A^{(0)} &= \partial_p \bar{H},\quad\zeta_q^{(1)} = -\int dt\, \partial_q\textbf{osc}(H),\\
    B^{(0)} &= -\partial_q \bar{H},\quad
    \zeta_p^{(1)} = \int dt\, \partial_q\textbf{osc}(H).
\end{split}
\end{align*}

To prove that the KB method can be recovered from the Lie approach, it is sufficient to prove the following claim.

\textit{Claim}: For the $\zeta_{q,p}$ described above, there exist invertible maps $\Phi_q, \Phi_p$ such that $S = \Phi_q(\zeta_q) = \Phi_p(\zeta_p)$ and \cref{eq:kb-eom-expand} is equivalent to 
\begin{align}
    \begin{cases}
    \dot{\mathfrak q} = \frac{\partial K(\mathfrak q, \mathfrak p)}{\partial \mathfrak p}    \\[3pt]
    \dot{\mathfrak p} = -\frac{\partial K(\mathfrak q, \mathfrak p)}{\partial \mathfrak p}
    \end{cases}
\end{align}
where $K = e^{L_S} H - \int_{0}^{1} d \epsilon \, e^{\epsilon L_S} \dot{S}$.

\textit{Proof:}
We construct the map $\Phi_{q,p}$ as 
\begin{align}
\begin{split}
    \zeta_q = \Phi^{-1}_q(S) = e^{L_S}\mathfrak q - \mathfrak q,\\ \zeta_p = \Phi^{-1}_p(S) = e^{L_S}\mathfrak p - \mathfrak p, 
\end{split}
\end{align}
where the Lie derivative $L_S = \{S(\mathfrak q ,\mathfrak p, t),\square \}$ and $\{\square, \square\}$ is the Poisson bracket over $\mathfrak q$ and $\mathfrak p$. Under this mapping,  we then have $q = e^{L_S} \mathfrak q$ and $p= e^{L_S} \mathfrak p$. Note that the exponential map here is $e^{L_S}$ instead of $e^{L_{-S}}$ because Hamilton's EOMs is formulated in the passive representation.

With this construction, we now prove the above claim at each order of $1/\omega$ by induction. 
For order zero, the first equation in \cref{eq:kb-eom-expand} is
\begin{gather*}
    \dot{\mathfrak q}^{(0)} + (\partial_t\zeta_q)^{(0)} = \{\mathfrak q, H(\mathfrak q,\mathfrak p, t)\}\\
    \Updownarrow\\
    \dot{\mathfrak q}^{(0)} + (\partial_te^{L_S}\mathfrak q-\partial_t\mathfrak q)^{(0)}=\{\mathfrak q, H(\mathfrak q,\mathfrak p, t)\}\\
    \Updownarrow\\
    \dot{\mathfrak q}^{(0)} + \{\dot{S}^{(1)}, \mathfrak q\} = \{\mathfrak q, H(\mathfrak q,\mathfrak p, t)\}\\
    \Updownarrow\\
    \dot{\mathfrak q}^{(0)} = \{\mathfrak q,  H(\mathfrak q,\mathfrak p, t) + \dot{S}^{(1)}\}\\
    \Updownarrow\\
     \dot{\mathfrak q}^{(0)} = \{\mathfrak q, K^{(0)}\},
\end{gather*}
where the superscript in a function $f^{(n)}$ should be now understood as a functional taking the $n^{\mathrm{th}}$ order of $f$. 

Similarly, the second equation in \cref{eq:kb-eom-expand} can be proven to be equivalent to $\dot{\mathfrak p}^{(0)} = -\{\mathfrak p, K^{(0)}\}$. Since $\mathfrak q$ and $\mathfrak p$ are conjugate variables, it can be readily verified that $\Phi_q(\zeta_q)$ and $\Phi_p(\zeta_p)$ will give the same $S^{(1)}$. This proves the claim to order zero.

Assuming that the claim holds to order $n$, order $n+1$ reads
\begin{gather*}
\begin{align*}
        \Big(\dot{\mathfrak q} + \frac{\partial\zeta_q}{\partial t}+ \frac{\partial\zeta_q}{\partial \mathfrak q}\dot{\mathfrak q} &+ \frac{\partial\zeta_q}{\partial \mathfrak p}\dot{\mathfrak p}\Big)^{(n+1)} \\
        &= \left(\partial_p {H}(\mathfrak q + \zeta_q, \mathfrak p + \zeta_p, t)\right)^{(n+1)}
\end{align*}\\[3pt]
    \Updownarrow\\[3pt]
    \begin{align*}
     \dot{\mathfrak q}^{(n+1)} + \left(\frac{\partial\zeta_q}{\partial t}\right)^{(n+1)} &+ \{\zeta_q, K\}^{(n+1)} \\
     &\,= e^{L_S}\{\mathfrak q, H(\mathfrak q,\mathfrak p,t)\}^{(n+1)}.  
    \end{align*}
\end{gather*}
On the right hand side of the last equation, we have used $e^{L_S}F(\mathfrak q,\mathfrak p) = F(e^{L_S}\mathfrak q,e^{L_S}\mathfrak p)$ for a general phase-space function $F$. We further write $\zeta_q$ and $K$ in terms of $S$ explicitly and after some manipulation we obtain
\begin{gather*}
\begin{align*}
    \dot{\mathfrak q}^{(n+1)} &+ \left(\partial_t(e^{L_S}\mathfrak q)\right)^{(n+1)} \\
    &+ \{e^{L_S}\mathfrak q, \int_{0}^{1} d \epsilon\, e^{\epsilon L_S} \dot{S}\}^{(n+1)} =\{\mathfrak q, K\}^{(n+1)}
\end{align*}\\
    \Updownarrow\\
\begin{align*}
    \dot{\mathfrak q}^{(n+1)} &+ \left(e^{L_S}e^{-L_S}\partial_t(e^{L_S}\mathfrak) \mathfrak q\right)^{(n+1)} \\
    &+ e^{L_S}\{\mathfrak q, \int_{0}^{1} d \epsilon\, e^{(\epsilon-1) L_S} \dot{S}\}^{(n+1)} =\{\mathfrak q, K\}^{(n+1)}
\end{align*}
\end{gather*}
One can then apply the identity \cref{eq:result} to the term $e^{-L_S}\partial_t(e^{L_S})$ in the above equation, and arrive at the desired result
\begin{gather*}
    \dot{\mathfrak q}^{(n+1)} = \{\mathfrak q, K^{(n+1)}\}.
\end{gather*}
Similarly, one can show that at order $n+1$ the second equation in \cref{eq:kb-eom-expand} is equivalent to $\dot{\mathfrak p}^{(n+1)} = \{\mathfrak p, K\}^{(n+1)}$. This completes the proof. 

We remark that, to prove the equivalence between KB method and the Lie method, three are the key steps: (1) identifying the relationship $\zeta_q = (e^{L_S} - I)\mathfrak q$, (2) rewriting the iterative procedure of KB in terms of Poisson brackets, and (3) using only general Lie algebra properties (instead of properties specific to Poisson brackets such as the explicit differentiation in \cref{eq:kb-eom-expand}) to prove the above claim. Without \textit{a priori} knowledge of the shared Lie structure underlying the two methods, their unification cannot be achieved. Order-by-order comparison between the consitutent components, i.e., $A^{(n)}$, $B^{(n)}$, and $\zeta_{q,p}^{(n)}$ in KB with $S^{(n)}$ and $K^{(n)}$ in Eq. (8), sheds no light on the connection between the methods.

\subsection*{Quantum higher-order rotating wave approximation}
We now brief on the quantum higher-order RWA method summarized in Sec. 2.1 of \cite{mirrahimi2015}. 
For a time-dependent Schr\"odinger equation $i\hbar \partial_t|\phi(t)\rangle = \hat{H}(t)|\phi(t)\rangle $ with a periodic Hamiltonian $ \hat{H}(t) =  \hat{H}(t+2\pi/\omega) $, this method assumes an ansatz of the form
\begin{align}
	|\phi(t)\rangle = |\varphi(t)\rangle + \hat{\delta}(t)|{\varphi(t)}\rangle,
\end{align}
where $|\varphi(t)\rangle$ is the solution to an ``averaged'' Schrodinger equation $i\hbar|\varphi(t)\rangle = \hat{K}_\text{RWA}|\varphi(t)\rangle$
and $\hat{\delta}(t) = \sum_{n>0} \hat{\delta}^{(n)}(t)$ is an operator periodic in $t$ with $\hat{\delta}^{(n)}$ of order $n$ in $1/\omega$ and $\bar{{\delta}} = 0$. With this ansatz, the Schr\"odinger equation transforms to
\begin{align}
\label{eq:KB-state}
	i\hbar \partial_t |\varphi\rangle + 	i\hbar (\partial_t \hat{\delta})|\varphi\rangle + \hat{\delta} i\hbar \partial_t|\varphi\rangle = (\hat{H}+\hat{H}\hat{\delta})|\varphi\rangle.
\end{align}
Then by collecting terms at each order one can compute $\hat{K}_\text{RWA}$ and $\hat{\delta}(t)$ iteratively by demanding the time dependence of \cref{eq:KB-state} to cancel out. 

Following a procedure similar to the one used for the classical KB method, one can prove the equivalence between the higher-order RWA and the Lie method by induction. For the sake of conciseness, here we only comment the relation between $\hat{\delta}$ and $\hat{S}$ and sketch the proof of the inductive step. 

In particular, we identify that $\hat{\delta}$ is related to $\hat{S}$ by $\hat{\delta} = e^{-\hat{S}/i\hbar}-1$. With this relation, the inductive step can be proven by
\begin{gather*}
\begin{align*}
    \Big(i\hbar \partial_t |\varphi\rangle + 	i\hbar (\partial_t \hat{\delta})|\varphi\rangle &+ \hat{\delta} i\hbar \partial_t|\varphi\rangle\Big)^{(n+1)}\\
    &= (\hat{H}+\hat{H}\hat{\delta})^{(n+1)}|\varphi\rangle 
\end{align*}\\
						\Updownarrow \nonumber\\
\begin{align*}
&i\hbar (\partial_t |\varphi\rangle)^{(n+1)}  + i\hbar(\partial_t e^{-\hat{S}/i\hbar})^{(n+1)}|\varphi\rangle \\&+ \Big((e^{-\hat{S}/i\hbar} -1) e^{\hat{S}/i\hbar}(\hat{H}-i\hbar\partial_t) e^{-\hat{S}/i\hbar} \Big)^{(n+1)}|\varphi\rangle \\
&\qquad\qquad\qquad\qquad\qquad\qquad=  (\hat{H}e^{-\hat{S}/i\hbar})^{(n+1)}|\varphi\rangle  
\end{align*}\\
								\Updownarrow \nonumber\\
	 		i\hbar( \partial_t |\varphi\rangle )^{(n+1)}  = \Big( e^{\hat{S}/i\hbar}(\hat{H}-i\hbar\partial_t) e^{-i\hat{S}}\Big)^{(n+1)} |\varphi\rangle\\
	 										\Updownarrow \nonumber\\
	 		i\hbar( \partial_t |\varphi\rangle )^{(n+1)}  = \hat{K}^{(n+1)} |\varphi\rangle,
\end{gather*}
where we used the fact that for matrix Lie groups $e^{L_S} H + \int_{0}^{1} d \epsilon \, e^{\epsilon L_S} \dot{S} = e^{\hat{S}/i\hbar}(\hat{H}-i\hbar\partial_t) e^{-\hat{S}/i\hbar}$.

\section{Closed form of the van Vleck expansion}
To illustrate the recursive formula, we consider as an example the widely employed quantum van Vleck perturbative expansion.
We have written a symbolic algebra algorithm, made available in \cite{xiao2021}, to compute it. Note that for code efficiency Eq. (8b) here is restated as $\hat{S}^{(n+1)} = -\int dt\, \mathrm{\mathbf{osc}}(L_{S^{(n)}}\hat{H}+\sum_{k\ne1}\hat{K}^{(n)}_{[k]})$. Below, we display the results of our algorithm up to order $5$. To recover the form of this operator-valued expansion as written in previous literature, the sub-routine of the algorithm needs to be adapted to the commutator \cite{eckardt2015,mikami2016,rahav2003}.
To our knowledge, this formula was only available in the literature to order 3 provided in \cite{mikami2016}.

For notational simplicity, in the following expressions we suppress the summation symbol. When a Fourier index $m_i$ appears in an expression it implies the summation over all valid $m_i\in \mathbb Z$. The (composite) Fourier index of a Hamiltonian term or that of a commutator (computed by summing the indices of the contained terms) should be non-zero unless it is zero by construction. The choices of $m_i$ violating this constraint are excluded. For example, under this convention the last terms in $\hat{K}^{(2)}$ and $\hat K^{(3)}$ should be understood as 
\begin{align*}
&\sum\limits_{m_1\ne0}\sum\limits_{\substack{m_2\ne0 \\ m_2\ne m_1}}\frac{[[\hat{H}_{m_2},\hat{H}_{m_1 - m_2}],\hat{H}_{-m_1}]}{3m_1m_2(\hbar\omega)^2}\\
&\sum\limits_{m_1\ne0}\sum\limits_{\substack{m_2\ne0 \\ m_2\ne m_1}}\sum\limits_{\substack{m_3\ne0 \\ m_3\ne m_1-m_2}}\!\!\!\!\!\!\!\frac{[[\hat{H}_{m_2},[\hat{H}_{m_3},\hat{H}_{m_1 - m_2 - m_3}]],\hat{H}_{-m_1}]}{24m_1m_2m_3(\hbar\omega)^3},
\end{align*}
respectively.

We note that the apparent discrepancies between our formula up to order three and that in \cite{mikami2016} are resolved by a trivial regrouping of terms stemming from a different choice in the range of the summation.

The expansions for $\hat K$ and $\hat S$ reads
\begin{widetext}
\begin{align*}
\hat{K}^{(0)} = &\;\hat{H}_0\\
\hat{K}^{(1)} = &\; \frac{[\hat{H}_{m_1},\hat{H}_{-m_1}]}{2m_1(\hbar\omega)}\\
\hat{K}^{(2)} = &\;\frac{[[\hat{H}_{m_1},\hat{H}_0],\hat{H}_{-m_1}]}{2m_1^2(\hbar\omega)^2}+\frac{[[\hat{H}_{m_2},\hat{H}_{m_1 - m_2}],\hat{H}_{-m_1}]}{3m_1m_2(\hbar\omega)^2}\\ 
\hat{K}^{(3)}=&\;\frac{[[[\hat{H}_{m_1},\hat{H}_0],\hat{H}_0],\hat{H}_{-m_1}]}{2m_1^3(\hbar\omega)^3}+\frac{[[[\hat{H}_{m_2},\hat{H}_0],\hat{H}_{m_1 - m_2}],\hat{H}_{-m_1}]}{3m_1m_2^2(\hbar\omega)^3}\\ 
&+\frac{[[[\hat{H}_{m_2},\hat{H}_{m_1 - m_2}],\hat{H}_0],\hat{H}_{-m_1}]}{4m_1^2m_2(\hbar\omega)^3}+\frac{[[[\hat{H}_{m_3},\hat{H}_{m_2 - m_3}],\hat{H}_{m_1 - m_2}],\hat{H}_{-m_1}]}{6m_1m_2m_3(\hbar\omega)^3}\\ 
&+\frac{[[\hat{H}_{m_1},\hat{H}_0],[\hat{H}_{m_2},\hat{H}_{-m_1 - m_2}]]}{12m_1^2m_2(\hbar\omega)^3}
+\frac{[[\hat{H}_{m_2},\hat{H}_{m_1 - m_2}],[\hat{H}_{m_3},\hat{H}_{-m_1 - m_3}]]}{24m_1m_2m_3(\hbar\omega)^3}\\
&+\frac{[[\hat{H}_{m_1},[\hat{H}_{m_2},\hat{H}_{-m_2}]],\hat{H}_{-m_1}]}{8m_1^2m_2(\hbar\omega)^3}+\frac{[[\hat{H}_{m_2},[\hat{H}_{m_3},\hat{H}_{m_1 - m_2 - m_3}]],\hat{H}_{-m_1}]}{24m_1m_2m_3(\hbar\omega)^3}\\
\hat{K}^{(4)}=
&\;\frac{[[[[\hat{H}_{m_1},\hat{H}_0],\hat{H}_0],\hat{H}_0],\hat{H}_{-m_1}]}{2m_1^4(\hbar\omega)^4}+\frac{[[[[\hat{H}_{m_2},\hat{H}_0],\hat{H}_0],\hat{H}_{m_1 - m_2}],\hat{H}_{-m_1}]}{3m_1m_2^3(\hbar\omega)^4}\\ 
&+\frac{[[[[\hat{H}_{m_2},\hat{H}_0],\hat{H}_{m_1 - m_2}],\hat{H}_0],\hat{H}_{-m_1}]}{4m_1^2m_2^2(\hbar\omega)^4}+\frac{[[[[\hat{H}_{m_3},\hat{H}_0],\hat{H}_{m_2 - m_3}],\hat{H}_{m_1 - m_2}],\hat{H}_{-m_1}]}{6m_1m_2m_3^2(\hbar\omega)^4}\\ 
&+\frac{[[[[\hat{H}_{m_2},\hat{H}_{m_1 - m_2}],\hat{H}_0],\hat{H}_0],\hat{H}_{-m_1}]}{4m_1^3m_2(\hbar\omega)^4}+\frac{[[[[\hat{H}_{m_3},\hat{H}_{m_2 - m_3}],\hat{H}_0],\hat{H}_{m_1 - m_2}],\hat{H}_{-m_1}]}{6m_1m_2^2m_3(\hbar\omega)^4}\\ 
&+\frac{[[[[\hat{H}_{m_3},\hat{H}_{m_2 - m_3}],\hat{H}_{m_1 - m_2}],\hat{H}_0],\hat{H}_{-m_1}]}{8m_1^2m_2m_3(\hbar\omega)^4}+\frac{[[[[\hat{H}_{m_4},\hat{H}_{m_3 - m_4}],\hat{H}_{m_2 - m_3}],\hat{H}_{m_1 - m_2}],\hat{H}_{-m_1}]}{12m_1m_2m_3m_4(\hbar\omega)^4}\\ 
&+\frac{[[[\hat{H}_{m_2},[\hat{H}_{m_3},\hat{H}_{m_1 - m_2 - m_3}]],\hat{H}_0],\hat{H}_{-m_1}]}{24m_1^2m_2m_3(\hbar\omega)^4}+\frac{[[[\hat{H}_{m_3},[\hat{H}_{m_4},\hat{H}_{m_2 - m_3 - m_4}]],\hat{H}_{m_1 - m_2}],\hat{H}_{-m_1}]}{36m_1m_2m_3m_4(\hbar\omega)^4}\\ 
&+\frac{[[[\hat{H}_{m_1},[\hat{H}_{m_2},\hat{H}_{-m_2}]],\hat{H}_0],\hat{H}_{-m_1}]}{6m_1^3m_2(\hbar\omega)^4}+\frac{[[[\hat{H}_{m_2},[\hat{H}_{m_3},\hat{H}_{-m_3}]],\hat{H}_{m_1 - m_2}],\hat{H}_{-m_1}]}{9m_1m_2^2m_3(\hbar\omega)^4}\\ 
&+\frac{[[[\hat{H}_{m_2},\hat{H}_0],[\hat{H}_{m_3},\hat{H}_{m_1 - m_2 - m_3}]],\hat{H}_{-m_1}]}{24m_1m_2^2m_3(\hbar\omega)^4}+\frac{[[[\hat{H}_{m_1},\hat{H}_0],[\hat{H}_{m_2},\hat{H}_{-m_2}]],\hat{H}_{-m_1}]}{8m_1^3m_2(\hbar\omega)^4}\\ 
&+\frac{[[[\hat{H}_{m_3},\hat{H}_{m_2 - m_3}],[\hat{H}_{m_4},\hat{H}_{m_1 - m_2 - m_4}]],\hat{H}_{-m_1}]}{48m_1m_2m_3m_4(\hbar\omega)^4}+\frac{[[[\hat{H}_{m_2},\hat{H}_{m_1 - m_2}],[\hat{H}_{m_3},\hat{H}_{-m_3}]],\hat{H}_{-m_1}]}{16m_1^2m_2m_3(\hbar\omega)^4}\\ 
&+\frac{[[\hat{H}_{m_2},[[\hat{H}_{m_3},\hat{H}_0],\hat{H}_{m_1 - m_2 - m_3}]],\hat{H}_{-m_1}]}{24m_1m_2m_3^2(\hbar\omega)^4}+\frac{[[\hat{H}_{m_2},[[\hat{H}_{m_4},\hat{H}_{m_3 - m_4}],\hat{H}_{m_1 - m_2 - m_3}]],\hat{H}_{-m_1}]}{48m_1m_2m_3m_4(\hbar\omega)^4}\\ 
&-\frac{[[\hat{H}_{m_2},[\hat{H}_{m_1 - m_2},[\hat{H}_{m_3},\hat{H}_{-m_3}]]],\hat{H}_{-m_1}]}{45m_1m_2m_3(m_1 - m_2)(\hbar\omega)^4}+\frac{[[\hat{H}_{m_1},[[\hat{H}_{m_2},\hat{H}_0],\hat{H}_{-m_2}]],\hat{H}_{-m_1}]}{8m_1^2m_2^2(\hbar\omega)^4}\\ 
&+\frac{3[[\hat{H}_{m_1},[[\hat{H}_{m_3},\hat{H}_{m_2 - m_3}],\hat{H}_{-m_2}]],\hat{H}_{-m_1}]}{40m_1^2m_2m_3(\hbar\omega)^4}+\frac{[[[\hat{H}_{m_1},\hat{H}_0],\hat{H}_0],[\hat{H}_{m_2},\hat{H}_{-m_1 - m_2}]]}{12m_1^3m_2(\hbar\omega)^4}\\ 
&+\frac{[[[\hat{H}_{m_2},\hat{H}_0],\hat{H}_{m_1 - m_2}],[\hat{H}_{m_3},\hat{H}_{-m_1 - m_3}]]}{12m_1m_2^2m_3(\hbar\omega)^4}+\frac{[[[\hat{H}_{m_2},\hat{H}_{m_1 - m_2}],\hat{H}_0],[\hat{H}_{m_3},\hat{H}_{-m_1 - m_3}]]}{24m_1^2m_2m_3(\hbar\omega)^4}\\ 
&+\frac{[[[\hat{H}_{m_3},\hat{H}_{m_2 - m_3}],\hat{H}_{m_1 - m_2}],[\hat{H}_{m_4},\hat{H}_{-m_1 - m_4}]]}{24m_1m_2m_3m_4(\hbar\omega)^4}+\frac{[[\hat{H}_{m_1},\hat{H}_0],[[\hat{H}_{m_2},\hat{H}_0],\hat{H}_{-m_1 - m_2}]]}{12m_1^2m_2^2(\hbar\omega)^4}\\ 
&+\frac{[[\hat{H}_{m_1},\hat{H}_0],[[\hat{H}_{m_3},\hat{H}_{m_2 - m_3}],\hat{H}_{-m_1 - m_2}]]}{24m_1^2m_2m_3(\hbar\omega)^4}+\frac{[[\hat{H}_{m_1},\hat{H}_0],[\hat{H}_{-m_1},[\hat{H}_{m_2},\hat{H}_{-m_2}]]]}{24m_1^3m_2(\hbar\omega)^4}\\ 
&+\frac{[[\hat{H}_{m_2},[\hat{H}_{m_3},\hat{H}_{m_1 - m_2 - m_3}]],[\hat{H}_{m_4},\hat{H}_{-m_1 - m_4}]]}{144m_1m_2m_3m_4(\hbar\omega)^4}+\frac{[[\hat{H}_{m_1},[\hat{H}_{m_2},\hat{H}_{-m_2}]],[\hat{H}_{m_3},\hat{H}_{-m_1 - m_3}]]}{144m_1^2m_2m_3(\hbar\omega)^4}\\ 
&-\frac{[\hat{H}_{m_1},[\hat{H}_{m_2},[\hat{H}_{m_3},[\hat{H}_{m_4},\hat{H}_{-m_1 - m_2 - m_3 - m_4}]]]]}{720m_1m_2m_3m_4(\hbar\omega)^4}
\end{align*}
\begin{align*}
&\hat{K}^{(5)} =\\
&\;\;\frac{[[[[[\hat{H}_{m_1},\hat{H}_0],\hat{H}_0],\hat{H}_0],\hat{H}_0],\hat{H}_{-m_1}]}{2m_1^5(\hbar\omega)^5}+\frac{[[[[[\hat{H}_{m_2},\hat{H}_0],\hat{H}_0],\hat{H}_0],\hat{H}_{m_1 - m_2}],\hat{H}_{-m_1}]}{3m_1m_2^4(\hbar\omega)^5}\\ 
&+\frac{[[[[[\hat{H}_{m_2},\hat{H}_0],\hat{H}_0],\hat{H}_{m_1 - m_2}],\hat{H}_0],\hat{H}_{-m_1}]}{4m_1^2m_2^3(\hbar\omega)^5}+\frac{[[[[[\hat{H}_{m_3},\hat{H}_0],\hat{H}_0],\hat{H}_{m_2 - m_3}],\hat{H}_{m_1 - m_2}],\hat{H}_{-m_1}]}{6m_1m_2m_3^3(\hbar\omega)^5}\\ 
&+\frac{[[[[[\hat{H}_{m_2},\hat{H}_0],\hat{H}_{m_1 - m_2}],\hat{H}_0],\hat{H}_0],\hat{H}_{-m_1}]}{4m_1^3m_2^2(\hbar\omega)^5}+\frac{[[[[[\hat{H}_{m_3},\hat{H}_0],\hat{H}_{m_2 - m_3}],\hat{H}_0],\hat{H}_{m_1 - m_2}],\hat{H}_{-m_1}]}{6m_1m_2^2m_3^2(\hbar\omega)^5}\\ 
&+\frac{[[[[[\hat{H}_{m_3},\hat{H}_0],\hat{H}_{m_2 - m_3}],\hat{H}_{m_1 - m_2}],\hat{H}_0],\hat{H}_{-m_1}]}{8m_1^2m_2m_3^2(\hbar\omega)^5}+\frac{[[[[[\hat{H}_{m_4},\hat{H}_0],\hat{H}_{m_3 - m_4}],\hat{H}_{m_2 - m_3}],\hat{H}_{m_1 - m_2}],\hat{H}_{-m_1}]}{12m_1m_2m_3m_4^2(\hbar\omega)^5}\\ 
&+\frac{[[[[[\hat{H}_{m_2},\hat{H}_{m_1 - m_2}],\hat{H}_0],\hat{H}_0],\hat{H}_0],\hat{H}_{-m_1}]}{4m_1^4m_2(\hbar\omega)^5}+\frac{[[[[[\hat{H}_{m_3},\hat{H}_{m_2 - m_3}],\hat{H}_0],\hat{H}_0],\hat{H}_{m_1 - m_2}],\hat{H}_{-m_1}]}{6m_1m_2^3m_3(\hbar\omega)^5}\\ 
&+\frac{[[[[[\hat{H}_{m_3},\hat{H}_{m_2 - m_3}],\hat{H}_0],\hat{H}_{m_1 - m_2}],\hat{H}_0],\hat{H}_{-m_1}]}{8m_1^2m_2^2m_3(\hbar\omega)^5}+\frac{[[[[[\hat{H}_{m_4},\hat{H}_{m_3 - m_4}],\hat{H}_0],\hat{H}_{m_2 - m_3}],\hat{H}_{m_1 - m_2}],\hat{H}_{-m_1}]}{12m_1m_2m_3^2m_4(\hbar\omega)^5}\\ 
&+\frac{[[[[[\hat{H}_{m_3},\hat{H}_{m_2 - m_3}],\hat{H}_{m_1 - m_2}],\hat{H}_0],\hat{H}_0],\hat{H}_{-m_1}]}{8m_1^3m_2m_3(\hbar\omega)^5}+\frac{[[[[[\hat{H}_{m_4},\hat{H}_{m_3 - m_4}],\hat{H}_{m_2 - m_3}],\hat{H}_0],\hat{H}_{m_1 - m_2}],\hat{H}_{-m_1}]}{12m_1m_2^2m_3m_4(\hbar\omega)^5}\\ 
&+\frac{[[[[[\hat{H}_{m_4},\hat{H}_{m_3 - m_4}],\hat{H}_{m_2 - m_3}],\hat{H}_{m_1 - m_2}],\hat{H}_0],\hat{H}_{-m_1}]}{16m_1^2m_2m_3m_4(\hbar\omega)^5}+\frac{[[[[[\hat{H}_{m_5},\hat{H}_{m_4 - m_5}],\hat{H}_{m_3 - m_4}],\hat{H}_{m_2 - m_3}],\hat{H}_{m_1 - m_2}],\hat{H}_{-m_1}]}{24m_1m_2m_3m_4m_5(\hbar\omega)^5}\\ 
&+\frac{[[[[\hat{H}_{m_2},[\hat{H}_{m_3},\hat{H}_{m_1 - m_2 - m_3}]],\hat{H}_0],\hat{H}_0],\hat{H}_{-m_1}]}{24m_1^3m_2m_3(\hbar\omega)^5}+\frac{[[[[\hat{H}_{m_3},[\hat{H}_{m_4},\hat{H}_{m_2 - m_3 - m_4}]],\hat{H}_0],\hat{H}_{m_1 - m_2}],\hat{H}_{-m_1}]}{36m_1m_2^2m_3m_4(\hbar\omega)^5}\\ 
&+\frac{[[[[\hat{H}_{m_3},[\hat{H}_{m_4},\hat{H}_{m_2 - m_3 - m_4}]],\hat{H}_{m_1 - m_2}],\hat{H}_0],\hat{H}_{-m_1}]}{48m_1^2m_2m_3m_4(\hbar\omega)^5}+\frac{[[[[\hat{H}_{m_4},[\hat{H}_{m_5},\hat{H}_{m_3 - m_4 - m_5}]],\hat{H}_{m_2 - m_3}],\hat{H}_{m_1 - m_2}],\hat{H}_{-m_1}]}{72m_1m_2m_3m_4m_5(\hbar\omega)^5}\\ 
&+\frac{[[[[\hat{H}_{m_1},[\hat{H}_{m_2},\hat{H}_{-m_2}]],\hat{H}_0],\hat{H}_0],\hat{H}_{-m_1}]}{6m_1^4m_2(\hbar\omega)^5}+\frac{[[[[\hat{H}_{m_2},[\hat{H}_{m_3},\hat{H}_{-m_3}]],\hat{H}_0],\hat{H}_{m_1 - m_2}],\hat{H}_{-m_1}]}{9m_1m_2^3m_3(\hbar\omega)^5}\\ 
&+\frac{[[[[\hat{H}_{m_2},[\hat{H}_{m_3},\hat{H}_{-m_3}]],\hat{H}_{m_1 - m_2}],\hat{H}_0],\hat{H}_{-m_1}]}{12m_1^2m_2^2m_3(\hbar\omega)^5}+\frac{[[[[\hat{H}_{m_3},[\hat{H}_{m_4},\hat{H}_{-m_4}]],\hat{H}_{m_2 - m_3}],\hat{H}_{m_1 - m_2}],\hat{H}_{-m_1}]}{18m_1m_2m_3^2m_4(\hbar\omega)^5}\\ 
&+\frac{[[[[\hat{H}_{m_2},\hat{H}_0],[\hat{H}_{m_3},\hat{H}_{m_1 - m_2 - m_3}]],\hat{H}_0],\hat{H}_{-m_1}]}{24m_1^2m_2^2m_3(\hbar\omega)^5}+\frac{[[[[\hat{H}_{m_3},\hat{H}_0],[\hat{H}_{m_4},\hat{H}_{m_2 - m_3 - m_4}]],\hat{H}_{m_1 - m_2}],\hat{H}_{-m_1}]}{36m_1m_2m_3^2m_4(\hbar\omega)^5}\\ 
&+\frac{[[[[\hat{H}_{m_1},\hat{H}_0],[\hat{H}_{m_2},\hat{H}_{-m_2}]],\hat{H}_0],\hat{H}_{-m_1}]}{6m_1^4m_2(\hbar\omega)^5}+\frac{[[[[\hat{H}_{m_2},\hat{H}_0],[\hat{H}_{m_3},\hat{H}_{-m_3}]],\hat{H}_{m_1 - m_2}],\hat{H}_{-m_1}]}{9m_1m_2^3m_3(\hbar\omega)^5}\\ 
&+\frac{[[[[\hat{H}_{m_3},\hat{H}_{m_2 - m_3}],[\hat{H}_{m_4},\hat{H}_{m_1 - m_2 - m_4}]],\hat{H}_0],\hat{H}_{-m_1}]}{48m_1^2m_2m_3m_4(\hbar\omega)^5}+\frac{[[[[\hat{H}_{m_4},\hat{H}_{m_3 - m_4}],[\hat{H}_{m_5},\hat{H}_{m_2 - m_3 - m_5}]],\hat{H}_{m_1 - m_2}],\hat{H}_{-m_1}]}{72m_1m_2m_3m_4m_5(\hbar\omega)^5}\\ 
&+\frac{[[[[\hat{H}_{m_2},\hat{H}_{m_1 - m_2}],[\hat{H}_{m_3},\hat{H}_{-m_3}]],\hat{H}_0],\hat{H}_{-m_1}]}{12m_1^3m_2m_3(\hbar\omega)^5}+\frac{[[[[\hat{H}_{m_3},\hat{H}_{m_2 - m_3}],[\hat{H}_{m_4},\hat{H}_{-m_4}]],\hat{H}_{m_1 - m_2}],\hat{H}_{-m_1}]}{18m_1m_2^2m_3m_4(\hbar\omega)^5}\\ 
&+\frac{[[[\hat{H}_{m_2},[[\hat{H}_{m_3},\hat{H}_0],\hat{H}_{m_1 - m_2 - m_3}]],\hat{H}_0],\hat{H}_{-m_1}]}{24m_1^2m_2m_3^2(\hbar\omega)^5}+\frac{[[[\hat{H}_{m_3},[[\hat{H}_{m_4},\hat{H}_0],\hat{H}_{m_2 - m_3 - m_4}]],\hat{H}_{m_1 - m_2}],\hat{H}_{-m_1}]}{36m_1m_2m_3m_4^2(\hbar\omega)^5}\\ 
&+\frac{[[[\hat{H}_{m_2},[[\hat{H}_{m_4},\hat{H}_{m_3 - m_4}],\hat{H}_{m_1 - m_2 - m_3}]],\hat{H}_0],\hat{H}_{-m_1}]}{48m_1^2m_2m_3m_4(\hbar\omega)^5}+\frac{[[[\hat{H}_{m_3},[[\hat{H}_{m_5},\hat{H}_{m_4 - m_5}],\hat{H}_{m_2 - m_3 - m_4}]],\hat{H}_{m_1 - m_2}],\hat{H}_{-m_1}]}{72m_1m_2m_3m_4m_5(\hbar\omega)^5}\\ 
&-\frac{[[[\hat{H}_{m_2},[\hat{H}_{m_1 - m_2},[\hat{H}_{m_3},\hat{H}_{-m_3}]]],\hat{H}_0],\hat{H}_{-m_1}]}{48m_1^2m_2m_3(m_1 - m_2)(\hbar\omega)^5}-\frac{[[[\hat{H}_{m_3},[\hat{H}_{m_2 - m_3},[\hat{H}_{m_4},\hat{H}_{-m_4}]]],\hat{H}_{m_1 - m_2}],\hat{H}_{-m_1}]}{72m_1m_2m_3m_4(m_2 - m_3)(\hbar\omega)^5}\\ 
&+\frac{[[[\hat{H}_{m_1},[[\hat{H}_{m_2},\hat{H}_0],\hat{H}_{-m_2}]],\hat{H}_0],\hat{H}_{-m_1}]}{6m_1^3m_2^2(\hbar\omega)^5}+\frac{[[[\hat{H}_{m_2},[[\hat{H}_{m_3},\hat{H}_0],\hat{H}_{-m_3}]],\hat{H}_{m_1 - m_2}],\hat{H}_{-m_1}]}{9m_1m_2^2m_3^2(\hbar\omega)^5}\\ 
&+\frac{5[[[\hat{H}_{m_1},[[\hat{H}_{m_3},\hat{H}_{m_2 - m_3}],\hat{H}_{-m_2}]],\hat{H}_0],\hat{H}_{-m_1}]}{48m_1^3m_2m_3(\hbar\omega)^5}+\frac{5[[[\hat{H}_{m_2},[[\hat{H}_{m_4},\hat{H}_{m_3 - m_4}],\hat{H}_{-m_3}]],\hat{H}_{m_1 - m_2}],\hat{H}_{-m_1}]}{72m_1m_2^2m_3m_4(\hbar\omega)^5}\\ 
&+\frac{[[[[\hat{H}_{m_2},\hat{H}_0],\hat{H}_0],[\hat{H}_{m_3},\hat{H}_{m_1 - m_2 - m_3}]],\hat{H}_{-m_1}]}{24m_1m_2^3m_3(\hbar\omega)^5}+\frac{[[[[\hat{H}_{m_1},\hat{H}_0],\hat{H}_0],[\hat{H}_{m_2},\hat{H}_{-m_2}]],\hat{H}_{-m_1}]}{8m_1^4m_2(\hbar\omega)^5}\\ 
&+\frac{[[[[\hat{H}_{m_3},\hat{H}_0],\hat{H}_{m_2 - m_3}],[\hat{H}_{m_4},\hat{H}_{m_1 - m_2 - m_4}]],\hat{H}_{-m_1}]}{48m_1m_2m_3^2m_4(\hbar\omega)^5}+\frac{[[[[\hat{H}_{m_2},\hat{H}_0],\hat{H}_{m_1 - m_2}],[\hat{H}_{m_3},\hat{H}_{-m_3}]],\hat{H}_{-m_1}]}{16m_1^2m_2^2m_3(\hbar\omega)^5}\\ 
&+\frac{[[[[\hat{H}_{m_3},\hat{H}_{m_2 - m_3}],\hat{H}_0],[\hat{H}_{m_4},\hat{H}_{m_1 - m_2 - m_4}]],\hat{H}_{-m_1}]}{48m_1m_2^2m_3m_4(\hbar\omega)^5}+\frac{[[[[\hat{H}_{m_2},\hat{H}_{m_1 - m_2}],\hat{H}_0],[\hat{H}_{m_3},\hat{H}_{-m_3}]],\hat{H}_{-m_1}]}{16m_1^3m_2m_3(\hbar\omega)^5}\\ 
&+\frac{[[[[\hat{H}_{m_4},\hat{H}_{m_3 - m_4}],\hat{H}_{m_2 - m_3}],[\hat{H}_{m_5},\hat{H}_{m_1 - m_2 - m_5}]],\hat{H}_{-m_1}]}{96m_1m_2m_3m_4m_5(\hbar\omega)^5}+\frac{[[[[\hat{H}_{m_3},\hat{H}_{m_2 - m_3}],\hat{H}_{m_1 - m_2}],[\hat{H}_{m_4},\hat{H}_{-m_4}]],\hat{H}_{-m_1}]}{32m_1^2m_2m_3m_4(\hbar\omega)^5}\\ 
&+\frac{[[[\hat{H}_{m_3},[\hat{H}_{m_4},\hat{H}_{m_2 - m_3 - m_4}]],[\hat{H}_{m_5},\hat{H}_{m_1 - m_2 - m_5}]],\hat{H}_{-m_1}]}{288m_1m_2m_3m_4m_5(\hbar\omega)^5}+\frac{[[[\hat{H}_{m_2},[\hat{H}_{m_3},\hat{H}_{m_1 - m_2 - m_3}]],[\hat{H}_{m_4},\hat{H}_{-m_4}]],\hat{H}_{-m_1}]}{96m_1^2m_2m_3m_4(\hbar\omega)^5}\\ 
&+\frac{[[[\hat{H}_{m_2},[\hat{H}_{m_3},\hat{H}_{-m_3}]],[\hat{H}_{m_4},\hat{H}_{m_1 - m_2 - m_4}]],\hat{H}_{-m_1}]}{72m_1m_2^2m_3m_4(\hbar\omega)^5}+\frac{[[[\hat{H}_{m_1},[\hat{H}_{m_2},\hat{H}_{-m_2}]],[\hat{H}_{m_3},\hat{H}_{-m_3}]],\hat{H}_{-m_1}]}{24m_1^3m_2m_3(\hbar\omega)^5}\\ 
&+\frac{[[[\hat{H}_{m_2},\hat{H}_0],[[\hat{H}_{m_3},\hat{H}_0],\hat{H}_{m_1 - m_2 - m_3}]],\hat{H}_{-m_1}]}{24m_1m_2^2m_3^2(\hbar\omega)^5}+\frac{[[[\hat{H}_{m_2},\hat{H}_0],[[\hat{H}_{m_4},\hat{H}_{m_3 - m_4}],\hat{H}_{m_1 - m_2 - m_3}]],\hat{H}_{-m_1}]}{48m_1m_2^2m_3m_4(\hbar\omega)^5}\\ 
&-\frac{[[[\hat{H}_{m_2},\hat{H}_0],[\hat{H}_{m_1 - m_2},[\hat{H}_{m_3},\hat{H}_{-m_3}]]],\hat{H}_{-m_1}]}{45m_1m_2^2m_3(m_1 - m_2)(\hbar\omega)^5}+\frac{[[[\hat{H}_{m_1},\hat{H}_0],[[\hat{H}_{m_2},\hat{H}_0],\hat{H}_{-m_2}]],\hat{H}_{-m_1}]}{8m_1^3m_2^2(\hbar\omega)^5}\\ 
&+\frac{3[[[\hat{H}_{m_1},\hat{H}_0],[[\hat{H}_{m_3},\hat{H}_{m_2 - m_3}],\hat{H}_{-m_2}]],\hat{H}_{-m_1}]}{40m_1^3m_2m_3(\hbar\omega)^5}+\frac{[[[\hat{H}_{m_3},\hat{H}_{m_2 - m_3}],[[\hat{H}_{m_4},\hat{H}_0],\hat{H}_{m_1 - m_2 - m_4}]],\hat{H}_{-m_1}]}{48m_1m_2m_3m_4^2(\hbar\omega)^5}\\ 
&+\frac{[[[\hat{H}_{m_3},\hat{H}_{m_2 - m_3}],[[\hat{H}_{m_5},\hat{H}_{m_4 - m_5}],\hat{H}_{m_1 - m_2 - m_4}]],\hat{H}_{-m_1}]}{96m_1m_2m_3m_4m_5(\hbar\omega)^5}-\frac{[[[\hat{H}_{m_3},\hat{H}_{m_2 - m_3}],[\hat{H}_{m_1 - m_2},[\hat{H}_{m_4},\hat{H}_{-m_4}]]],\hat{H}_{-m_1}]}{90m_1m_2m_3m_4(m_1 - m_2)(\hbar\omega)^5}\\ 
&+\frac{[[[\hat{H}_{m_2},\hat{H}_{m_1 - m_2}],[[\hat{H}_{m_3},\hat{H}_0],\hat{H}_{-m_3}]],\hat{H}_{-m_1}]}{16m_1^2m_2m_3^2(\hbar\omega)^5}+\frac{3[[[\hat{H}_{m_2},\hat{H}_{m_1 - m_2}],[[\hat{H}_{m_4},\hat{H}_{m_3 - m_4}],\hat{H}_{-m_3}]],\hat{H}_{-m_1}]}{80m_1^2m_2m_3m_4(\hbar\omega)^5}\\ 
&+\frac{[[\hat{H}_{m_2},[[[\hat{H}_{m_3},\hat{H}_0],\hat{H}_0],\hat{H}_{m_1 - m_2 - m_3}]],\hat{H}_{-m_1}]}{24m_1m_2m_3^3(\hbar\omega)^5}+\frac{[[\hat{H}_{m_2},[[[\hat{H}_{m_4},\hat{H}_0],\hat{H}_{m_3 - m_4}],\hat{H}_{m_1 - m_2 - m_3}]],\hat{H}_{-m_1}]}{48m_1m_2m_3m_4^2(\hbar\omega)^5}\\ 
&+\frac{[[\hat{H}_{m_2},[[[\hat{H}_{m_4},\hat{H}_{m_3 - m_4}],\hat{H}_0],\hat{H}_{m_1 - m_2 - m_3}]],\hat{H}_{-m_1}]}{48m_1m_2m_3^2m_4(\hbar\omega)^5}+\frac{[[\hat{H}_{m_2},[[[\hat{H}_{m_5},\hat{H}_{m_4 - m_5}],\hat{H}_{m_3 - m_4}],\hat{H}_{m_1 - m_2 - m_3}]],\hat{H}_{-m_1}]}{96m_1m_2m_3m_4m_5(\hbar\omega)^5}\\ 
&+\frac{[[\hat{H}_{m_2},[[\hat{H}_{m_4},[\hat{H}_{m_5},\hat{H}_{m_3 - m_4 - m_5}]],\hat{H}_{m_1 - m_2 - m_3}]],\hat{H}_{-m_1}]}{288m_1m_2m_3m_4m_5(\hbar\omega)^5}+\frac{[[\hat{H}_{m_2},[[\hat{H}_{m_3},[\hat{H}_{m_4},\hat{H}_{-m_4}]],\hat{H}_{m_1 - m_2 - m_3}]],\hat{H}_{-m_1}]}{72m_1m_2m_3^2m_4(\hbar\omega)^5}\\ 
&-\frac{[[\hat{H}_{m_2},[[\hat{H}_{m_1 - m_2},\hat{H}_0],[\hat{H}_{m_3},\hat{H}_{-m_3}]]],\hat{H}_{-m_1}]}{45m_1m_2m_3(m_1 - m_2)^2(\hbar\omega)^5}-\frac{[[\hat{H}_{m_2},[[\hat{H}_{m_3},\hat{H}_{m_1 - m_2 - m_3}],[\hat{H}_{m_4},\hat{H}_{-m_4}]]],\hat{H}_{-m_1}]}{90m_1m_2m_3m_4(m_1 - m_2)(\hbar\omega)^5}\\ 
&-\frac{[[\hat{H}_{m_2},[\hat{H}_{m_1 - m_2},[[\hat{H}_{m_3},\hat{H}_0],\hat{H}_{-m_3}]]],\hat{H}_{-m_1}]}{45m_1m_2m_3^2(m_1 - m_2)(\hbar\omega)^5}-\frac{11[[\hat{H}_{m_2},[\hat{H}_{m_1 - m_2},[[\hat{H}_{m_4},\hat{H}_{m_3 - m_4}],\hat{H}_{-m_3}]]],\hat{H}_{-m_1}]}{720m_1m_2m_3m_4(m_1 - m_2)(\hbar\omega)^5}\\ 
&+\frac{[[\hat{H}_{m_1},[[[\hat{H}_{m_2},\hat{H}_0],\hat{H}_0],\hat{H}_{-m_2}]],\hat{H}_{-m_1}]}{8m_1^2m_2^3(\hbar\omega)^5}+\frac{3[[\hat{H}_{m_1},[[[\hat{H}_{m_3},\hat{H}_0],\hat{H}_{m_2 - m_3}],\hat{H}_{-m_2}]],\hat{H}_{-m_1}]}{40m_1^2m_2m_3^2(\hbar\omega)^5}\\ 
&+\frac{[[\hat{H}_{m_1},[[[\hat{H}_{m_3},\hat{H}_{m_2 - m_3}],\hat{H}_0],\hat{H}_{-m_2}]],\hat{H}_{-m_1}]}{16m_1^2m_2^2m_3(\hbar\omega)^5}+\frac{3[[\hat{H}_{m_1},[[[\hat{H}_{m_4},\hat{H}_{m_3 - m_4}],\hat{H}_{m_2 - m_3}],\hat{H}_{-m_2}]],\hat{H}_{-m_1}]}{80m_1^2m_2m_3m_4(\hbar\omega)^5}\\ 
&+\frac{7[[\hat{H}_{m_1},[[\hat{H}_{m_3},[\hat{H}_{m_4},\hat{H}_{m_2 - m_3 - m_4}]],\hat{H}_{-m_2}]],\hat{H}_{-m_1}]}{720m_1^2m_2m_3m_4(\hbar\omega)^5}+\frac{5[[\hat{H}_{m_1},[[\hat{H}_{m_2},[\hat{H}_{m_3},\hat{H}_{-m_3}]],\hat{H}_{-m_2}]],\hat{H}_{-m_1}]}{144m_1^2m_2^2m_3(\hbar\omega)^5}\\ 
&+\frac{[[\hat{H}_{m_1},[[\hat{H}_{m_2},\hat{H}_0],[\hat{H}_{m_3},\hat{H}_{-m_2 - m_3}]]],\hat{H}_{-m_1}]}{80m_1^2m_2^2m_3(\hbar\omega)^5}+\frac{[[\hat{H}_{m_1},[[\hat{H}_{m_3},\hat{H}_{m_2 - m_3}],[\hat{H}_{m_4},\hat{H}_{-m_2 - m_4}]]],\hat{H}_{-m_1}]}{160m_1^2m_2m_3m_4(\hbar\omega)^5}\\ 
&-\frac{[[\hat{H}_{m_2},[\hat{H}_{m_3},[\hat{H}_{m_4},[\hat{H}_{m_5},\hat{H}_{m_1 - m_2 - m_3 - m_4 - m_5}]]]],\hat{H}_{-m_1}]}{1440m_1m_2m_3m_4m_5(\hbar\omega)^5}+\frac{[[[[\hat{H}_{m_1},\hat{H}_0],\hat{H}_0],\hat{H}_0],[\hat{H}_{m_2},\hat{H}_{-m_1 - m_2}]]}{12m_1^4m_2(\hbar\omega)^5}\\ 
&+\frac{[[[[\hat{H}_{m_2},\hat{H}_0],\hat{H}_0],\hat{H}_{m_1 - m_2}],[\hat{H}_{m_3},\hat{H}_{-m_1 - m_3}]]}{12m_1m_2^3m_3(\hbar\omega)^5}+\frac{[[[[\hat{H}_{m_2},\hat{H}_0],\hat{H}_{m_1 - m_2}],\hat{H}_0],[\hat{H}_{m_3},\hat{H}_{-m_1 - m_3}]]}{24m_1^2m_2^2m_3(\hbar\omega)^5}\\ 
&+\frac{[[[[\hat{H}_{m_3},\hat{H}_0],\hat{H}_{m_2 - m_3}],\hat{H}_{m_1 - m_2}],[\hat{H}_{m_4},\hat{H}_{-m_1 - m_4}]]}{24m_1m_2m_3^2m_4(\hbar\omega)^5}+\frac{[[[[\hat{H}_{m_2},\hat{H}_{m_1 - m_2}],\hat{H}_0],\hat{H}_0],[\hat{H}_{m_3},\hat{H}_{-m_1 - m_3}]]}{24m_1^3m_2m_3(\hbar\omega)^5}\\ 
&+\frac{[[[[\hat{H}_{m_3},\hat{H}_{m_2 - m_3}],\hat{H}_0],\hat{H}_{m_1 - m_2}],[\hat{H}_{m_4},\hat{H}_{-m_1 - m_4}]]}{24m_1m_2^2m_3m_4(\hbar\omega)^5}+\frac{[[[[\hat{H}_{m_3},\hat{H}_{m_2 - m_3}],\hat{H}_{m_1 - m_2}],\hat{H}_0],[\hat{H}_{m_4},\hat{H}_{-m_1 - m_4}]]}{48m_1^2m_2m_3m_4(\hbar\omega)^5}\\ 
&+\frac{[[[[\hat{H}_{m_4},\hat{H}_{m_3 - m_4}],\hat{H}_{m_2 - m_3}],\hat{H}_{m_1 - m_2}],[\hat{H}_{m_5},\hat{H}_{-m_1 - m_5}]]}{48m_1m_2m_3m_4m_5(\hbar\omega)^5}+\frac{[[[\hat{H}_{m_2},[\hat{H}_{m_3},\hat{H}_{m_1 - m_2 - m_3}]],\hat{H}_0],[\hat{H}_{m_4},\hat{H}_{-m_1 - m_4}]]}{144m_1^2m_2m_3m_4(\hbar\omega)^5}\\ 
&+\frac{[[[\hat{H}_{m_3},[\hat{H}_{m_4},\hat{H}_{m_2 - m_3 - m_4}]],\hat{H}_{m_1 - m_2}],[\hat{H}_{m_5},\hat{H}_{-m_1 - m_5}]]}{144m_1m_2m_3m_4m_5(\hbar\omega)^5}+\frac{[[[\hat{H}_{m_1},[\hat{H}_{m_2},\hat{H}_{-m_2}]],\hat{H}_0],[\hat{H}_{m_3},\hat{H}_{-m_1 - m_3}]]}{36m_1^3m_2m_3(\hbar\omega)^5}\\ 
&+\frac{[[[\hat{H}_{m_2},[\hat{H}_{m_3},\hat{H}_{-m_3}]],\hat{H}_{m_1 - m_2}],[\hat{H}_{m_4},\hat{H}_{-m_1 - m_4}]]}{36m_1m_2^2m_3m_4(\hbar\omega)^5}+\frac{[[[\hat{H}_{m_1},\hat{H}_0],\hat{H}_0],[[\hat{H}_{m_2},\hat{H}_0],\hat{H}_{-m_1 - m_2}]]}{12m_1^3m_2^2(\hbar\omega)^5}\\ 
&+\frac{[[[\hat{H}_{m_1},\hat{H}_0],\hat{H}_0],[[\hat{H}_{m_3},\hat{H}_{m_2 - m_3}],\hat{H}_{-m_1 - m_2}]]}{24m_1^3m_2m_3(\hbar\omega)^5}+\frac{[[[\hat{H}_{m_1},\hat{H}_0],\hat{H}_0],[\hat{H}_{-m_1},[\hat{H}_{m_2},\hat{H}_{-m_2}]]]}{24m_1^4m_2(\hbar\omega)^5}\\ 
&+\frac{[[[\hat{H}_{m_2},\hat{H}_0],\hat{H}_{m_1 - m_2}],[[\hat{H}_{m_3},\hat{H}_0],\hat{H}_{-m_1 - m_3}]]}{24m_1m_2^2m_3^2(\hbar\omega)^5}+\frac{[[[\hat{H}_{m_2},\hat{H}_0],\hat{H}_{m_1 - m_2}],[[\hat{H}_{m_4},\hat{H}_{m_3 - m_4}],\hat{H}_{-m_1 - m_3}]]}{24m_1m_2^2m_3m_4(\hbar\omega)^5}\\ 
&+\frac{[[[\hat{H}_{m_2},\hat{H}_{m_1 - m_2}],\hat{H}_0],[[\hat{H}_{m_3},\hat{H}_0],\hat{H}_{-m_1 - m_3}]]}{24m_1^2m_2m_3^2(\hbar\omega)^5}+\frac{[[[\hat{H}_{m_2},\hat{H}_{m_1 - m_2}],\hat{H}_0],[[\hat{H}_{m_4},\hat{H}_{m_3 - m_4}],\hat{H}_{-m_1 - m_3}]]}{48m_1^2m_2m_3m_4(\hbar\omega)^5}\\ 
&+\frac{[[[\hat{H}_{m_2},\hat{H}_{m_1 - m_2}],\hat{H}_0],[\hat{H}_{-m_1},[\hat{H}_{m_3},\hat{H}_{-m_3}]]]}{48m_1^3m_2m_3(\hbar\omega)^5}+\frac{[[[\hat{H}_{m_3},\hat{H}_{m_2 - m_3}],\hat{H}_{m_1 - m_2}],[[\hat{H}_{m_5},\hat{H}_{m_4 - m_5}],\hat{H}_{-m_1 - m_4}]]}{96m_1m_2m_3m_4m_5(\hbar\omega)^5}\\ 
&+\frac{[[\hat{H}_{m_2},[\hat{H}_{m_3},\hat{H}_{m_1 - m_2 - m_3}]],[\hat{H}_{-m_1},[\hat{H}_{m_4},\hat{H}_{-m_4}]]]}{288m_1^2m_2m_3m_4(\hbar\omega)^5}+\frac{[[\hat{H}_{m_1},[\hat{H}_{m_2},\hat{H}_{-m_2}]],[\hat{H}_{-m_1},[\hat{H}_{m_3},\hat{H}_{-m_3}]]]}{72m_1^3m_2m_3(\hbar\omega)^5}\\ 
&+\frac{[[\hat{H}_{m_1},\hat{H}_0],[[[\hat{H}_{m_2},\hat{H}_0],\hat{H}_0],\hat{H}_{-m_1 - m_2}]]}{12m_1^2m_2^3(\hbar\omega)^5}+\frac{[[\hat{H}_{m_1},\hat{H}_0],[[[\hat{H}_{m_3},\hat{H}_0],\hat{H}_{m_2 - m_3}],\hat{H}_{-m_1 - m_2}]]}{24m_1^2m_2m_3^2(\hbar\omega)^5}\\ 
&+\frac{[[\hat{H}_{m_1},\hat{H}_0],[[[\hat{H}_{m_3},\hat{H}_{m_2 - m_3}],\hat{H}_0],\hat{H}_{-m_1 - m_2}]]}{24m_1^2m_2^2m_3(\hbar\omega)^5}+\frac{[[\hat{H}_{m_1},\hat{H}_0],[[[\hat{H}_{m_4},\hat{H}_{m_3 - m_4}],\hat{H}_{m_2 - m_3}],\hat{H}_{-m_1 - m_2}]]}{48m_1^2m_2m_3m_4(\hbar\omega)^5}\\ 
&+\frac{[[\hat{H}_{m_1},\hat{H}_0],[[\hat{H}_{m_3},[\hat{H}_{m_4},\hat{H}_{m_2 - m_3 - m_4}]],\hat{H}_{-m_1 - m_2}]]}{144m_1^2m_2m_3m_4(\hbar\omega)^5}+\frac{[[\hat{H}_{m_1},\hat{H}_0],[[\hat{H}_{m_2},[\hat{H}_{m_3},\hat{H}_{-m_3}]],\hat{H}_{-m_1 - m_2}]]}{36m_1^2m_2^2m_3(\hbar\omega)^5}\\ 
&-\frac{[[\hat{H}_{m_1},\hat{H}_0],[[\hat{H}_{-m_1},\hat{H}_0],[\hat{H}_{m_2},\hat{H}_{-m_2}]]]}{24m_1^4m_2(\hbar\omega)^5}+\frac{[[\hat{H}_{m_1},\hat{H}_0],[[\hat{H}_{m_2},\hat{H}_{-m_1 - m_2}],[\hat{H}_{m_3},\hat{H}_{-m_3}]]]}{48m_1^3m_2m_3(\hbar\omega)^5}\\ 
&-\frac{[[\hat{H}_{m_1},\hat{H}_0],[\hat{H}_{m_2},[\hat{H}_{-m_1 - m_2},[\hat{H}_{m_3},\hat{H}_{-m_3}]]]]}{720m_1^2m_2m_3(-m_1 - m_2)(\hbar\omega)^5}+\frac{[[\hat{H}_{m_1},\hat{H}_0],[\hat{H}_{-m_1},[[\hat{H}_{m_2},\hat{H}_0],\hat{H}_{-m_2}]]]}{24m_1^3m_2^2(\hbar\omega)^5}\\ 
&+\frac{7[[\hat{H}_{m_1},\hat{H}_0],[\hat{H}_{-m_1},[[\hat{H}_{m_3},\hat{H}_{m_2 - m_3}],\hat{H}_{-m_2}]]]}{240m_1^3m_2m_3(\hbar\omega)^5}+\frac{[[[\hat{H}_{m_2},\hat{H}_0],[\hat{H}_{m_3},\hat{H}_{m_1 - m_2 - m_3}]],[\hat{H}_{m_4},\hat{H}_{-m_1 - m_4}]]}{144m_1m_2^2m_3m_4(\hbar\omega)^5}\\ 
&+\frac{[[[\hat{H}_{m_1},\hat{H}_0],[\hat{H}_{m_2},\hat{H}_{-m_2}]],[\hat{H}_{m_3},\hat{H}_{-m_1 - m_3}]]}{144m_1^3m_2m_3(\hbar\omega)^5}+\frac{[[[\hat{H}_{m_3},\hat{H}_{m_2 - m_3}],[\hat{H}_{m_4},\hat{H}_{m_1 - m_2 - m_4}]],[\hat{H}_{m_5},\hat{H}_{-m_1 - m_5}]]}{288m_1m_2m_3m_4m_5(\hbar\omega)^5}\\ 
&+\frac{[[[\hat{H}_{m_2},\hat{H}_{m_1 - m_2}],[\hat{H}_{m_3},\hat{H}_{-m_3}]],[\hat{H}_{m_4},\hat{H}_{-m_1 - m_4}]]}{288m_1^2m_2m_3m_4(\hbar\omega)^5}+\frac{[[\hat{H}_{m_2},[[\hat{H}_{m_3},\hat{H}_0],\hat{H}_{m_1 - m_2 - m_3}]],[\hat{H}_{m_4},\hat{H}_{-m_1 - m_4}]]}{144m_1m_2m_3^2m_4(\hbar\omega)^5}\\ 
&+\frac{[[\hat{H}_{m_2},[[\hat{H}_{m_4},\hat{H}_{m_3 - m_4}],\hat{H}_{m_1 - m_2 - m_3}]],[\hat{H}_{m_5},\hat{H}_{-m_1 - m_5}]]}{288m_1m_2m_3m_4m_5(\hbar\omega)^5}-\frac{[[\hat{H}_{m_2},[\hat{H}_{m_1 - m_2},[\hat{H}_{m_3},\hat{H}_{-m_3}]]],[\hat{H}_{m_4},\hat{H}_{-m_1 - m_4}]]}{240m_1m_2m_3m_4(m_1 - m_2)(\hbar\omega)^5}\\ 
&+\frac{[[\hat{H}_{m_1},[[\hat{H}_{m_2},\hat{H}_0],\hat{H}_{-m_2}]],[\hat{H}_{m_3},\hat{H}_{-m_1 - m_3}]]}{144m_1^2m_2^2m_3(\hbar\omega)^5}+\frac{[[\hat{H}_{m_1},[[\hat{H}_{m_3},\hat{H}_{m_2 - m_3}],\hat{H}_{-m_2}]],[\hat{H}_{m_4},\hat{H}_{-m_1 - m_4}]]}{360m_1^2m_2m_3m_4(\hbar\omega)^5}\\ 
&+\frac{[[[\hat{H}_{m_2},\hat{H}_0],\hat{H}_{m_1 - m_2}],[\hat{H}_{m_3},[\hat{H}_{m_4},\hat{H}_{-m_1 - m_3 - m_4}]]]}{144m_1m_2^2m_3m_4(\hbar\omega)^5}-\frac{[[[\hat{H}_{m_2},\hat{H}_0],\hat{H}_{m_1 - m_2}],[\hat{H}_{-m_1},[\hat{H}_{m_3},\hat{H}_{-m_3}]]]}{144m_1^2m_2^2m_3(\hbar\omega)^5}\\ 
&+\frac{[[[\hat{H}_{m_3},\hat{H}_{m_2 - m_3}],\hat{H}_{m_1 - m_2}],[\hat{H}_{m_4},[\hat{H}_{m_5},\hat{H}_{-m_1 - m_4 - m_5}]]]}{288m_1m_2m_3m_4m_5(\hbar\omega)^5}-\frac{[[[\hat{H}_{m_3},\hat{H}_{m_2 - m_3}],\hat{H}_{m_1 - m_2}],[\hat{H}_{-m_1},[\hat{H}_{m_4},\hat{H}_{-m_4}]]]}{288m_1^2m_2m_3m_4(\hbar\omega)^5}\\ 
&-\frac{[[\hat{H}_{m_1},\hat{H}_0],[\hat{H}_{m_2},[\hat{H}_{m_3},[\hat{H}_{m_4},\hat{H}_{-m_1 - m_2 - m_3 - m_4}]]]]}{720m_1^2m_2m_3m_4(\hbar\omega)^5}-\frac{[[\hat{H}_{m_2},\hat{H}_{m_1 - m_2}],[\hat{H}_{m_3},[\hat{H}_{m_4},[\hat{H}_{m_5},\hat{H}_{-m_1 - m_3 - m_4 - m_5}]]]]}{1440m_1m_2m_3m_4m_5(\hbar\omega)^5}\\ 
&-\frac{[\hat{H}_{m_1},[[\hat{H}_{m_2},\hat{H}_0],[\hat{H}_{m_3},[\hat{H}_{m_4},\hat{H}_{-m_1 - m_2 - m_3 - m_4}]]]]}{720m_1m_2^2m_3m_4(\hbar\omega)^5}-\frac{[\hat{H}_{m_1},[[\hat{H}_{m_3},\hat{H}_{m_2 - m_3}],[\hat{H}_{m_4},[\hat{H}_{m_5},\hat{H}_{-m_1 - m_2 - m_4 - m_5}]]]]}{1440m_1m_2m_3m_4m_5(\hbar\omega)^5}\\ 
&-\frac{[\hat{H}_{m_1},[\hat{H}_{m_2},[[\hat{H}_{m_3},\hat{H}_0],[\hat{H}_{m_4},\hat{H}_{-m_1 - m_2 - m_3 - m_4}]]]]}{720m_1m_2m_3^2m_4(\hbar\omega)^5}-\frac{[\hat{H}_{m_1},[\hat{H}_{m_2},[[\hat{H}_{m_4},\hat{H}_{m_3 - m_4}],[\hat{H}_{m_5},\hat{H}_{-m_1 - m_2 - m_3 - m_5}]]]]}{1440m_1m_2m_3m_4m_5(\hbar\omega)^5}\\ 
&-\frac{[\hat{H}_{m_1},[\hat{H}_{m_2},[\hat{H}_{m_3},[[\hat{H}_{m_4},\hat{H}_0],\hat{H}_{-m_1 - m_2 - m_3 - m_4}]]]]}{720m_1m_2m_3m_4^2(\hbar\omega)^5}-\frac{[\hat{H}_{m_1},[\hat{H}_{m_2},[\hat{H}_{m_3},[[\hat{H}_{m_5},\hat{H}_{m_4 - m_5}],\hat{H}_{-m_1 - m_2 - m_3 - m_4}]]]]}{1440m_1m_2m_3m_4m_5(\hbar\omega)^5}
\end{align*}
\begin{align*}
\frac{\hat{S}^{(1)}}{i\hbar}=&\frac{\hat{H}_{m_1}}{m_1(\hbar\omega)}e^{im_1\omega t}\\
\frac{\hat{S}^{(2)}}{i\hbar}= &\;\frac{[\hat{H}_{m_1},\hat{H}_0]}{m_1^2(\hbar\omega)^2}e^{im_1\omega t}+\frac{[\hat{H}_{m_2},\hat{H}_{m_1 - m_2}]}{2m_1m_2(\hbar\omega)^2}e^{im_1\omega t}\\ 
\frac{\hat{S}^{(3)}}{i\hbar} =&\;\frac{[[\hat{H}_{m_1},\hat{H}_0],\hat{H}_0]}{m_1^3(\hbar\omega)^3}e^{im_1\omega t}+\frac{[[\hat{H}_{m_2},\hat{H}_0],\hat{H}_{m_1 - m_2}]}{2m_1m_2^2(\hbar\omega)^3}e^{im_1\omega t}\\ 
&-\frac{[[\hat{H}_{-m_2},\hat{H}_{m_1 + m_2}],\hat{H}_0]}{2m_1^2m_2(\hbar\omega)^3}e^{im_1\omega t}-\frac{[[\hat{H}_{-m_3},\hat{H}_{m_2 + m_3}],\hat{H}_{m_1 - m_2}]}{4m_1m_2m_3(\hbar\omega)^3}e^{im_1\omega t}\\ 
&+\frac{[\hat{H}_{m_1},[\hat{H}_{m_2},\hat{H}_{-m_2}]]}{3m_1^3(\hbar\omega)^3}e^{im_1\omega t}-\frac{[\hat{H}_{m_2},[\hat{H}_{-m_3},\hat{H}_{m_1 - m_2 + m_3}]]}{12m_1m_2m_3(\hbar\omega)^3}e^{im_1\omega t}\\ 
\frac{\hat{S}^{(4)}}{i\hbar} =&\;\frac{[[[\hat{H}_{m_1},\hat{H}_0],\hat{H}_0],\hat{H}_0]}{m_1^4(\hbar\omega)^4}e^{im_1\omega t}+\frac{[[[\hat{H}_{m_2},\hat{H}_0],\hat{H}_0],\hat{H}_{m_1 - m_2}]}{2m_1m_2^3(\hbar\omega)^4}e^{im_1\omega t}\\ 
&+\frac{[[[\hat{H}_{m_2},\hat{H}_0],\hat{H}_{m_1 - m_2}],\hat{H}_0]}{2m_1^2m_2^2(\hbar\omega)^4}e^{im_1\omega t}+\frac{[[[\hat{H}_{m_3},\hat{H}_0],\hat{H}_{m_2 - m_3}],\hat{H}_{m_1 - m_2}]}{4m_1m_2m_3^2(\hbar\omega)^4}e^{im_1\omega t}\\ 
&+\frac{[[[\hat{H}_{m_2},\hat{H}_{m_1 - m_2}],\hat{H}_0],\hat{H}_0]}{2m_1^3m_2(\hbar\omega)^4}e^{im_1\omega t}+\frac{[[[\hat{H}_{m_3},\hat{H}_{m_2 - m_3}],\hat{H}_0],\hat{H}_{m_1 - m_2}]}{4m_1m_2^2m_3(\hbar\omega)^4}e^{im_1\omega t}\\ 
&+\frac{[[[\hat{H}_{m_3},\hat{H}_{m_2 - m_3}],\hat{H}_{m_1 - m_2}],\hat{H}_0]}{4m_1^2m_2m_3(\hbar\omega)^4}e^{im_1\omega t}+\frac{[[[\hat{H}_{m_4},\hat{H}_{m_3 - m_4}],\hat{H}_{m_2 - m_3}],\hat{H}_{m_1 - m_2}]}{8m_1m_2m_3m_4(\hbar\omega)^4}e^{im_1\omega t}\\ 
&+\frac{[[\hat{H}_{m_2},[\hat{H}_{m_3},\hat{H}_{m_1 - m_2 - m_3}]],\hat{H}_0]}{12m_1^2m_2m_3(\hbar\omega)^4}e^{im_1\omega t}+\frac{[[\hat{H}_{m_3},[\hat{H}_{m_4},\hat{H}_{m_2 - m_3 - m_4}]],\hat{H}_{m_1 - m_2}]}{24m_1m_2m_3m_4(\hbar\omega)^4}e^{im_1\omega t}\\ 
&+\frac{[[\hat{H}_{m_1},[\hat{H}_{m_2},\hat{H}_{-m_2}]],\hat{H}_0]}{3m_1^3m_2(\hbar\omega)^4}e^{im_1\omega t}+\frac{[[\hat{H}_{m_2},[\hat{H}_{m_3},\hat{H}_{-m_3}]],\hat{H}_{m_1 - m_2}]}{6m_1m_2^2m_3(\hbar\omega)^4}e^{im_1\omega t}\\ 
&+\frac{[[\hat{H}_{m_2},\hat{H}_0],[\hat{H}_{m_3},\hat{H}_{m_1 - m_2 - m_3}]]}{12m_1m_2^2m_3(\hbar\omega)^4}e^{im_1\omega t}+\frac{[[\hat{H}_{m_1},\hat{H}_0],[\hat{H}_{m_2},\hat{H}_{-m_2}]]}{3m_1^3m_2(\hbar\omega)^4}e^{im_1\omega t}\\ 
&+\frac{[[\hat{H}_{m_3},\hat{H}_{m_2 - m_3}],[\hat{H}_{m_4},\hat{H}_{m_1 - m_2 - m_4}]]}{24m_1m_2m_3m_4(\hbar\omega)^4}e^{im_1\omega t}+\frac{[[\hat{H}_{m_2},\hat{H}_{m_1 - m_2}],[\hat{H}_{m_3},\hat{H}_{-m_3}]]}{6m_1^2m_2m_3(\hbar\omega)^4}e^{im_1\omega t}\\ 
&+\frac{[\hat{H}_{m_2},[[\hat{H}_{m_3},\hat{H}_0],\hat{H}_{m_1 - m_2 - m_3}]]}{12m_1m_2m_3^2(\hbar\omega)^4}e^{im_1\omega t}+\frac{[\hat{H}_{m_2},[[\hat{H}_{m_4},\hat{H}_{m_3 - m_4}],\hat{H}_{m_1 - m_2 - m_3}]]}{24m_1m_2m_3m_4(\hbar\omega)^4}e^{im_1\omega t}\\ 
&-\frac{[\hat{H}_{m_2},[\hat{H}_{m_1 - m_2},[\hat{H}_{m_3},\hat{H}_{-m_3}]]]}{24m_1m_2m_3(m_1 - m_2)(\hbar\omega)^4}e^{im_1\omega t}+\frac{[\hat{H}_{m_1},[[\hat{H}_{m_2},\hat{H}_0],\hat{H}_{-m_2}]]}{3m_1^2m_2^2(\hbar\omega)^4}e^{im_1\omega t}\\ 
&+\frac{5[\hat{H}_{m_1},[[\hat{H}_{m_3},\hat{H}_{m_2 - m_3}],\hat{H}_{-m_2}]]}{24m_1^2m_2m_3(\hbar\omega)^4}e^{im_1\omega t}\\
\frac{\hat{S}^{(5)}}{i\hbar} = &\;\frac{[[[[\hat{H}_{m_1},\hat{H}_0],\hat{H}_0],\hat{H}_0],\hat{H}_0]}{m_1^5(\hbar\omega)^5}e^{im_1\omega t}+\frac{[[[[\hat{H}_{m_2},\hat{H}_0],\hat{H}_0],\hat{H}_0],\hat{H}_{m_1 - m_2}]}{2m_1m_2^4(\hbar\omega)^5}e^{im_1\omega t}\\ 
&+\frac{[[[[\hat{H}_{m_2},\hat{H}_0],\hat{H}_0],\hat{H}_{m_1 - m_2}],\hat{H}_0]}{2m_1^2m_2^3(\hbar\omega)^5}e^{im_1\omega t}+\frac{[[[[\hat{H}_{m_3},\hat{H}_0],\hat{H}_0],\hat{H}_{m_2 - m_3}],\hat{H}_{m_1 - m_2}]}{4m_1m_2m_3^3(\hbar\omega)^5}e^{im_1\omega t}\\ 
&+\frac{[[[[\hat{H}_{m_2},\hat{H}_0],\hat{H}_{m_1 - m_2}],\hat{H}_0],\hat{H}_0]}{2m_1^3m_2^2(\hbar\omega)^5}e^{im_1\omega t}+\frac{[[[[\hat{H}_{m_3},\hat{H}_0],\hat{H}_{m_2 - m_3}],\hat{H}_0],\hat{H}_{m_1 - m_2}]}{4m_1m_2^2m_3^2(\hbar\omega)^5}e^{im_1\omega t}\\ 
&+\frac{[[[[\hat{H}_{m_3},\hat{H}_0],\hat{H}_{m_2 - m_3}],\hat{H}_{m_1 - m_2}],\hat{H}_0]}{4m_1^2m_2m_3^2(\hbar\omega)^5}e^{im_1\omega t}+\frac{[[[[\hat{H}_{m_4},\hat{H}_0],\hat{H}_{m_3 - m_4}],\hat{H}_{m_2 - m_3}],\hat{H}_{m_1 - m_2}]}{8m_1m_2m_3m_4^2(\hbar\omega)^5}e^{im_1\omega t}\\ 
&+\frac{[[[[\hat{H}_{m_2},\hat{H}_{m_1 - m_2}],\hat{H}_0],\hat{H}_0],\hat{H}_0]}{2m_1^4m_2(\hbar\omega)^5}e^{im_1\omega t}+\frac{[[[[\hat{H}_{m_3},\hat{H}_{m_2 - m_3}],\hat{H}_0],\hat{H}_0],\hat{H}_{m_1 - m_2}]}{4m_1m_2^3m_3(\hbar\omega)^5}e^{im_1\omega t}\\ 
&+\frac{[[[[\hat{H}_{m_3},\hat{H}_{m_2 - m_3}],\hat{H}_0],\hat{H}_{m_1 - m_2}],\hat{H}_0]}{4m_1^2m_2^2m_3(\hbar\omega)^5}e^{im_1\omega t}+\frac{[[[[\hat{H}_{m_4},\hat{H}_{m_3 - m_4}],\hat{H}_0],\hat{H}_{m_2 - m_3}],\hat{H}_{m_1 - m_2}]}{8m_1m_2m_3^2m_4(\hbar\omega)^5}e^{im_1\omega t}\\ 
&+\frac{[[[[\hat{H}_{m_3},\hat{H}_{m_2 - m_3}],\hat{H}_{m_1 - m_2}],\hat{H}_0],\hat{H}_0]}{4m_1^3m_2m_3(\hbar\omega)^5}e^{im_1\omega t}+\frac{[[[[\hat{H}_{m_4},\hat{H}_{m_3 - m_4}],\hat{H}_{m_2 - m_3}],\hat{H}_0],\hat{H}_{m_1 - m_2}]}{8m_1m_2^2m_3m_4(\hbar\omega)^5}e^{im_1\omega t}\\ 
&+\frac{[[[[\hat{H}_{m_4},\hat{H}_{m_3 - m_4}],\hat{H}_{m_2 - m_3}],\hat{H}_{m_1 - m_2}],\hat{H}_0]}{8m_1^2m_2m_3m_4(\hbar\omega)^5}e^{im_1\omega t}+\frac{[[[[\hat{H}_{m_5},\hat{H}_{m_4 - m_5}],\hat{H}_{m_3 - m_4}],\hat{H}_{m_2 - m_3}],\hat{H}_{m_1 - m_2}]}{16m_1m_2m_3m_4m_5(\hbar\omega)^5}e^{im_1\omega t}\\ 
&+\frac{[[[\hat{H}_{m_2},[\hat{H}_{m_3},\hat{H}_{m_1 - m_2 - m_3}]],\hat{H}_0],\hat{H}_0]}{12m_1^3m_2m_3(\hbar\omega)^5}e^{im_1\omega t}+\frac{[[[\hat{H}_{m_3},[\hat{H}_{m_4},\hat{H}_{m_2 - m_3 - m_4}]],\hat{H}_0],\hat{H}_{m_1 - m_2}]}{24m_1m_2^2m_3m_4(\hbar\omega)^5}e^{im_1\omega t}\\ 
&+\frac{[[[\hat{H}_{m_3},[\hat{H}_{m_4},\hat{H}_{m_2 - m_3 - m_4}]],\hat{H}_{m_1 - m_2}],\hat{H}_0]}{24m_1^2m_2m_3m_4(\hbar\omega)^5}e^{im_1\omega t}+\frac{[[[\hat{H}_{m_4},[\hat{H}_{m_5},\hat{H}_{m_3 - m_4 - m_5}]],\hat{H}_{m_2 - m_3}],\hat{H}_{m_1 - m_2}]}{48m_1m_2m_3m_4m_5(\hbar\omega)^5}e^{im_1\omega t}\\ 
&+\frac{[[[\hat{H}_{m_1},[\hat{H}_{m_2},\hat{H}_{-m_2}]],\hat{H}_0],\hat{H}_0]}{3m_1^4m_2(\hbar\omega)^5}e^{im_1\omega t}+\frac{[[[\hat{H}_{m_2},[\hat{H}_{m_3},\hat{H}_{-m_3}]],\hat{H}_0],\hat{H}_{m_1 - m_2}]}{6m_1m_2^3m_3(\hbar\omega)^5}e^{im_1\omega t}\\ 
&+\frac{[[[\hat{H}_{m_2},[\hat{H}_{m_3},\hat{H}_{-m_3}]],\hat{H}_{m_1 - m_2}],\hat{H}_0]}{6m_1^2m_2^2m_3(\hbar\omega)^5}e^{im_1\omega t}+\frac{[[[\hat{H}_{m_3},[\hat{H}_{m_4},\hat{H}_{-m_4}]],\hat{H}_{m_2 - m_3}],\hat{H}_{m_1 - m_2}]}{12m_1m_2m_3^2m_4(\hbar\omega)^5}e^{im_1\omega t}\\ 
&+\frac{[[[\hat{H}_{m_2},\hat{H}_0],[\hat{H}_{m_3},\hat{H}_{m_1 - m_2 - m_3}]],\hat{H}_0]}{12m_1^2m_2^2m_3(\hbar\omega)^5}e^{im_1\omega t}+\frac{[[[\hat{H}_{m_3},\hat{H}_0],[\hat{H}_{m_4},\hat{H}_{m_2 - m_3 - m_4}]],\hat{H}_{m_1 - m_2}]}{24m_1m_2m_3^2m_4(\hbar\omega)^5}e^{im_1\omega t}\\ 
&+\frac{[[[\hat{H}_{m_1},\hat{H}_0],[\hat{H}_{m_2},\hat{H}_{-m_2}]],\hat{H}_0]}{3m_1^4m_2(\hbar\omega)^5}e^{im_1\omega t}+\frac{[[[\hat{H}_{m_2},\hat{H}_0],[\hat{H}_{m_3},\hat{H}_{-m_3}]],\hat{H}_{m_1 - m_2}]}{6m_1m_2^3m_3(\hbar\omega)^5}e^{im_1\omega t}\\ 
&+\frac{[[[\hat{H}_{m_3},\hat{H}_{m_2 - m_3}],[\hat{H}_{m_4},\hat{H}_{m_1 - m_2 - m_4}]],\hat{H}_0]}{24m_1^2m_2m_3m_4(\hbar\omega)^5}e^{im_1\omega t}+\frac{[[[\hat{H}_{m_4},\hat{H}_{m_3 - m_4}],[\hat{H}_{m_5},\hat{H}_{m_2 - m_3 - m_5}]],\hat{H}_{m_1 - m_2}]}{48m_1m_2m_3m_4m_5(\hbar\omega)^5}e^{im_1\omega t}\\ 
&+\frac{[[[\hat{H}_{m_2},\hat{H}_{m_1 - m_2}],[\hat{H}_{m_3},\hat{H}_{-m_3}]],\hat{H}_0]}{6m_1^3m_2m_3(\hbar\omega)^5}e^{im_1\omega t}+\frac{[[[\hat{H}_{m_3},\hat{H}_{m_2 - m_3}],[\hat{H}_{m_4},\hat{H}_{-m_4}]],\hat{H}_{m_1 - m_2}]}{12m_1m_2^2m_3m_4(\hbar\omega)^5}e^{im_1\omega t}\\ 
&+\frac{[[\hat{H}_{m_2},[[\hat{H}_{m_3},\hat{H}_0],\hat{H}_{m_1 - m_2 - m_3}]],\hat{H}_0]}{12m_1^2m_2m_3^2(\hbar\omega)^5}e^{im_1\omega t}+\frac{[[\hat{H}_{m_3},[[\hat{H}_{m_4},\hat{H}_0],\hat{H}_{m_2 - m_3 - m_4}]],\hat{H}_{m_1 - m_2}]}{24m_1m_2m_3m_4^2(\hbar\omega)^5}e^{im_1\omega t}\\ 
&+\frac{[[\hat{H}_{m_2},[[\hat{H}_{m_4},\hat{H}_{m_3 - m_4}],\hat{H}_{m_1 - m_2 - m_3}]],\hat{H}_0]}{24m_1^2m_2m_3m_4(\hbar\omega)^5}e^{im_1\omega t}+\frac{[[\hat{H}_{m_3},[[\hat{H}_{m_5},\hat{H}_{m_4 - m_5}],\hat{H}_{m_2 - m_3 - m_4}]],\hat{H}_{m_1 - m_2}]}{48m_1m_2m_3m_4m_5(\hbar\omega)^5}e^{im_1\omega t}\\ 
&+-\frac{[[\hat{H}_{m_2},[\hat{H}_{m_1 - m_2},[\hat{H}_{m_3},\hat{H}_{-m_3}]]],\hat{H}_0]}{24m_1^2m_2m_3(m_1 - m_2)(\hbar\omega)^5}e^{im_1\omega t}+-\frac{[[\hat{H}_{m_3},[\hat{H}_{m_2 - m_3},[\hat{H}_{m_4},\hat{H}_{-m_4}]]],\hat{H}_{m_1 - m_2}]}{48m_1m_2m_3m_4(m_2 - m_3)(\hbar\omega)^5}e^{im_1\omega t}\\ 
&+\frac{[[\hat{H}_{m_1},[[\hat{H}_{m_2},\hat{H}_0],\hat{H}_{-m_2}]],\hat{H}_0]}{3m_1^3m_2^2(\hbar\omega)^5}e^{im_1\omega t}+\frac{[[\hat{H}_{m_2},[[\hat{H}_{m_3},\hat{H}_0],\hat{H}_{-m_3}]],\hat{H}_{m_1 - m_2}]}{6m_1m_2^2m_3^2(\hbar\omega)^5}e^{im_1\omega t}\\ 
&+\frac{5[[\hat{H}_{m_1},[[\hat{H}_{m_3},\hat{H}_{m_2 - m_3}],\hat{H}_{-m_2}]],\hat{H}_0]}{24m_1^3m_2m_3(\hbar\omega)^5}e^{im_1\omega t}+\frac{5[[\hat{H}_{m_2},[[\hat{H}_{m_4},\hat{H}_{m_3 - m_4}],\hat{H}_{-m_3}]],\hat{H}_{m_1 - m_2}]}{48m_1m_2^2m_3m_4(\hbar\omega)^5}e^{im_1\omega t}\\ 
&+\frac{[[[\hat{H}_{m_2},\hat{H}_0],\hat{H}_0],[\hat{H}_{m_3},\hat{H}_{m_1 - m_2 - m_3}]]}{12m_1m_2^3m_3(\hbar\omega)^5}e^{im_1\omega t}+\frac{[[[\hat{H}_{m_1},\hat{H}_0],\hat{H}_0],[\hat{H}_{m_2},\hat{H}_{-m_2}]]}{3m_1^4m_2(\hbar\omega)^5}e^{im_1\omega t}\\ 
&+\frac{[[[\hat{H}_{m_3},\hat{H}_0],\hat{H}_{m_2 - m_3}],[\hat{H}_{m_4},\hat{H}_{m_1 - m_2 - m_4}]]}{24m_1m_2m_3^2m_4(\hbar\omega)^5}e^{im_1\omega t}+\frac{[[[\hat{H}_{m_2},\hat{H}_0],\hat{H}_{m_1 - m_2}],[\hat{H}_{m_3},\hat{H}_{-m_3}]]}{6m_1^2m_2^2m_3(\hbar\omega)^5}e^{im_1\omega t}\\ 
&+\frac{[[[\hat{H}_{m_3},\hat{H}_{m_2 - m_3}],\hat{H}_0],[\hat{H}_{m_4},\hat{H}_{m_1 - m_2 - m_4}]]}{24m_1m_2^2m_3m_4(\hbar\omega)^5}e^{im_1\omega t}+\frac{[[[\hat{H}_{m_2},\hat{H}_{m_1 - m_2}],\hat{H}_0],[\hat{H}_{m_3},\hat{H}_{-m_3}]]}{6m_1^3m_2m_3(\hbar\omega)^5}e^{im_1\omega t}\\ 
&+\frac{[[[\hat{H}_{m_4},\hat{H}_{m_3 - m_4}],\hat{H}_{m_2 - m_3}],[\hat{H}_{m_5},\hat{H}_{m_1 - m_2 - m_5}]]}{48m_1m_2m_3m_4m_5(\hbar\omega)^5}e^{im_1\omega t}+\frac{[[[\hat{H}_{m_3},\hat{H}_{m_2 - m_3}],\hat{H}_{m_1 - m_2}],[\hat{H}_{m_4},\hat{H}_{-m_4}]]}{12m_1^2m_2m_3m_4(\hbar\omega)^5}e^{im_1\omega t}\\ 
&+\frac{[[\hat{H}_{m_3},[\hat{H}_{m_4},\hat{H}_{m_2 - m_3 - m_4}]],[\hat{H}_{m_5},\hat{H}_{m_1 - m_2 - m_5}]]}{144m_1m_2m_3m_4m_5(\hbar\omega)^5}e^{im_1\omega t}+\frac{[[\hat{H}_{m_2},[\hat{H}_{m_3},\hat{H}_{m_1 - m_2 - m_3}]],[\hat{H}_{m_4},\hat{H}_{-m_4}]]}{36m_1^2m_2m_3m_4(\hbar\omega)^5}e^{im_1\omega t}\\ 
&+\frac{[[\hat{H}_{m_2},[\hat{H}_{m_3},\hat{H}_{-m_3}]],[\hat{H}_{m_4},\hat{H}_{m_1 - m_2 - m_4}]]}{36m_1m_2^2m_3m_4(\hbar\omega)^5}e^{im_1\omega t}+\frac{[[\hat{H}_{m_1},[\hat{H}_{m_2},\hat{H}_{-m_2}]],[\hat{H}_{m_3},\hat{H}_{-m_3}]]}{9m_1^3m_2m_3(\hbar\omega)^5}e^{im_1\omega t}\\ 
&+\frac{[[\hat{H}_{m_2},\hat{H}_0],[[\hat{H}_{m_3},\hat{H}_0],\hat{H}_{m_1 - m_2 - m_3}]]}{12m_1m_2^2m_3^2(\hbar\omega)^5}e^{im_1\omega t}+\frac{[[\hat{H}_{m_2},\hat{H}_0],[[\hat{H}_{m_4},\hat{H}_{m_3 - m_4}],\hat{H}_{m_1 - m_2 - m_3}]]}{24m_1m_2^2m_3m_4(\hbar\omega)^5}e^{im_1\omega t}\\ 
&+-\frac{[[\hat{H}_{m_2},\hat{H}_0],[\hat{H}_{m_1 - m_2},[\hat{H}_{m_3},\hat{H}_{-m_3}]]]}{24m_1m_2^2m_3(m_1 - m_2)(\hbar\omega)^5}e^{im_1\omega t}+\frac{[[\hat{H}_{m_1},\hat{H}_0],[[\hat{H}_{m_2},\hat{H}_0],\hat{H}_{-m_2}]]}{3m_1^3m_2^2(\hbar\omega)^5}e^{im_1\omega t}\\ 
&+\frac{5[[\hat{H}_{m_1},\hat{H}_0],[[\hat{H}_{m_3},\hat{H}_{m_2 - m_3}],\hat{H}_{-m_2}]]}{24m_1^3m_2m_3(\hbar\omega)^5}e^{im_1\omega t}+\frac{[[\hat{H}_{m_3},\hat{H}_{m_2 - m_3}],[[\hat{H}_{m_4},\hat{H}_0],\hat{H}_{m_1 - m_2 - m_4}]]}{24m_1m_2m_3m_4^2(\hbar\omega)^5}e^{im_1\omega t}\\ 
&+\frac{[[\hat{H}_{m_3},\hat{H}_{m_2 - m_3}],[[\hat{H}_{m_5},\hat{H}_{m_4 - m_5}],\hat{H}_{m_1 - m_2 - m_4}]]}{48m_1m_2m_3m_4m_5(\hbar\omega)^5}e^{im_1\omega t}+-\frac{[[\hat{H}_{m_3},\hat{H}_{m_2 - m_3}],[\hat{H}_{m_1 - m_2},[\hat{H}_{m_4},\hat{H}_{-m_4}]]]}{48m_1m_2m_3m_4(m_1 - m_2)(\hbar\omega)^5}e^{im_1\omega t}\\ 
&+\frac{[[\hat{H}_{m_2},\hat{H}_{m_1 - m_2}],[[\hat{H}_{m_3},\hat{H}_0],\hat{H}_{-m_3}]]}{6m_1^2m_2m_3^2(\hbar\omega)^5}e^{im_1\omega t}+\frac{5[[\hat{H}_{m_2},\hat{H}_{m_1 - m_2}],[[\hat{H}_{m_4},\hat{H}_{m_3 - m_4}],\hat{H}_{-m_3}]]}{48m_1^2m_2m_3m_4(\hbar\omega)^5}e^{im_1\omega t}\\ 
&+\frac{[\hat{H}_{m_2},[[[\hat{H}_{m_3},\hat{H}_0],\hat{H}_0],\hat{H}_{m_1 - m_2 - m_3}]]}{12m_1m_2m_3^3(\hbar\omega)^5}e^{im_1\omega t}+\frac{[\hat{H}_{m_2},[[[\hat{H}_{m_4},\hat{H}_0],\hat{H}_{m_3 - m_4}],\hat{H}_{m_1 - m_2 - m_3}]]}{24m_1m_2m_3m_4^2(\hbar\omega)^5}e^{im_1\omega t}\\ 
&+\frac{[\hat{H}_{m_2},[[[\hat{H}_{m_4},\hat{H}_{m_3 - m_4}],\hat{H}_0],\hat{H}_{m_1 - m_2 - m_3}]]}{24m_1m_2m_3^2m_4(\hbar\omega)^5}e^{im_1\omega t}+\frac{[\hat{H}_{m_2},[[[\hat{H}_{m_5},\hat{H}_{m_4 - m_5}],\hat{H}_{m_3 - m_4}],\hat{H}_{m_1 - m_2 - m_3}]]}{48m_1m_2m_3m_4m_5(\hbar\omega)^5}e^{im_1\omega t}\\ 
&+\frac{[\hat{H}_{m_2},[[\hat{H}_{m_4},[\hat{H}_{m_5},\hat{H}_{m_3 - m_4 - m_5}]],\hat{H}_{m_1 - m_2 - m_3}]]}{144m_1m_2m_3m_4m_5(\hbar\omega)^5}e^{im_1\omega t}+\frac{[\hat{H}_{m_2},[[\hat{H}_{m_3},[\hat{H}_{m_4},\hat{H}_{-m_4}]],\hat{H}_{m_1 - m_2 - m_3}]]}{36m_1m_2m_3^2m_4(\hbar\omega)^5}e^{im_1\omega t}\\ 
&+-\frac{[\hat{H}_{m_2},[[\hat{H}_{m_1 - m_2},\hat{H}_0],[\hat{H}_{m_3},\hat{H}_{-m_3}]]]}{24m_1m_2m_3(m_1 - m_2)^2(\hbar\omega)^5}e^{im_1\omega t}+-\frac{[\hat{H}_{m_2},[[\hat{H}_{m_3},\hat{H}_{m_1 - m_2 - m_3}],[\hat{H}_{m_4},\hat{H}_{-m_4}]]]}{48m_1m_2m_3m_4(m_1 - m_2)(\hbar\omega)^5}e^{im_1\omega t}\\ 
&+-\frac{[\hat{H}_{m_2},[\hat{H}_{m_3},[\hat{H}_{m_1 - m_2 - m_3},[\hat{H}_{m_4},\hat{H}_{-m_4}]]]]}{720m_1m_2m_3m_4(m_1 - m_2 - m_3)(\hbar\omega)^5}e^{im_1\omega t}+-\frac{[\hat{H}_{m_2},[\hat{H}_{m_1 - m_2},[[\hat{H}_{m_3},\hat{H}_0],\hat{H}_{-m_3}]]]}{24m_1m_2m_3^2(m_1 - m_2)(\hbar\omega)^5}e^{im_1\omega t}\\ 
&+-\frac{7[\hat{H}_{m_2},[\hat{H}_{m_1 - m_2},[[\hat{H}_{m_4},\hat{H}_{m_3 - m_4}],\hat{H}_{-m_3}]]]}{240m_1m_2m_3m_4(m_1 - m_2)(\hbar\omega)^5}e^{im_1\omega t}+\frac{[\hat{H}_{m_1},[[[\hat{H}_{m_2},\hat{H}_0],\hat{H}_0],\hat{H}_{-m_2}]]}{3m_1^2m_2^3(\hbar\omega)^5}e^{im_1\omega t}\\ 
&+\frac{5[\hat{H}_{m_1},[[[\hat{H}_{m_3},\hat{H}_0],\hat{H}_{m_2 - m_3}],\hat{H}_{-m_2}]]}{24m_1^2m_2m_3^2(\hbar\omega)^5}e^{im_1\omega t}+\frac{[\hat{H}_{m_1},[[[\hat{H}_{m_3},\hat{H}_{m_2 - m_3}],\hat{H}_0],\hat{H}_{-m_2}]]}{6m_1^2m_2^2m_3(\hbar\omega)^5}e^{im_1\omega t}\\ 
&+\frac{5[\hat{H}_{m_1},[[[\hat{H}_{m_4},\hat{H}_{m_3 - m_4}],\hat{H}_{m_2 - m_3}],\hat{H}_{-m_2}]]}{48m_1^2m_2m_3m_4(\hbar\omega)^5}e^{im_1\omega t}+\frac{19[\hat{H}_{m_1},[[\hat{H}_{m_3},[\hat{H}_{m_4},\hat{H}_{m_2 - m_3 - m_4}]],\hat{H}_{-m_2}]]}{720m_1^2m_2m_3m_4(\hbar\omega)^5}e^{im_1\omega t}\\ 
&+\frac{4[\hat{H}_{m_1},[[\hat{H}_{m_2},[\hat{H}_{m_3},\hat{H}_{-m_3}]],\hat{H}_{-m_2}]]}{45m_1^2m_2^2m_3(\hbar\omega)^5}e^{im_1\omega t}+\frac{[\hat{H}_{m_1},[[\hat{H}_{m_2},\hat{H}_0],[\hat{H}_{m_3},\hat{H}_{-m_2 - m_3}]]]}{24m_1^2m_2^2m_3(\hbar\omega)^5}e^{im_1\omega t}\\ 
&+\frac{[\hat{H}_{m_1},[[\hat{H}_{m_3},\hat{H}_{m_2 - m_3}],[\hat{H}_{m_4},\hat{H}_{-m_2 - m_4}]]]}{48m_1^2m_2m_3m_4(\hbar\omega)^5}e^{im_1\omega t}+-\frac{[\hat{H}_{m_2},[\hat{H}_{m_3},[\hat{H}_{m_4},[\hat{H}_{m_5},\hat{H}_{m_1 - m_2 - m_3 - m_4 - m_5}]]]]}{720m_1m_2m_3m_4m_5(\hbar\omega)^5}e^{im_1\omega t} 
\end{align*}
\end{widetext}

\bibliography{bibsupp}